\documentclass[a4paper,11pt]{article}
\pdfoutput=1 
\usepackage{jheppub}
\usepackage[T1]{fontenc} 
\usepackage[all]{xy}
\usepackage{rotating}
\usepackage{float}
\usepackage{tikz}
\usepackage{circuitikz}
\usepackage{tikz-network}
\usepackage{diagbox}
\usepackage{braket}
\usepackage{empheq}
\usepackage{tabularx}
\usepackage{multirow}
\usepackage{slashed}
\usepackage[normalem]{ulem}
\usepackage{mathrsfs}

\usepackage{tikz}
\usepackage{tikz-cd}
\usepackage{diagbox}
\usetikzlibrary{positioning}
\usetikzlibrary{calc}
\usetikzlibrary{decorations.pathreplacing,calligraphy}
\usetikzlibrary{arrows.meta}

\usepackage{xstring}
\usetikzlibrary{decorations.pathmorphing} 
\usetikzlibrary{decorations.markings} 
\usetikzlibrary{arrows} 
\usetikzlibrary{shapes} 
\usetikzlibrary{matrix} 
\usetikzlibrary{positioning} 
\usepackage[english]{babel} 
\usepackage[autostyle]{csquotes}
\usetikzlibrary{positioning,fit}
\usepackage[dvipsnames]{xcolor}
\tikzset{->-/.style={decoration={
  markings,
  mark=at position #1 with {\arrow{>}}},postaction={decorate}}}

\renewcommand{\b}{\overline }
\renewcommand{\d}{\text{d}}

\newcommand{\diag}{\text{diag}}

\newcommand{\be}{\begin{equation}} 
\newcommand{\ee}{\end{equation}} 
\newcommand{\bea}{\begin{equation} \begin{aligned}} \newcommand{\eea}{\end{aligned} \end{equation}}
\newcommand{\bes}{\begin{equation*}}
\newcommand{\ees}{\end{equation*}}

\newcommand{\ov}{\over}

\newcommand{\Tor}{{\rm Tor}}

\newcommand{\nn}{\nonumber}

\newcommand{\Q}{\mathbb{Q}}
\newcommand{\CA}{\mathcal{A}} 
\newcommand{\CB}{\mathcal{B}} 
\newcommand{\CC}{\mathcal{C}}

\newcommand{\CF}{\mathcal{F}} 
\newcommand{\CG}{\mathcal{G}} 
\newcommand{\CH}{\mathcal{H}}
\newcommand{\CI}{\mathcal{I}}
\newcommand{\CJ}{\mathcal{J}}

\newcommand{\CM}{\mathcal{M}}  
\newcommand{\CN}{\mathcal{N}}
\newcommand{\CO}{\mathcal{O}} 
\newcommand{\CP}{\mathcal{P}}

\newcommand{\CR}{\mathcal{R}}
\newcommand{\CS}{\mathcal{S}}
\newcommand{\CT}{\mathcal{T}} 
\newcommand{\CU}{\mathcal{U}}

\newcommand{\CW}{\mathcal{W}}

\newcommand{\Tr}{\text{Tr}}

\newcommand{\Z}{\mathbb{Z}}

\newcommand{\R}{\mathbb{R}}

\newcommand{\half}{{\frac{1}{2}}}

\newcommand{\Fg}{\mathfrak{g}}

\newcommand{\q}{\boldsymbol{q}}
\newcommand{\p}{\boldsymbol{p}}

\newcommand{\SL}{{\mathscr{L}}}

\renewcommand{\l}{\mathsf{l}}
\usetikzlibrary{fit,backgrounds}
\numberwithin{equation}{section}  
\allowdisplaybreaks  
\numberwithin{table}{section}

\usepackage{lscape}

\let\a=\alpha \let\b=\beta \let\g=\gamma \let\d=\delta 
  \let\th=\theta  \let\k=\kappa
\let\l=\lambda \let\m=\mu \let\n=\nu \let\x=\xi 
\let\r=\rho
\let\s=\sigma     
   \let\G=\Gamma \let\D=\Delta  \let\L=\Lambda
  
\let\S=\CC 

\def\diag{\mathop{\rm diag}\nolimits}

\def\a{\alpha}
\def\b{\beta}

\def\CA{{\cal A}}
\def\CB{{\cal B}}
\def\CC{{\cal C}}
\def\CD{{\cal D}}

\def\CF{{\cal F}}
\def\CG{{\cal G}}
\def\CH{{\cal H}}
\def\CI{{\cal I}}
\def\CJ{{\cal J}}

\def\CM{{\cal M}}
\def\CN{{\cal N}}
\def\CO{{\cal O}}
\def\CP{{\cal P}}

\def\CR{{\cal R}}
\def\CS{{\cal S}}
\def\CT{{\cal T}}
\def\CU{{\cal U}}

\def\CW{{\cal W}}

\def\Wa{\widetilde{a}}

\def\Wk{\widetilde{k}}

\def\beq#1\eeq{\begin{align}#1\end{align}}

\makeatletter
\newcommand*{\rom}[1]{\expandafter\romannumeral #1}

\makeatother

\title{Three-dimensional TQFTs 
from Argyres--Douglas theories
via the 3d/3d correspondence 
}

\abstract{We study the twisted dimensional reduction of 4d $\mathcal{N}=2$ superconformal field theories (SCFTs) of Argyres--Douglas type using the 3d/3d correspondence, focussing on the $(A_1, A_{2n})$ theory realised as the class-$\mathcal{S}$ theory $T[\mathcal{C}]$ associated to a curve $\mathcal{C}$ with one irregular puncture. We argue that the resulting 3d $\CN=4$ rank-0 SCFT can be obtained as the infrared fixed point of a Dimofte--Gaiotto--Gukov (DGG) 3d $\CN=2$ abelian Chern--Simons-matter (ACSM) theory $T[M_3^{(k)}]$ associated to the three-manifold  $M_3^{(k)}= L(2n+3, 2k)$, a lens space obtained by fibering $\CC$ over the circle with $k \in \Z_{2n+3}^\times$ the twist parameter. The triangulation of $M_3^{(k)}$ is derived from the 4d BPS spectrum and determines the ACSM theory, including its superpotential. We find that $T[M_3^{(k)}]$ flows to the expected 3d $\CN=4$ SCFT when the monopole superpotential is near-maximal in a precise sense, while the maximal superpotential gives us an ACSM theory $T_A[M_{3}^{(k)}]$ which flows directly to a non-unitary topological quantum field theory (TQFT), the topological $A$-twist of the 3d $\CN=4$ fixed point. Geometrically, the 3d SCFT and TQFT points are related by shifting the location of unavoidable conical singularities between distinct edges of the triangulation of $M_3^{(k)}$. We propose a precise relation between singular edges of the triangulation and SCFT points, which correctly predicts when the 3d $\CN=4$ SCFT is actually a unitary TQFT. 

The 3d TQFTs obtained from $M_3^{(k)}$ can support two-dimensional vertex operator algebras (VOAs) on their holomorphic boundary, including the Schur-sector VOA of the 4d SCFT for $k=1$. 
For $n=k=1$, our construction reproduces the  minimal 3d $\CN=4$ SCFT of Gang and Yamazaki starting from the lens space $L(5,2)$, in which case the TQFT $T_A[M_3^{(k)}]$ is the Yang--Lee TQFT which supports the minimal model $M(2,5)$ on its boundary. We perform detailed checks of our proposals, including by reconstructing the modular structure of the infrared TQFTs from supersymmetric partition functions on Seifert manifolds. We carefully match the phases of the partition functions between supersymmetric and TQFT schemes in order to determine the VOA central charges $c_{\rm 2d}$ mod $8$ from the 3d $\CN=2$ ACSM data, which includes the gravitational Chern--Simons level.

}

\author[a]{Cyril Closset,}
\author[a]{Adam Keyes,}
\author[b]{Sungjoon Kim}

\affiliation[a]{School of Mathematics, University of Birmingham, Watson Building, Edgbaston, Birmingham B15 2TT, UK}
\affiliation[b]{Korea Institute for Advanced Study, 85 Hoegiro, Dongdaemun-Gu, Seoul 02455, Korea}

\emailAdd{c.closset@bham.ac.uk}
\emailAdd{axk1352@student.bham.ac.uk}
\emailAdd{sungjoon@kias.re.kr}

\begin{document} 
\maketitle
\flushbottom



\section{Introduction}

Most if not all supersymmetric quantum field theories (SQFTs) appear to be secretly ruled by M5-branes. Indeed,  the study of partial compactifications of the 6d $\CN=(2,0)$ SCFT of type $\mathfrak{g}$ has led to many deep results about strongly-coupled SQFTs in various space-time dimensions. In this work, we focus on the following three conjectural equalities:
\be\label{equalities intro}
6= 4+2~,\qquad \qquad 6= 3+3~, \qquad\qquad 4-1=2+1~.
\ee
We also focus entirely on the case of two parallel M5-branes --- that is, on compactifications of the $\Fg=A_1$ 6d $(2,0)$ theory. We consider the 6d SCFT on
\be\label{6d background intro}
\CN_3\times S^1 \times \CC~,
\ee
where $\CN_3$ is a ``supersymmetric'' three-manifold and the 6d field theory is topologically twisted along the curve $\CC$. 
The first equality in~\eqref{equalities intro} refers to the class-$\CS$ construction~\cite{Gaiotto:2009we}, wherein the compactification of the 6d theory on the Gaiotto curve $\CC$ defines a 4d SCFT $T[\CC]$ on the four-manifold $\CN_3\times S^1$. The second equality in~\eqref{equalities intro} stands for the 3d/3d correspondence and the Dimofte--Gaiotto--Gukov (DGG) construction~\cite{Dimofte:2011ju, Dimofte:2011py}, in which case a 3d $\CN=2$ supersymmetric abelian gauge theory $T[M_3]$ on $\CN_3$ is constructed from the data of a ``topological'' three-manifold $M_3$ endowed with an ideal triangulation. The last equality in~\eqref{equalities intro} states our expectation that one can study the circle compactification of the 4d SCFT $T[\CC]$ as a DGG theory defined from the product three-manifold $M_3\cong S^1\times \CC$, as is apparent from~\eqref{6d background intro}. More precisely, we will consider a twisted circle compactification on $S^1$, whereby we introduce a non-trivial holonomy for the superconformal $r$-symmetry  $U(1)_r$ along the circle, so that we may consider $M_3^{(k)}\cong S^1_r\times_k \CC$ and obtain the infrared equivalence:
\be\label{4d to 3d intro}
T[M_3^{(k)}] \text{ on } \CN_3\quad  \cong \quad  T[\CC] \text{ on } \CN_3\times S^1_{r}~,
\ee
where  the integer $k$ determines the twist. This approach builds on many recent works on the twisted circle compactification of 4d SCFTs to obtain 3d topological quantum field theories (TQFTs) which admit a 2d chiral algebra --- {\it i.e.}~a vertex operator algebra (VOA) --- at the boundary of 3d space-time~\cite{Dedushenko:2018bpp, Dedushenko:2023cvd,ArabiArdehali:2024ysy,Gaiotto:2024ioj,ArabiArdehali:2024vli,Kim:2024dxu,Go:2025ixu,Kim:2025klh,Kim:2025rog,Nishinaka:2025ytu,Hamachika:2026whv,Yoshida:2026wvb,Jeong:2026dzz}. 
 For $k=1$, the boundary VOA is expected to be precisely the celebrated Beem {\it et al.} VOA~\cite{Beem:2013sza} which captures the Schur sector of the 4d SCFT $T[\CC]$~\cite{Cordova:2015nma,Dedushenko:2018bpp, Dedushenko:2023cvd}. Note that the twisted dimensional reduction preserves all the supersymmetries of the 4d $\CN=2$ SCFT, and gives us a 3d $\CN=4$ SQFT in the infrared.

This paper studies specifically the 4d Argyres--Douglas (AD) theory $\CT_{A_{2n}}$~\cite{Argyres:1995jj, Argyres:1995xn, Eguchi:1996vu} --- also known as the $(A_1, A_{2n})$ AD theory~\cite{Cecotti:2010fi}--- realised as a type-$A_1$ class-$\CS$ theory obtained from a genus-zero Riemann surface $\CC$ with a single irregular puncture~\cite{Gaiotto:2009hg,Xie:2012hs}. The corresponding Schur-sector VOA is the chiral Virasoro minimal model $M(2, 2n+3)$~\cite{Beem:2014cca,Beem:2014rka,Cordova:2015nma,Beem:2017ooy}. We leave an extension of our approach to other Argyres--Douglas theories as an exciting challenge for future work.

The 4d SCFT $\CT_{A_{2n}}$ has no Higgs branch, which simplifies matters significantly. Indeed, its $k=1$ twisted compactification gives rise to a 3d $\CN=4$ rank-0 SCFT~\cite{Gang:2018huc, Gang:2021hrd}, which is a non-trivial fixed point with neither Higgs nor Coulomb branch. In this work, we construct this twisted compactification as a DGG theory $T[M_3^{(k)}]$ for any $k$. In order to do so, we first construct $M_3^{(k)}$ as a triangulated three-manifold by using the mutation and $r$-flow method in a simple BPS chamber of the 4d SCFT~\cite{Cecotti:2010fi, Cecotti:2011iy, Gaiotto:2024ioj}. We will show that this three-manifold is simply a lens space:
\be\label{M3k lens space intro}
M_3^{(k)} \cong L(2n+3, 2k)~,
\ee
where we need to keep track of some unavoidable conical singularity which, morally speaking, is the 3d uplift of the irregular singularity of the Gaiotto curve $\CC$.

Given this understanding of $M_3^{(k)}$, we then use the 3d/3d correspondence and the DGG construction to write down a 3d $\CN=2$ abelian Chern--Simons-matter (ACSM) theory which must flow to a 3d $\CN=4$ SQFT $T[M_3^{(k)}]$. Note that our construction predicts that these 3d $\CN=2$ ACSM theories actually flow to 3d $\CN=4$ supersymmetric theories, since the twisted compactification preserves $8$ supercharges --- we will provide many explicit consistency checks of this supersymmetry enhancement by studying supersymmetric observables. Our construction also naturally gives us a 3d $\CN=2$ UV completion of the non-unitary TQFT $T_A[M_3^{(k)}]$ from the geometry~\eqref{M3k lens space intro}. Indeed, the most naive application of the DGG rules directly lands us on the TQFT point instead of the non-trivial rank-0 SCFT --- this is closely related to many previous observations, see~{\it e.g.}~\cite{Cho:2020ljj,Choi:2022dju,Gang:2024tlp,Bonetti:2024cvq,Gang:2025ykf}. One of the main results of this paper is a new prescription to construct an ACSM theory from the geometry~\eqref{M3k lens space intro} that flows directly to the non-trivial 3d $\CN=4$ fixed point.

Note that, throughout the paper and by a common abuse of notation, we use the notation $T[M_3^{(k)}]$ and $T_A[M_3^{(k)}]$ to denote the 3d SCFTs and TQFTs, respectively, as well as the corresponding 3d $\CN=2$ ACSM theories that are their 3d $\CN=2$ UV completions.

\medskip
\noindent
{\bf Summary of results.} Let us describe in some more details the main results of this paper, focussing on the most novel aspects: 

\begin{itemize}

\item We derive explicitly the lens space geometry~\eqref{M3k lens space intro}  as a triangulated three-manifold obtained by gluing $L=4n|k|$ tetrahedra, corresponding to the simplest BPS chamber of the 4d AD theory with $2n$ BPS particles. The ACSM theory has a $U(1)^L$ gauge group coupled to $L$ chiral multiplets, and it depends also on a choice of polarisation of the tetrahedra in the DGG construction.  It is expected that different choices of triangulation and polarisation of $M_3^{(k)}$ result in ACSM theories which are infrared-dual to each other --- we have verified this in many examples. Compared to previous approaches to study the dimensional reduction to 3d of AD theories, one key advantage of our method is that the monopole superpotential is given as an integral part of the construction.%
\footnote{For $k=1$ and for a specific choice of polarisation of our $4n$ tetrahedra, we obtain the same ACSM theory that was obtained by Gaiotto and Kim~\cite{Gaiotto:2024ioj} using the trace of the monodromy operator in the minimal BPS chamber --- not too surprisingly since our methods are closely related.}

\item We revisit the relationship between the Coulomb branch index of $\CT_{A_{2n}}$ and the $S^3$ partition function of the 3d TQFT $T_A[M_3^{(1)}]$, which was an early indication of the existence of an intermediate 3d TQFT in the SCFT/VOA correspondence~\cite{Fredrickson:2017jcf, Fredrickson:2017yka, Dedushenko:2018bpp}. We pay particular attention to the overall factor in the partition function and we identify a Casimir-like factor to be added to the Coulomb branch  index which depends on the combination $a_{\rm 4d}- c_{\rm 4d}$ of 4d central charges --- see equation~\eqref{CB index with ac}, which should hold more generally for any 4d SCFT. 

\item We find an interesting connection between 3d $\CN=4$ SCFTs and conical singularities of the DGG construction for the lens space~\eqref{M3k lens space intro}. Essentially, while a conical singularity is unavoidable in our triangulation with one vertex and $L+1$ edges, there are various ways to distribute deficit angles throughout the edges.  Besides an `external' edge $C_\infty$ which is the uplift of the regularised boundary of the Gaiotto curve $\CC$, there can exist distinct `internal' edges $C_I$ that are isotopic to a one-cycle $C_0\cong S^1_r$, where the one-cycles $C_\infty$ and $C_0$ are the cores of the two solid tori in the Heegaard decomposition of the lens space. The condition for the edge $C_I$ to be isotopic to $C_0$ is found very concretely in terms of the `chord distance' $s_I$ for this edge projected to $\CC$. The necessary and sufficient condition is $s_I= \pm 2k$ mod $2n+3$. We then find the correspondence:
\be
C_\infty \quad\longleftrightarrow \quad {\rm TQFT} \; T_A[M_3^{(k)}]~, \qquad\quad
C_I \quad\longleftrightarrow \quad {\rm SCFT} \; T[M_3^{(k)}]~,
\ee
depending on whether the conical singularity is entirely at $C_\infty$ or entirely at $C_I\cong C_0$. One can also continuously interpolate between the two constructions, corresponding to the tuning of the $R$-charge in the 3d $\CN=2$ ACSM theory along an `axial' symmetry. Operationally, our DGG theory directly flows to the 3d TQFT $T_A[M_3^{(k)}]$ if we consider the maximal monopole superpotential which breaks all topological symmetries, while the non-trivial SCFT $T[M_3^{(k)}]$ is reached by considering the same ACSM theory with one fewer superpotential term, freeing up the axial symmetry, as dictated by the choice of the internal edge $C_I\cong C_0$. This gives us a clear understanding of the relationship between the SCFT and the $A$-twisted TQFT from the 3d/3d perspective, while also opening up new research directions; for instance, edges $C_I$ which are not isotopic to $C_0$ apparently lead us to interesting 3d $\CN=2$ SCFTs for torus knot complements.%
\footnote{See~\cite{Choi:2022dju,Gang:2024tlp,Gang:2025ykf, Gang:2026iem} for related constructions of class-$\CR$ rank-0 SCFTs, where {\it e.g.}~torus knot complements realise the Virasoro minimal models.}

\item We use the full machinery of supersymmetric localisation for 3d $\CN=2$ gauge theories on Seifert manifolds $\CN_3$~\cite{Closset:2018ghr, Closset:2019hyt} to better investigate the infrared 3d TQFT $T_A[M_3^{(k)}]$ or its boundary VOA by deriving its modular data --- that is, its $S$ and $T$ matrices --- using supersymmetric methods. Compared to much of the previous literature on the subject, we take special care in keeping track of the phases of supersymmetric partition functions, since they are physical observables~\cite{Witten:1988hf, Closset:2012vp}. In particular, this allows us to read off the central charge $c_{\rm 2d}$ of the 3d TQFT directly from $Z_{\CN_3}$; see~\cite{Closset:2025lqt} for a closely related discussion. Here, it is important to note that $c_{\rm 2d}$ is determined from the 3d bulk. Hence, as a 3d TQFT central charge, it is only defined mod $8$. We therefore determine $c_{\rm 2d}$ mod 8 independently of the choice of boundary condition. In general, further work would be needed to fully understand boundary conditions in $T_A[M_3^{(k)}]$ and then fully determine $c_{\rm 2d}$ as a VOA central charge for any specific choice of holomorphic boundary condition.

\end{itemize}

\noindent
Note also that the field theories at fixed $n$ and distinct $k$ are closely related to each other, since they arise from the same underlying 4d SCFT. We study some elementary aspects of the Galois relations between the TQFTs $T_A[M_3^{(k)}]$. We also note that the 3d rank-0 $\CN=4$ SCFTs have a $F$-function which is $k$-independent --- hence they carry the same `number of degrees of freedom'~\cite{Jafferis:2011zi} even though they are distinct SCFTs.

\subsection{\texorpdfstring{$L(5,2)$}{L(5,2)} and the minimal 3d \texorpdfstring{$\CN=4$}{N=4} rank-0 theory}

As an illustration of our methods, let us preview key results in the case of the twisted dimensional reduction of the AD theory  $\CT_{A_2}$ for $k=1$. For this rank-one 4d SCFT, we have the three-manifold
\be\label{L52}
M_3^{(1)}= L(5,2)~,
\ee
and $T[L(5,2)]$ should give us the minimal 3d $\CN=4$ rank-0 theory of Gang and Yamazaki (GY)~\cite{Gang:2018huc,Dedushenko:2023cvd,ArabiArdehali:2024ysy,Gaiotto:2024ioj}, whose topological $A$-twist admits the minimal model $M(2,5)$ on its boundary~\cite{Gang:2023rei,Ferrari:2023fez}, the latter being the VOA associated to $\CT_{A_2}$ by the SCFT/VOA correspondence~\cite{Beem:2013sza,Beem:2014cca,Beem:2014rka,Cordova:2015nma,Beem:2017ooy}. 
In the minimal BPS chamber of $\CT_{A_2}$, there are two BPS particles and two BPS anti-particles, hence we construct a triangulation of~\eqref{L52} with four tetrahedra, which is displayed in figure~\ref{fig: 5gon flip} (section~\ref{sec: 3d 3d}) and results in a $U(1)^4$ ACSM theory with a specific monopole superpotential. This $U(1)^4$ theory is indeed infrared dual to the GY theory, which was originally presented as a $U(1)_{3\ov 2}$ theory coupled to one chiral multiplet. We confirm this duality by standard methods, including by matching the superconformal index. At the 3d A-twisted point, $T_A[M_3^{(1)}]$ gives us a 3d TQFT whose modular data coincide with that of $M(2,5)$. Under certain polarisation and boundary condition choices, the 3d $\CN=2$ half-index of $T_A[M_3^{(1)}]$ reproduces the vacuum character of $M(2,5)$.

\begin{figure}[tbp]
\centering
\begin{tikzpicture}[scale=1.0, rotate=0]
\usetikzlibrary{calc}

\coordinate (P0) at (3,2);
\coordinate (P1) at (0.7,0);
\coordinate (P2) at (4,-0.7);
\coordinate (P3) at (5,1);

\node at (3,2.2) {$v_0$};
\node at (0.69,-0.2) {$v_1$};
\node at (4.1,-0.9) {$v_2$};
\node at (5.2,1) {$v_3$};

\draw[dotted,line width=1pt,line join=round] (P1) -- (P3) ;
\draw[line width=1pt,line join=round] (P0) -- (P1) -- (P2) -- (P3) -- (P0) ;
\draw[line width=1pt,line join=round] (P0) -- (P2) ;

  \coordinate (A) at (2.5,0.3);
  \coordinate (B) at (2.8,1);
  \draw[dashed,red,line width=1pt,<->]  (A) to[ out=190, in=150, distance=2.2cm] (B);
  \node at (2.6,0.3) {\textcolor{red}{$\circ$}};
  \node at (2.9,1) {\textcolor{red}{$\times$}};

  \coordinate (C) at (3.5,0.2);
  \coordinate (D) at (4.2,0.7);
  \draw[dashed,blue,line width=1pt,<->]  (C) to[out=-20, in=10, distance=2.2cm] (D);
  \node at (3.4,0.2) {\textcolor{blue}{$\times$}};
  \node at (4.1,0.7) {\textcolor{blue}{$\circ$}};
  
\end{tikzpicture}
\caption{\label{fig: GY tetra}The lens space $L(5,2)$ can be triangulated with a single tetrahedron whose four faces, denoted by $\circ$ and $\times$ for the front and back ones, respectively, are identified in pairs as \textcolor{red}{$(v_0 v_1 v_2) \sim (v_3 v_0 v_1)$} and \textcolor{blue}{$(v_1 v_2 v_3) \sim (v_3 v_0 v_2)$}, where each triple denotes an ordered 2-simplex.
}
\end{figure}
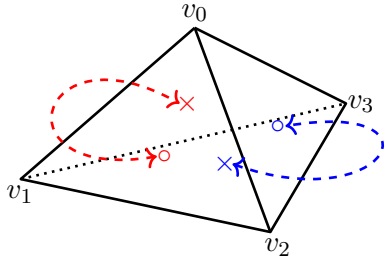

Interestingly, the same three-manifold~\eqref{L52} can be triangulated with a single tetrahedron by the self-gluing depicted in figure \ref{fig: GY tetra}, where the four vertices $v_i$ are identified as a single vertex which essentially corresponds to the irregular singularity of the Gaiotto curve. The resulting DGG theory gives rise to the original 3d $\CN=2$ $U(1)_{3\ov 2}$ gauge theory with a single chiral multiplet of Gang and Yamazaki~\cite{Gang:2018huc}. Our 3d/3d construction therefore provides a simple and elegant M5-brane construction of the minimal 3d $\CN=4$ SCFT~\cite{Gang:2018huc,Gang:2021hrd} in its various duality frames.
 This will be discussed further in~\cite{Kim:2026ghk}.

\subsection{Outlook}
The present work demonstrates that the equalities~\eqref{equalities intro} indeed hold for the simplest class of 4d SCFTs of class-$\CS$ with fractional Coulomb-branch dimensions. The twisted circle reduction indeed gives us a rank-0 3d $\CN=4$ SCFT which is nicely captured by the 3d/3d correspondence:
\be
T[\CC]= \CT_{A_{2n}} \quad \overset{S^1_r}{\longrightarrow} \quad\,\text{rank-0 3d $\CN=4$ SCFT} \, =\, T[M_3^{(k)}]~. 
\ee
The striking statement that those 3d rank-0 theories can be obtained from M5-branes wrapping the lens space~\eqref{M3k lens space intro} deserves further study. The key challenge, at the more conceptual level,  is that the compactification of the 6d SCFT on $M_3^{(k)}$ must involve some singularities which are roughly the three-manifold manifestation of the irregular singularity that appears in the class-$\CS$ description. In principle, one should understand precisely how fibering the Gaiotto curve over the circle implies some natural boundary conditions on $PSL(2, \mathbb{C})$ gauge fields on $M_3^{(k)}$, which would also presumably connect to the wild character variety approach~\cite{Fredrickson:2017jcf,Dedushenko:2018bpp}. While we shall make some preliminary comments on matching the 3d $\CN=2$ Bethe vacua of $T_A[M_3^{(k)}]$ to flat connections on $M_3^{(k)}$, a fuller account of both sides of the 3d/3d correspondence, in the present setup, is left entirely for future work.

Interestingly, there exist simpler triangulations of $M_3^{(k)}$, with fewer tetrahedra than the ones we study in this paper, which do not obviously come from spanning the full BPS spectrum. In particular, there exists a triangulation of $M_3^{(1)}=L(2n+3, 2)$ with $n$ tetrahedra~\cite{Matveev:1990,Jaco:2006,Jaco:2009}, in which case the DGG construction recovers the $U(1)^n$ ACSM theory first discussed by Gang, Kim and Stubbs~\cite{Gang:2023rei}; for $n=1$, this is of course the GY theory obtained from the triangulation of figure~\ref{fig: GY tetra}. These simpler DGG theories will be discussed elsewhere~\cite{Kim:2026ghk}

Many of our computations can be generalised in various directions. One can consider more general SCFTs that have a class-$\CS$ origin from the $A_1$ 6d SCFT, such as the $(A_1, A_{2n+1})$ AD theories. The new ingredient, in all such cases, is the presence of non-trivial flavour symmetries and of a non-trivial Higgs branch of the 4d SCFT, which corresponds to the associated VOA not being lisse~\cite{Beem:2017ooy}. Another important generalisation would be to consider the $A_k$ 6d SCFT which gives rises, for instance, to the AD theories $(A_k, A_l)$.

\medskip
\noindent
This paper is organised as follows. In section~\ref{sec: 3d 3d}, we discuss the circle compactification of the $A_{2n}$ AD theory from the 3d/3d perspective, reviewing known results and presenting our expectations for the modular matrices of the 3d TQFT $T_A[M_3^{(k)}]$. We then construct the three-manifold $M_3^{(k)}$, study its topology, and discuss our main proposal on how to obtain the SCFT $T[M_3^{(k)}]$ from the geometry. In section~\ref{sec: susy partition functions}, we review relevant tools to compute supersymmetric partition functions. In section~\ref{sec: AD to TQFT}, we study the DGG construction of ACSM theories from $M_3^{(k)}$ and we compute many explicit examples that flow to rank-0 SCFTs and/or to 3d TQFTs that admit interesting boundary VOAs. For completeness, useful materials on BPS quivers from the Gaiotto curve are summarised in appendix~\ref{app:Gaiotto curve and quiver}, and a discussion of Galois orbits of the $S$ matrix is given in appendix~\ref{app:galois}.


\section{Circle compactification of the \texorpdfstring{$A_{2n}$}{(A2n)} Argyres--Douglas theory}
\label{sec: 3d 3d}

A class-$\CS$ theory $T[\CC]$ is a 4d $\CN=2$ SQFT obtained by compactifying the 6d $\CN=(2,0)$ SCFT on the Gaiotto curve  $\CC$, a Riemann surface with punctures, from which one can reconstruct the Seiberg--Witten curve as a ramified cover~\cite{Gaiotto:2009we}. For completeness, we review the key aspects of this construction most relevant to us in appendix~\ref{app:Gaiotto curve and quiver}.  In the following, $\CC$ denotes the Riemann sphere with one irregular puncture at $z=\infty$, corresponding to the quadratic differential 
\be\label{phi GC SCFT} 
\phi=  z^{2n+1} dz\otimes dz~.
\ee 
This engineers the $A_{2n}$ Argyres--Douglas theory ---  also called the $(A_1, A_{2n})$ theory --- which we denote by:
\be\label{CTA2n def}
T[\CC]= \CT_{A_{2n}}~.
\ee
In this section, we explore various aspects of the circle compactification of $\CT_{A_{2n}}$ to three dimensions using the 3d/3d correspondence.

\medskip
\noindent
{\bf On the circle with a twist.} 
 Let us consider the 4d $\CN=2$ SCFT compactified on a circle $S^1_r$ with coordinate $\theta\in [0,2\pi)$ and with a non-trivial $U(1)_r$ twist --- that is, we turn on a non-trivial holonomy
\be\label{def t U1r}
\sqrt{\mathfrak{t}} \equiv  \exp\left({i \int_{S^1_r} A^{U(1)_r}}\right) = e^{\pi i k}~,
\ee
where $r$ is the $U(1)_r$ $R$-symmetry and we choose $k\in \Z$ in order to preserve all supersymmetry.%
\footnote{Note that we choose the anti-periodic (NS) spin structure on $S^1_r$ if $k$ is odd, while for $k$ even we must choose the periodic (R) spin structure. The twisted compactification then preserves all the supercharges.} We normalised the $r$-charge so that any Coulomb-branch operator $\CO$ has $U(1)_r$ charge equal to twice its conformal dimension, $\D= {r\ov 2}$. Moving along the $S^1_r$ circle by an angle $\th$ then corresponds to a $U(1)_r$ rotation
\be
\CO \to e^{ i \D k \th}  \CO~.
\ee
If all operators of the SCFT had integer $r$-charges, as in the case of  theories with 4d $\CN=2$ supersymmetric Lagrangian descriptions, the twist would be trivial. In general, however, the Coulomb-branch spectrum is rational --- see~{\it e.g.}~\cite{Caorsi:2018zsq} --- and the twist~\eqref{def t U1r} partially lifts the Coulomb branch. In our case, the theory~\eqref{CTA2n def} has  the Coulomb-branch spectrum:
\be\label{CB spec A2n}
\CS_{\rm CB} = \left\{\Delta= {2\ov 2n+3}(2n+1-s)\right\}_{s=0}^{n-1}~,
\ee
namely  $\D\in \frac{2}{N} \Z$ with $N=2n+3$, and the entire Coulomb branch is lifted by the twisted compactification as long as $N$ and $k$ are coprime.

Our main goal in this section is to construct and understand the three-manifold $M_3^{(k)}$ such that the 3d supersymmetric field theory $T[M_3^{(k)}]$ of the 3d/3d correspondence realises the twisted compactification of the 4d $\CN=2$ SCFT. That is, we expect a relation
\be
T[M_3^{(k)}] \text{ on $\R^3$}\quad \cong\quad  T[\CC]   \text{ on $\R^3\times S^1_r$.} 
\ee
in the deep infrared. While the theory $T[M_3]$ generally only preserves 3d $\CN=2$ supersymmetry, here we should actually recover the full 3d $\CN=4$ supersymmetry in the infrared; we will find much evidence that this is indeed the case in later sections.

\subsection{The Coulomb branch index revisited}\label{subsec:CB rev}
Before turning to the explicit construction of $M_3^{(k)}$, let us first review interesting aspects of the circle compactification and how some 3d TQFTs can appear in the infrared, following the original discussion in~\cite{Dedushenko:2018bpp}. We consider the 6d $\CN=(2,0)$ SCFT of type $A_1$ on 
\be
L(p,-1)\times S^1 \times \CC~,
\ee
with $L(p,-1)\cong S^3/\Z_p$ a lens space. 
The supersymmetric partition function of $T[\CC]$ on $L(p,-1)\times S^1$ computes the lens space index $\CI^{(p)}(\p, \q, t)$~\cite{Benini:2011nc}, a close cousin of the 4d $\CN=2$ superconformal index~\cite{Kinney:2005ej} which one recovers by setting $p=1$. The so-called Coulomb-branch (CB) index $\CI^{(p)}_{\rm CB}(\mathfrak{t})$ is obtained from the lens space index in the limit $\p,\q,t\rightarrow 0$ with fixed $\mathfrak{t}\equiv  \p\q/t$~\cite{Gadde:2011uv, Gukov:2016lki}. For $p=1$, the CB index simply counts the Coulomb-branch operators. It has been argued that the CB index computes some equivariant Verlinde formula for the $PSL(2,\mathbb{C})$ complex Chern--Simons theory at real level $p$~\cite{Gukov:2015sna}:
\be\label{CB equiv Verlinde}
\CI_{\rm CB}^{(p)}(\mathfrak{t}) = {\rm dim}_{\mathfrak{t}} \, \CH[PSL(2,\mathbb{C})_p \; \text{ $\mathbb{C}$CS on } \CC]~.
\ee
On the 4d $\CN=2$ side, the parameter $\mathfrak{t}$ is a fugacity for the $U(1)_r$ $R$-symmetry.%
\footnote{It enters the index as $\mathfrak{t}^{{r\ov 2}-R}$, but we will consider theories where all contributing states satisfy $R=0$.} On the complex CS side, $\mathfrak{t}$ corresponds to a non-trivial grading $\CH\cong \oplus_j \CH_j$ of the infinite-dimensional Hilbert space of the $\mathbb{C}$CS theory on $\CC$, with:
\be
{\rm dim}_{\mathfrak{t}} \, \CH = \sum_{j=0}^\infty \mathfrak{t}^{j\ov 2n+3}\,  {\rm dim}\, \CH_j~.
\ee
For our purpose, one particularly useful way to understand~\eqref{CB equiv Verlinde} is the following. Any 4d $\CN=2$ theory $\CT$ compactified on a circle can be described as a 3d $\CN=4$ supersymmetric field theory viewed as a Kaluza--Klein (KK) theory, denoted by $D_{\rm KK}\CT$. By the 3d/3d correspondence, on the one hand, the compactification of the 6d $(2,0)$ SCFT on the supersymmetric background $L(p,-1)$ is described by the complex CS theory at real level $p$~\cite{Dimofte:2014zga}. On the other hand, the compactification on the three-manifold $S^1\times \CC$ with a topological twist gives us the KK theory,
\be\label{TS1C def KK}
T[S^1\times \CC] = D_{\rm KK}\CT_{A_{2n}}~.
\ee
Note that this $S^1\times \CC$ is not a smooth three manifold --- there is a singular circle corresponding to the irregular singularity on the Gaiotto curve $\CC$, which will become a central point in our story.  
The relation~\eqref{CB equiv Verlinde} corresponds to the equality between the lens-space partition function of the 3d $\CN=4$ KK theory and the $S^1\times \CC$ partition function of the complex CS theory:%
\footnote{The question of whether we have the $SL(2,\mathbb{C})$ or $PSL(2,\mathbb{C})$ gauge group is a slightly subtle one, and it is not essential for the present discussion;  the wild character variety is the same in either case~\protect\cite{Fredrickson:2017yka}. }
\be\label{ZLeqZSCC}
Z_{L(p,-1)}[D_{\rm KK}\CT_{A_{2n}}] = Z_{S^1\times \CC}[PSL(2,\mathbb{C})_p]~.
\ee
The Coulomb branch of the 3d theory~\eqref{TS1C def KK} is a hyperk\"ahler manifold denoted by $\CM_{2, 2n+1}$. It can be described either as a Hitchin moduli space of Higgs bundles over $\CC$~\cite{Gaiotto:2009hg, Fredrickson:2017jcf} or as a wild character variety parameterising $PSL(2,\mathbb{C})$ flat connections on $\CC$ with prescribed boundary conditions at the irregular puncture~\cite{biquard2004wild}. Furthermore, the Hilbert space of the $PSL(2,\mathbb{C})_p$ theory on $\CC$ is obtained, in principle, through a geometric quantisation of $\CM_{2, 2n+1}$ at level $p$~\cite{Gukov:2016lki, Fredrickson:2017yka}. 

The key fact, for our purpose, is that there exists a circle action on~$\CM_{2, 2n+1}$,  generalising the Hitchin action on the Hitchin moduli space of a smooth curve, which is a consequence of the $U(1)_r$ symmetry of the 4d SCFT. The right-hand-side of~\eqref{CB equiv Verlinde} can then be computed using equivariant localisation for this action, with $\mathfrak{t}$ being the equivariant parameter. There are exactly $n+1$ isolated fixed points~\cite{Fredrickson:2017jcf} and one obtains the CB index as a sum over them~\cite{Fredrickson:2017yka}:
\be\label{CBindex expansion}
\CI_{\rm CB}^{(p)}(\mathfrak{t}) = \sum_{\alpha=0}^n \tilde\CF_\alpha(\mathfrak{t})^{-p } \, \tilde{\CH}_\alpha(\mathfrak{t})^{-1}~.
\ee
Here we introduced the quantities:
\bea\label{F and H of t}
&\tilde\CF_\alpha(\mathfrak{t})= \mathfrak{t}^{  \frac{\a(2n+1-\a)}{2(2n+3)}- {n(6n+5)\ov 12(2n+3)}}~,\\
&\tilde{\CH}_\alpha(\mathfrak{t})\equiv   \prod_{l=1}^{n-\alpha}\left(1-\mathfrak{t}^{\frac{2(n+l+1)}{2n+3}}\right)
\left(1-\mathfrak{t}^{-\frac{2l-1}{2n+3}}\right)
\prod_{l=n-\alpha+1}^{n}\left(1-\mathfrak{t}^{\frac{2l+1}{2n+3}}\right)
\left(1-\mathfrak{t}^{\frac{2(n-l+1)}{2n+3}}\right)~,
\eea
which can be computed in terms of the fixed point data~\cite{Fredrickson:2017jcf}. At $p=1$, one can check that this reproduces the ordinary CB index of the AD theory:
\be\label{ICB Deltas}
\CI_{\rm CB}^{(1)}(\mathfrak{t})  = \mathfrak{t}^{-{n\ov 6}\left(\Delta_{\rm max}+{5\ov 2}\right)} \prod_{\Delta\in \CS_{\rm CB}} { \mathfrak{t}^{\Delta\ov 2} \ov 1- \mathfrak{t}^\Delta}~,
\ee
where $\Delta_{\rm max} = 2(2n+1)/(2n+3)$ is the highest CB  scaling dimension and the CB spectrum was given in~\eqref{CB spec A2n}. 
We also note that the ratio $\CI_{\rm CB}^{(p)}(\mathfrak{t})/\CI_{\rm CB}^{(1)}(\mathfrak{t})$ is always a Laurent polynomial in $\mathfrak{t}^{1\ov 2n+3}$. The index~\eqref{ICB Deltas} can also be computed using the 4d $\CN=1$ gauge-theory UV completion of the 4d SCFT~\cite{Maruyoshi:2016aim}. While the CB index is generally normalised so that it equals $1$ at $\mathfrak{t}=0$, we chose a different overall normalisation which we will explain shortly. 

\medskip
\noindent
{\bf Connection to 2d VOA and supersymmetric Casimir energy.}
Strikingly, the expansion~\eqref{CBindex expansion} can be interpreted as a sum over the chiral primaries of the Schur-sector VOA~\cite{Fredrickson:2017jcf, Fredrickson:2017yka}, which is the Virasoro minimal model
\be
{\rm VOA}[\CT_{A_{2n}}]= M(2, 2n+3)~, \qquad c_{\rm 2d}^{M(2,2n+3)}=-12  c_{\rm 4d}~. 
\ee
This is a non-unitary rational conformal field theory (RCFT) with $n+1$ chiral primaries and whose conformal dimensions and central charge are given by:
\be
h_\alpha^{M(2,2n+3)} = -\frac{\a(2n+1-\a)}{2(2n+3)}\quad (\a= 0, \cdots, n), \qquad
c_{\rm 2d}^{M(2,2n+3)}  =1 - \frac{3(1+2n)^2}{3+2n}~,
\ee
with $\a=0$ corresponding to the identity operator. 
The VOA characters --- that is, the torus one-point functions --- read~\cite{DiFrancesco:1997nk}:
\bea
&\chi^{M(2,N)}_{\alpha}(\q) &=&\; \frac{1}{\eta(\tau)} \sum_{m\in\mathbb{Z}}\left( \q^{\frac{(2Nm + N/2 - \alpha - 1)^{2}}{2N}} - \q^{\frac{(2Nm + N/2 + \alpha + 1)^{2}}{2N}} \right)\\
&&=&\; \q^{h_\alpha-{c_{\rm 2d}\ov 24}} \Big(1+ \sum_{l=1}^\infty d_l\, \q^l  \Big)~,
\eea
with $d_l$ some non-negative integers, $\q=e^{2\pi i \tau}$ the torus complex structure, $\eta(\tau)$ the Dedekind function, and $N \equiv 2n+3$. The vacuum character $\chi_0^{M(2,N)}$ reproduces the Schur index of $\CT_{A_{2n}}$~\cite{Cordova:2015nma}, with the Casimir energy prefactor being reproduced by the appropriate limit of the 4d supersymmetric Casimir energy~\cite{Bobev:2015kza}. Under modular transformation of $\tau$ (with $T:\tau\rightarrow \tau+1$ and $S: \tau\rightarrow -1/\tau$), the characters transform in the $(n+1)$-dimensional representation of ${\rm SL(2,\Z)}$ given explicitly by:
\bea\label{ST VOAkeq1}
& T_{\a\b}^{M(2,N)}= \delta_{\a\b}\,   \exp{\left(2\pi i \left(h_\a^{M(2,N)}-{1\ov 24} c_{\rm 2d}^{M(2,N)}\right)\right)}~, \\
&S_{\a\b}^{M(2,N)} =  {2 (-1)^{\a+\b+n}\ov \sqrt{N}} \sin\left({2\pi  (\a+1)(\b+1) \ov N}\right)~,
\eea
Note that the expression for $\tilde\CF_\alpha$ in~\eqref{F and H of t} is given by:
\be\label{tF with h c}
\tilde\CF_\alpha(\mathfrak{t}) = \mathfrak{t}^{-h_\a^{M(2,N)}+{1\ov 24} c_{\rm 2d}^{M(2,N)}}~,
\ee
which is precisely normalised such that:
\be
\tilde\CF_\alpha\big(\mathfrak{t}=e^{2\pi i}\big)^{-1}= T_{\alpha\alpha}^{M(2,N)}~,
\ee
where it is understood that $\mathfrak{t}^{1\ov N}$ becomes $\zeta_N\equiv e^{2\pi i \ov N}$ upon substituting $\mathfrak{t}$ with $e^{2\pi i}$. 
Given the normalisation~\eqref{tF with h c}, we find that the CB index~\eqref{ICB Deltas} takes the simple form:
\be\label{CB index with ac}
\CI_{\rm CB}^{(1)}(\mathfrak{t})  = \mathfrak{t}^{2(a_{\rm 4d}- c_{\rm 4d})}\prod_{\Delta\in \CS_{\rm CB}} { 1 \ov 1- \mathfrak{t}^\Delta}~,
\ee
where the prefactor is written in terms of the 4d conformal anomalies~\cite{Xie:2012hs}:
\be
a_{\rm 4d}= \frac{n (24 n+19)}{24 (2 n+3)}~, \qquad  c_{\rm 4d}=\frac{n (6 n+5)}{6 (2 n+3)}~, \qquad 
2(a_{\rm 4d}- c_{\rm 4d})= -{n\ov 12(2n+3)}~,
\ee
which satisfy~\cite{Shapere:2008zf,Xie:2015rpa}:
\be
2(a_{\rm 4d}- c_{\rm 4d})= \half \left(\sum_{\Delta\in \CS_{\rm CB}}\Delta\right) - {n\ov 6} \Delta_{\rm max} -{5n\ov 12}~.
\ee
We interpret this as a supersymmetric Casimir energy for the Coulomb-branch index, similarly to known results for the full $S^3$ index~\cite{Bobev:2015kza}. Indeed, the factor $\half \Delta$ for each Coulomb-branch operator is the natural zero-point energy contribution, and one can also obtain the precise factor $\mathfrak{t}^{2(a_{\rm 4d}- c_{\rm 4d})}$ as a formal limit of the Casimir energy for the 4d $\CN=2$ superconformal index~\cite{Bobev:2015kza}. We conjecture that, for any 4d $\CN=2$ SCFT, the prefactor in~\eqref{CB index with ac} is the correct form for the supersymmetric Casimir energy contribution to the CB index (up to contributions proportional to the flavour-$U(1)_r$ anomalies in theories with a non-trivial flavour symmetry group).%
\footnote{Unlike the Schur limit $\q=t$, which is an unambiguous locus in fugacity space, the CB limit ($\p,\q,t\rightarrow 0$ at fixed $\mathfrak{t}\equiv  \p\q/t$) is a subtle scaling limit. Taking it naively on the $\CN=2$ supersymmetric Casimir energy derived in~\protect\cite{Bobev:2015kza} leads to an ambiguous limit which depends on the $S^3_b$ squashing parameter $b^2= \log\p/\log\q$~\protect\cite{Imamura:2011uw, Closset:2013vra}. Our proposed Casimir energy for the CB index corresponds to $b^2=-1$. While it would be interesting to understand this Casimir factor better, we consider the intricate relationships to the VOA quantities to be the best evidence for including this prefactor in~\protect\eqref{CB index with ac}. We checked those relationships for many other AD theories as well.}

\subsection{Twisted compactification and 3d TQFT partition functions}\label{subsec:twisted comp}
Let us now consider the 4d SCFT compactified on $S^1$ with the $U(1)_r$ twist~\eqref{def t U1r}, which fully preserves supersymmetry. In the deep infrared, we must then obtain some 3d $\CN=4$ SQFT which is either a non-trivial fixed point --- that is, a rank-0 3d $\CN=4$ SCFT --- or a unitary 3d TQFT; the latter is trivially supersymmetric, and it must be unitary because the compactification preserves reflection positivity in 3d. We will denote the infrared 3d SQFT by $T[M_3^{(k)}]$, with $M_3 = \CC\times_k S^1_r$. 

Any 3d $\CN=4$ SQFT can be topologically twisted using some $SU(2)$ subgroup of the $SU(2)_C\times SU(2)_H$ $R$-symmetry acting on the Coulomb and Higgs branch, respectively --- in the case at hand, these are related to the 4d $R$-symmetry as 
\be
U(1)_r\subset SU(2)_C~,\qquad 
SU(2)_R = SU(2)_H~.
\ee
We are mainly interested in the 3d $\CN=4$ $A$-twist of $T[M_3^{(k)}]$ --- that is, the topological twist of the 3d spin by $SU(2)_H$, which gives us a 3d TQFT denoted by $T_A[M_3^{(k)}]$. Note that $T_A[M_3^{(k)}]= T[M_3^{(k)}]$ if $T[M_3^{(k)}]$ is already a unitary TQFT, as the topological twist is then a trivial operation.%
\footnote{From now on, unless otherwise stated, $T[M_3^{(k)}]$ denotes a non-trivial SCFT while $T_A[M_3^{(k)}]$ is a 3d TQFT, unitary or non-unitary. The reason for this convention will become clearer soon. } 
By contrast, when $T[M_3^{(k)}]$ is a non-trivial rank-0 SCFT, one expects that $T_A[M_3^{(k)}]$ is a {\it non-unitary} TQFT, essentially because the topological twist does not preserve reflection positivity.  Supersymmetric partition functions of either $T[M_3^{(k)}]$ or $T_A[M_3^{(k)}]$ can be computed as a sum over Bethe vacua in the $A$-model formalism, as we will review further in section~\ref{sec: susy partition functions}.

\medskip
\noindent {\bf Lens space partition function and modular matrices.}  Focusing on the 3d TQFT $T_A[M_3^{(k)}]$ for now, its lens space partition function takes the form:
\be
Z_{L(p,-1)}\big[T_A[M_3^{(k)}]\big] = \sum_{\alpha=0}^n \CF_\alpha^{-p}  \CH_\alpha^{-1} =  \sum_{\alpha=0}^n \left(T_{\a\a}^{(k)}\right)^{p}\left(S_{0\a}^{(k)}\right)^{2} ~,
\ee
where $\alpha$ indexes the 3d Bethe vacua; in the second equality, we used the relations $\CH_\alpha = S_{0\alpha}^{-2}$ and $\CF_\alpha^{-1}= T_{\alpha\alpha}$ between the 3d $A$-model quantities and the modular $S$ and $T$ matrices of the 3d TQFT --- see~{\it e.g.}~\cite{Closset:2025lqt} and section~\ref{sec: susy partition functions} below. This partition function can be obtained from the lens space index~\eqref{CBindex expansion} by the substitution $\mathfrak{t}=e^{2\pi i k}$. More precisely, we claim that 
\be\label{lens space Z from CB}
Z_{L(p,-1)}\big[T_A[M_3^{(k)}]\big] = e^{-{\pi i \ov 6} p n(k-1)}\, \CI_{\rm CB}^{(p)}(e^{2\pi i k})~,
\ee
which is equivalent to the identities:
\be\label{F and H identities gen k}
\CF_\alpha[T_A[M_3^{(k)}]] = e^{{\pi i \ov 6} n(k-1)} \tilde\CF_\alpha(e^{2\pi i k})~,\qquad
\CH_\alpha[T_A[M_3^{(k)}]] = \tilde\CH_\alpha(e^{2\pi i k})~,
\ee
with $\tilde\CF_\alpha(\mathfrak{t})$ and  $\tilde\CH_\alpha(\mathfrak{t})$ given in~\eqref{F and H of t}. 
The prefactor $ e^{{\pi i \ov 6} n(k-1)}$ in the 3d fibering operator can be understood as a contribution $\Delta K_g=2n(k-1)$ to the gravitational Chern--Simons level $K_g$~\cite{Closset:2012vg, Closset:2017zgf}, to be discussed momentarily.

For $k=1$, the $S$ and $T$ matrices are given in~\eqref{ST VOAkeq1}. The 3d TQFTs for distinct values of $k$ coprime to $N=2n+3$ can be obtained as Galois conjugate TQFTs; in particular, the modular matrices are given by:
\bea\label{ST VOAk}
& T_{\a\b}^{(k)}= \delta_{\a\b}\,   \exp{\left(2\pi i \left(h_\a(k)-{c_{\rm 2d}(k)\ov 24}\right)\right)}~, \\
&S_{\a\b}^{(k)} =  (-1)^{n(k-1)} \genfrac{(}{)}{}{}{k}{2n+3} \, {2 (-1)^{\a+\b+n}\ov \sqrt{2n+3}} \sin\left({2\pi k (\a+1)(\b+1) \ov 2n+3}\right)~,
\eea
where $\genfrac{(}{)}{}{}{k}{N}\in \{\pm 1\}$ denotes the Jacobi symbol, and we have introduced the quantities:
\be
h_\a(k)=  k\, h_\alpha^{M(2,2n+3)}\quad ({\rm mod}\; 1)~,
\qquad 
c_{\rm 2d}(k) = k\, c_{\rm 2d}^{M(2,2n+3)}  + 2n (k-1) \quad ({\rm mod}\; 8)~.
\label{eq: conj c2d}
\ee
 By direct computation using the orthogonality properties of the sine function, one can check that the matrices~\eqref{ST VOAk} satisfy the $SL(2,\Z)$ relations
\be\label{ST rels gen k}
(S^{(k)})^2= \mathbf{1}~, \qquad\quad (S^{(k)}T^{(k)})^3= \mathbf{1}~,
\ee
for every allowed choice of $k$. 
The contribution  $\Delta K_g =2n (k-1)$ to $c_{\rm 2d}(k)$ (mod 8) is necessary for the modular group relations~\eqref{ST rels gen k} to hold, and we will present finer checks of this 2d central charge in later sections. For any rank-$n$ 4d SCFT, this same formula 
\be\label{c2d MTC}
c_{\rm 2d}(k)= -12 k\, c_{\rm 4d}+ 2 n (k-1)
\ee
 was first conjectured in~\cite{Kim:2024dxu} based on explicit computations with the 4d quantum monodromy operator. Note that the MTC central charge~\eqref{c2d MTC} is only defined mod $8$, which leaves us with an ambiguity in identifying the boundary central charge from the 3d bulk data alone. As we show in appendix~\ref{app:galois}, the $S^{(k)}$ matrices~\eqref{ST VOAk} can be understood as spanning the Galois orbit of the $k=1$ $S$ matrix. In particular, one can check that
\be\label{Gk appearing}
S^{(k)} = G_{(k)} S^{(1)}~,
\ee
with $G_{(k)}$ a permutation matrix given in~\eqref{Gk explicit}. We easily see that the first equality in~\eqref{F and H identities gen k} holds, while the second equality is equivalent to
\be\label{identity S to H k}
\tilde\CH_\alpha(e^{2\pi ik}) = \left(S_{0\a}^{(k)}\right)^{-2}~.
\ee
Using the notation $\omega_k=e^{2\pi i  k\ov N}$ for that primitive $N$-th root of unity, we indeed have
\be\nn
\tilde\CH_\alpha(e^{2\pi ik}) 
     = \prod_{\substack{l=1\\ l\neq 2\alpha+2, N-2\alpha-2}}^{N-1} (1- \omega_k^{l}) =\frac{N}{(1-\omega_k^{2(\a+1)})(1-\omega_k^{-2(\a+1)})}= {N\ov 4 \sin^2\left(2\pi k{\a+1\ov N}\right)}~,
\ee
where we used the simple identity $\prod_{l=1}^{N-1}(1-\omega_k^l)= N$. This proves~\eqref{identity S to H k}.

\medskip
\noindent
{\bf Quasi-periodicity of the modular matrices under $k\rightarrow k+N$.}  Since our 3d theories are obtained from twisted dimensional reduction with the $U(1)_r$ twist~\eqref{def t U1r}, one would expect to find a periodicity under the shift $k\rightarrow k+N$, with $N=2n+3$ the common denominator of the Coulomb-branch dimensions~\eqref{CB spec A2n}. We first note that
\be
h_\alpha(k+N)= h_\alpha(k) \; ({\rm mod}\; 1)~, \qquad\quad  c_{\rm 2d}(k+N) = c_{\rm 2d}(k) - 4n(2n+1)~.
\ee 
Since we can always shift the central charge $c_{\rm 2d}$ by a shift of the gravitational Chern--Simons level, 
\be\label{c and K shift}
c_{\rm 2d}\rightarrow c_{\rm 2d}+ \Delta K_g~, \qquad\quad K_g \rightarrow K_g+\Delta K_g~,
\ee
we see that the $T$ matrix~\eqref{ST VOAk} is periodic up to such a 3d counterterm $e^{-{\pi i \Delta K_g\ov 12}}$. For $n$ even, it follows from the discussion in appendix~\ref{app:cyclotimic} that the $S$ matrix is also periodic, while it incurs an overall sign for $n$ odd. In summary: 
\be
T_{\a\b}^{(k+N)}= e^{{\pi i n(2n+1) \ov 3}} T_{\a\b}^{(k)}~,\qquad 
S_{\a\b}^{(k+N)}=(-1)^n S_{\a\b}^{(k)}~.
\ee
It would be highly desirable to have an understanding of these `anomalous' transformations from the 4d physics.

\medskip
\noindent
{\bf Unitary TQFTs and $F$-function.} The modular tensor categories (MTC) with $S$ matrices~\eqref{ST VOAk} have the quantum dimensions
\be
d_\a \equiv {S_{0\a}\ov S_{00}} = (-1)^\a \sin\left({2\pi k (\a+1)\ov 2n+3}\right)\Big/\sin\left({2\pi k\ov 2n+3}\right)~.
\ee
A necessary condition for having a unitary TQFT is that $d_\a\geq 1$ --- it is also likely to be a sufficient condition, see~{\it e.g}~\cite{Buican:2021axn} and references therein for a more detailed discussion. We then find that $T_A[M_3^{(k)}]$ must be a non-unitary TQFT except in the special cases $k=n+1$ or $k=n+2$ --- that is, $k=\pm(n+1)$ mod $N$. In the latter case, we have 
\be\label{da unitaryTQFT}
d_\alpha=  \sin\left({\pi  (\a+1)\ov 2n+3}\right)\Big/\sin\left({\pi \ov 2n+3}\right) \geq 1 \quad \forall \a~, \qquad k=\pm(n+1)~,
\ee
 and we apparently find a unitary TQFT. We then also expect that $T_A[M_3^{(k)}]=T[M_3^{(k)}]$ for $k=\pm(n+1)$. 
 We can also compute the $F$-function of these TQFT~\cite{Jafferis:2011zi}, which reads:
 \be
 F= -\log \left|Z_{S^3}\right| = -\log \left|\sum_\a S^2_{0\a}T_{\a\a} \right|= -\log \left| S_{00}\right| = -\log  \left| \frac{2 \sin \left(\frac{2 \pi  k}{2 n+3}\right)}{\sqrt{2 n+3}}\right|~. 
 \ee
 Note that it depends on $k$. However, for $k \neq \pm (n+1)$ mod $N$, we are considering the 3-sphere partition function of a non-unitary TQFT and therefore $F$ cannot be straightforwardly interpreted as a measure of the degrees of freedom which always decreases under RG flow. For the unitary TQFT at $k=\pm(n+1)$, on the other hand, we find:
 \be\label{Fscft}
 F_{\rm SCFT}  = - \log\left({2\sin(\pi/N)\ov \sqrt{N}}\right)~.
 \ee
 As we will show in section~\ref{sec: AD to TQFT}, this is actually the value of $F$ for the 3d $\CN=4$ rank-0 SCFT $T[M_3^{(k)}]$, for any $k$, in agreement with previous discussions~\cite{Gang:2021hrd}. In particular, for $n=1$, one finds $F_{\rm SCFT}=-\frac{1}{2} \log \left(\frac{1}{10} \left(5-\sqrt{5}\right)\right)$ for the Gang--Yamazaki SCFT~\cite{Gang:2018huc, Gang:2021hrd}.

\subsection{From polygons to lens spaces: constructing \texorpdfstring{$M_3^{(k)}$}{M3k} from tetrahedra}\label{subsec:M3 from Delta}

We now turn to the question of constructing $M_3^{(k)}$ as a triangulated three-manifold, to which one can associate a 3d $\CN=2$ abelian Chern--Simons matter theory of DGG type~\cite{Dimofte:2011ju} that flows to the 3d $\CN=4$  rank-0  theory $T[M_3^{(k)}]$ in the infrared. We first review some general aspects of the BPS states of the 4d SCFT $T[\CC]$, and then construct the three-manifold $M_3^{(k)} \cong S_r^1  \times_k \CC$ by the method of $U(1)_r$ twisting~\cite{Alim:2011ae, Cecotti:2011iy, Gaiotto:2024ioj}.

\medskip
\noindent
{\bf BPS quivers and mutations.}
 The BPS states of the 4d SCFT are famously encoded in BPS quivers~\cite{Alim:2011ae}. In appendix~\ref{app:Gaiotto curve and quiver}, we review how one can associate a BPS quiver to each allowed triangulation of the Gaiotto curve $\CC$~\cite{Gaiotto:2009hg}. Distinct triangulations correspond to distinct BPS quivers which are, nevertheless, mutation equivalent~\cite{Fomin:2007rcq}. In particular, any flip of the triangulation corresponds to a mutation of the BPS quiver --- see figure~\ref{fig: flip}. 

\begin{figure}[tbp]
\centering
\begin{tikzpicture}[scale=1, rotate=0]
\usetikzlibrary{calc}

    \coordinate (P1) at (1,1);
    \coordinate (P2) at (-1,1);
    \coordinate (P3) at (-1,-1);
    \coordinate (P4) at (1,-1);

    \coordinate (Q1) at (4+1,1);
    \coordinate (Q2) at (4-1,1);
    \coordinate (Q3) at (4-1,-1);
    \coordinate (Q4) at (4+1,-1);

\draw[black, line width=1pt] (P2) -- (P4);
\draw[black, line width=1pt] (P1) -- (P2) -- (P3) -- (P4) -- (P1);

\draw[black, line width=1pt] (Q1) -- (Q3);
\draw[black, line width=1pt] (Q1) -- (Q2) -- (Q3) -- (Q4) -- (Q1);

\node at (2,0) {$\overset{\text{flip}}{\longleftrightarrow}$};

\end{tikzpicture}
\caption{\label{fig: flip} A flip of an internal edge in an ideal triangulation of a Gaiotto curve $\CC$  corresponds to a quiver mutation at the node associated to the edge being flipped, as discussed in appendix~\protect\ref{app:Gaiotto curve and quiver}. }
\end{figure}
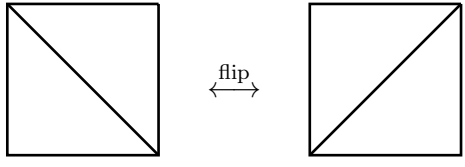
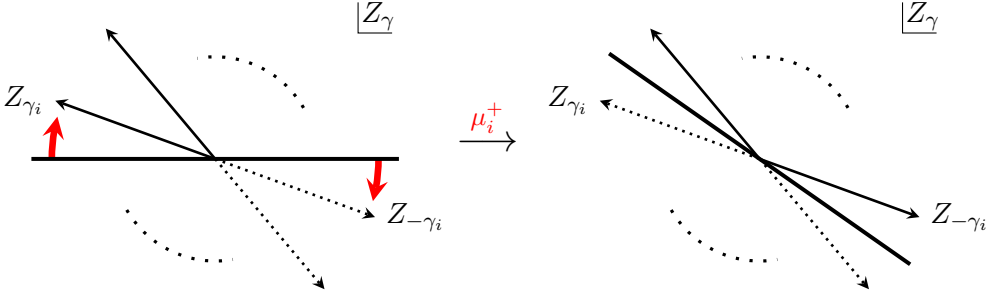
\begin{figure}[tbp]
    \centering
    \begin{tikzpicture}[>=stealth, scale=0.9]

\draw[->, red, line width=2.5pt] (2.4,0) 
  arc[start angle=0, end angle=-15, radius=2.4];

\draw[->, red, line width=2.5pt] (-2.4,0) 
  arc[start angle=180, end angle=165, radius=2.4];
\draw[line width=1.5pt] (-2.7,0) -- (2.7,0);

\coordinate (O) at (0,0);
\draw[->, line width=1pt] (O) -- (160:2.5) node[left] {$Z_{\g_i}$};
\draw[->, line width=1pt] (O) -- (130:2.5) ;

\draw[->, line width=1pt, dotted] (O) -- (-20:2.5) node[right] {$Z_{-\g_i}$};
\draw[->, line width=1pt, dotted] (O) -- (-50:2.5) ;

\draw[loosely dotted, line width=1.2pt] (30:1.5) 
  arc[start angle=30, end angle=100, radius=1.5];

\draw[loosely dotted, line width=1.2pt] (210:1.5) 
  arc[start angle=210, end angle=280, radius=1.5];

\draw (2.1,2.3) -- (2.1,1.85) -- (2.6,1.85);
\node at (2.4,2.1) {$Z_{\gamma}$};

\node at (4,0.5) {\Large $\overset{\textcolor{red}{\mu_i^+}}{\longrightarrow}$};

\begin{scope}[shift={(8,0)}]
\draw[line width=1.5pt,rotate=-35] (-2.7,0) -- (2.7,0);
\coordinate (O) at (0,0);
\draw[->, line width=1pt,dotted] (O) -- (160:2.5) node[left] {$Z_{\g_i}$};
\draw[->, line width=1pt] (O) -- (130:2.5) ;

\draw[->, line width=1pt] (O) -- (-20:2.5) node[right] {$Z_{-\g_i}$};
\draw[->, line width=1pt, dotted] (O) -- (-50:2.5) ;

\draw[loosely dotted, line width=1.2pt] (30:1.5) 
  arc[start angle=30, end angle=100, radius=1.5];

\draw[loosely dotted, line width=1.2pt] (210:1.5) 
  arc[start angle=210, end angle=280, radius=1.5];

\draw (2.1,2.3) -- (2.1,1.85) -- (2.6,1.85);
\node at (2.4,2.1) {$Z_{\gamma}$};
\end{scope}

\end{tikzpicture}
    \caption{\label{fig: cone mutation} The right mutation $\mu_i^+$ as a rotation of the upper-half $Z$-plane. When the plane rotates clockwise, the left boundary of the cone of the BPS particles, $Z_{\g_i}$, escapes the plane, while its CPT conjugate, $Z_{-\g_{i}}$ enters, becoming a new right boundary of the cone. The inverse left mutation similarly corresponds to an anti-clockwise rotation.}
\end{figure}

The AD theory $\CT_{A_{2n}}$ has a very simple BPS structure. In every BPS chamber on the Coulomb branch, the BPS spectrum consists of a finite number of  hypermultiplets. This  `complete' SQFT is then most amenable to the mutation method~\cite{Alim:2011ae}, wherein rotating the central-charge plane by $2\pi$ induces a finite sequence of $2P$ quiver mutations. Here $P$ is the total number of BPS particles in a fixed BPS chamber, and the sequence of mutations is equivalently described as a sequence of $2P$ flips of the triangulation of $\CC$. Note that the BPS spectrum depends on the fixed Coulomb branch location, which in turn fixes the stability parameters (FI parameters) of the BPS quiver; in a given chamber, only a subset of the full set of BPS quivers in the full mutation class are reachable by the mutation method, which selects a very specific sequence of mutations. In the following, we will focus on the simplest BPS chamber with $P= 2n$, but the same computations can be carried out in any finite chamber.

\medskip
\noindent
{\bf $U(1)_r$ twist from $Z$ to $\CC$.} The key observation is that the central charge $Z$ has charge $2$ under the superconformal $U(1)_r$ symmetry, while the local coordinate $z$ on $\CC$ that appears in~\eqref{phi GC SCFT} has  charge $r[z]= 4/N$. The two quantities are closely related by the Seiberg--Witten geometry: 
\be
Z_\gamma = \int_\gamma \sqrt{\phi}~.
\ee
The twist~\eqref{def t U1r} then implies that the central charges and the Gaiotto curve rotate as
\begin{align}
    Z_\g \; \to \; e^{i k \theta} Z_\g~, \qquad\qquad z\; \to \;  e^{i {2k \theta\ov N}}  z~,
\end{align}
respectively, as we move along $S_r^1$ by an angle $\theta$. If we then declare that all the vectors $Z_{\gamma}$ in the upper-half $Z$-plane correspond to the BPS particles --- and thus $Z_{-\gamma}$ in the lower-half plane are the anti-particles ---, rotating the $Z$-plane by $2\pi k$ in the anti-clockwise direction (where $k<0$ corresponds to a $2\pi |k|$ in the clockwise direction) gives us a series of $4 n|k|$ quiver mutations. Here, the assumption that we are in a finite BPS chamber is of course crucial; for $\CT_{A_{2n}}$ all BPS chambers are finite.

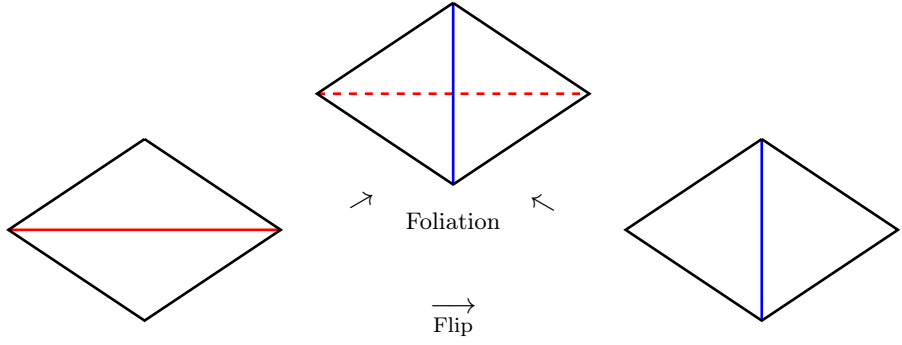
\begin{figure}[tbp]
\centering
\begin{tikzpicture}[scale=1.2, rotate=0]
\usetikzlibrary{calc}

    \coordinate (P1) at (0,1);
    \coordinate (P2) at (0,-1);
    \coordinate (P3) at (-1.5,0);
    \coordinate (P4) at (1.5,0);

    \coordinate (Q1) at (-3.4,1-1.5);
    \coordinate (Q2) at (-3.4,-1-1.5);
    \coordinate (Q3) at (-3.4-1.5,0-1.5);
    \coordinate (Q4) at (-3.4+1.5,0-1.5);

    \coordinate (R1) at (3.4+0,1-1.5);
    \coordinate (R2) at (3.4+0,-1-1.5);
    \coordinate (R3) at (3.4+-1.5,0-1.5);
    \coordinate (R4) at (3.4+1.5,0-1.5);

\draw[red, line width=1pt, dashed] (P3) -- (P4);
\draw[blue, line width=1pt] (P1) -- (P2);
\draw[black, line width=1pt] (P1) -- (P3) -- (P2) -- (P4) -- (P1);

\draw[red, line width=1pt] (Q3) -- (Q4);
\draw[black, line width=1pt] (Q1) -- (Q3) -- (Q2) -- (Q4) -- (Q1);

\draw[blue, line width=1pt] (R1) -- (R2);
\draw[black, line width=1pt] (R1) -- (R3) -- (R2) -- (R4) -- (R1);

\node at (0,-2.5) {$\underset{\text{Flip}}{\longrightarrow}$};
\node[rotate=30] at (-1,-1.2) {$\to$};
\node[rotate=150] at (1,-1.2) {$\to$};
\node at (0,-1.4) {\footnotesize Foliation};

\end{tikzpicture}
\caption{\label{fig: foliation} Each flip of an internal edge in the ideal triangulation can be interpreted as a formation of a tetrahedron. The internal edges for the flip are colored as red and blue, which are colored the same for the corresponding edges in the tetrahedron. }
\end{figure}

\medskip
\noindent
{\bf Constructing \texorpdfstring{$M_3^{(k)}$}{M3k} from tetrahedra.} 
Given a series of mutations realising the $U(1)_r$ twist on the $Z$-plane, we have the corresponding series of triangulations of the Gaiotto curve $\CC$ related by flips. Recall that $\CC$ is realised as a $(2n+3)$-sided polygon, where the boundary polygon is the ideal circle boundary of the irregular singularity at $z=\infty$; see appendix~\ref{app:Gaiotto curve and quiver}. We wish to construct an explicit 3d triangulation of the twisted product
\be\label{M3 foliated}
M_3^{(k)} \cong S^1_r\times_k \CC
\ee
associated to this series of flips. The idea is to foliate $M_3^{(k)}$ along the $S^1_r$ direction, where successive flips correspond to the formation of tetrahedra~\cite{Cecotti:2011iy}; see figure~\ref{fig: foliation}. One can then glue $2P|k|$ tetrahedra together one at a time following the periodic mutation sequence. 

For definiteness, let us choose the minimal chamber, with $P=2n$ BPS particles, corresponding to the following ideal triangulation of $\CC$~\cite{Gaiotto:2009hg,Cecotti:2011iy}:
\begin{equation}
 \vcenter{\hbox{%
 \begin{tikzpicture}[scale=2, rotate=90]
\usetikzlibrary{calc}
\foreach \i in {1,...,11} {
    \coordinate (P\i) at ({cos(360/11*(\i-1))},{sin(360/11*(\i-1))});
}

\draw[line width=1pt] (P1)
\foreach \i in {2,...,11} { -- (P\i) } -- cycle;

\draw[line width=1pt,line join=round] (P3) -- (P5) -- (P2) -- (P6) -- (P1) -- (P7) -- (P11) -- (P8) -- (P10);

\node at (0,1.7) {\Large$\CC\;\;\cong$};
\node at (-0.3,0) {$\cdots$};

\node at (-0.15,0.93) {\scalebox{0.5}{ $\g_1$}};
\node at (-0.14,-0.9) {\scalebox{0.5}{ $\g_{2n}$}};

\node at (-0.15,0.5) {\scalebox{0.5}{ $\g_{3}$}};
\node at (-0.3,-0.52) {\scalebox{0.5}{ $\g_{2n\texttt{-}2}$}};

\node at (0.25,0.73) {\scalebox{0.5}{ $\g_{2}$}};
\node at (0.25,-0.75) {\scalebox{0.5}{ $\g_{2n\texttt{-}1}$}};

\node at (0.25,0.23) {\scalebox{0.5}{ $\g_{4}$}};
\node at (0.25,-0.24) {\scalebox{0.5}{ $\g_{2n\texttt{-}3}$}};

\end{tikzpicture}
    \label{eq: (2n+3)-gon}}}
\end{equation}
Here,  the $2n$ charges $\g_i$ for the  stable hypermultiplets are associated to the $2n$ internal edges, and one obtains the BPS quiver:
\begin{equation}
    \vcenter{\hbox{%
    \begin{tikzpicture}[>=stealth]

\def\d{1.5}

\node[circle, draw, minimum size=4mm, line width=1pt] (g1) at (0,0) {};
\node[circle, draw, minimum size=4mm, line width=1pt] (g2) at (\d,0) {};
\node[circle, draw, minimum size=4mm, line width=1pt] (g3) at (2*\d,0) {};

\node at (3*\d,0) (dots) {$\cdots$};

\node[circle, draw, minimum size=4mm, line width=1pt] (g2n1) at (4*\d,0) {};
\node[circle, draw, minimum size=4mm, line width=1pt] (g2n) at (5*\d,0) {};

\node at ($(g1)+(0,6mm)$) {$\gamma_1$};
\node at ($(g2)+(0,6mm)$) {$\gamma_2$};
\node at ($(g3)+(0,6mm)$) {$\gamma_3$};
\node at ($(g2n1)+(0,6mm)$) {$\gamma_{2n-1}$};
\node at ($(g2n)+(0,6mm)$) {$\gamma_{2n}$};

\draw[->, line width=1pt] (g1) -- (g2);
\draw[<-, line width=1pt] (g2) -- (g3);
\draw[->, line width=1pt] (g3) -- (dots);

\draw[<-, line width=1pt] (dots) -- (g2n1);
\draw[->, line width=1pt] (g2n1) -- (g2n);

\end{tikzpicture}}}
\end{equation}
Since the $A_{2n}$ theory is a complete 4d $\CN=2$ SQFT, we can order the central charges at will~\cite{Alim:2011ae}. In the minimal chamber,  the central charges $Z_{\g_i} \in \mathbb{C}$ are arranged in the upper-half $Z$-plane as shown in figure~\ref{fig: central charge}, with the phases for $\gamma_j$ with $j$ even larger than those for $j$ odd; the relative order among the odd- or even-labeled central charges is irrelevant since their mutual Dirac pairings are trivial. 
\begin{figure}[tbp]
\centering
\begin{tikzpicture}[>=stealth, scale=0.9]
\draw[line width=1.5pt] (-4,0) -- (4,0);
\coordinate (O) at (0,0);
\draw[->, thick] (O) -- (160:3) node[left] {$Z_{\gamma_{2n}}$};
\draw[->, thick] (O) -- (130:3) node[left] {$Z_{\gamma_2}$};

\draw[->, thick] (O) -- (20:3) node[right] {$Z_{\gamma_{1}}$};
\draw[->, thick] (O) -- (50:3) node[right] {$Z_{\gamma_{2n-1}}$};

\draw[->, thick, dotted] (O) -- (-20:3) node[right] {$Z_{-\gamma_{2n}}$};
\draw[->, thick, dotted] (O) -- (-50:3) node[right] {$Z_{-\gamma_{2}}$};

\draw[->, thick, dotted] (O) -- (-160:3) node[left] {$Z_{-\gamma_{1}}$};
\draw[->, thick, dotted] (O) -- (-130:3) node[left] {$Z_{-\gamma_{2n-1}}$};

\draw[dotted, line width=1.2pt] (137:2.0) arc[start angle=137, end angle=153, radius=2.0];
\node at (147:2.4) {\footnotesize even};

\draw[dotted, line width=1.2pt] (27:2.0) arc[start angle=27, end angle=43, radius=2.0];
\node at (33:2.4) {\footnotesize odd};

\draw[dotted, line width=1.2pt] (-43:2.0) arc[start angle=-43, end angle=-27, radius=2.0];
\node at (-33:2.4) {\footnotesize even};

\draw[dotted, line width=1.2pt] (-153:2.0) arc[start angle=-153, end angle=-137, radius=2.0];
\node at (-147:2.4) {\footnotesize odd};

\draw (3.5,2.75) -- (3.5,2.3) -- (4,2.3);
\node at (3.8,2.54) {$Z_{\gamma}$};

\end{tikzpicture}
\caption{\label{fig: central charge} Central charges $Z_{\g_i}$ on the $Z$-plane for the $A_{2n}$ theory in the minimal chamber. The BPS particles and anti-particles are denoted by solid and dashed arrows, respectively. Here, the phases of $\text{arg}(Z_{\g_i})$ for $i$ odd are chosen to be smaller than those of $\text{arg}(Z_{\g_j})$ for  $j$ even. }
\end{figure}
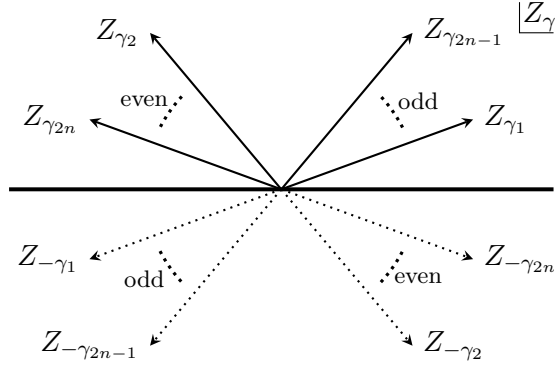

As we go around $S_r^1$ once, the central charge plane is rotated by an angle $2\pi k$. In particular, it is rotated once anti-clockwise for $k=1$, which amounts to a full sequence of mutations in the BPS quiver:
\bea
    &{\g_{2n}
    \to
    \cdots
    \to
    \g_{2}
    \to
    \;\;\g_{2n-1}
    \to
    \cdots
    \to
    \g_{1}
    \to}
    \\
    &{\;\;-\g_{2n}
    \to
    \cdots
    \to
    -\g_{2}
    \to
    \;\;-\g_{2n-1}
    \to
    \cdots
    \to
    -\g_{1}
    \,.}
    \label{eq: A2n mutation}
\eea
The corresponding $4n$ successive flips of the internal edges in the triangulation~\eqref{eq: (2n+3)-gon} produce $4n$ sheets of triangulated Gaiotto curves foliated along the $S_r^1$ direction. Each of the transitions between two consecutive planar triangulations can be represented by a tetrahedron, such that a pair of diagonal edges of the tetrahedron are identified with the edges {\it before} and {\it after} the flip, respectively, as described in figure~\ref{fig: foliation}. The successive tetrahedra must be glued together while respecting the identification of the internal edges and the orientation. Importantly, one can easily see that the triangulated Gaiotto curves at $\theta=0$ and $\theta=2\pi$ are identical up to a rotation of the polygon by $4\pi/N$ (with $N=2n+3$). We can therefore glue the two curves with the appropriate twist, which is the geometric realisation of the $U(1)_r$ twist. This is precisely the geometry $M_3^{(k=1)}$ triangulated by $4n$ tetrahedra. 

\begin{figure}[tbp]
\centering
\begin{tikzpicture}[scale=1.5, rotate=0]
\usetikzlibrary{calc}
	
   \draw[gray!90!white,line width=0.5pt] (-1,0)--(-1+7.8,0);
   \draw[gray!90!white,line width=0.5pt] (-0.5,1)--(-0.5+7.8,1);
   \draw[gray!90!white,line width=0.5pt] (0.2,0.6)--(0.2+7.8,0.6);
   \draw[gray!90!white,line width=0.5pt] (0.2,-0.6)--(0.2+7.8,-0.6);
   \draw[gray!90!white,line width=0.5pt] (-0.5,-1)--(-0.5+7.8,-1);

  \begin{scope}[shift={(0,-2.5)}]
   \draw[gray!90!white,line width=0.5pt] (-0.3,0)--(7.6,0);
   \draw[gray!90!white,line width=0.5pt] (-0.3,1)--(7.6,1);
   \draw[gray!90!white,line width=0.5pt] (-0.3,0.6)--(7.6,0.6);
   \draw[gray!90!white,line width=0.5pt] (-0.3,-0.6)--(7.6,-0.6);
   \draw[gray!90!white,line width=0.5pt] (-0.3,-1)--(7.6,-1);
   
	\draw[|->,line width=1pt] (-0.3,-1.6)--(-0.2+7.8,-1.6);
	\node at (-0.3,-1.9) {$\th = 0$};
	\node at (-0.2+7.8,-1.9) {$\th = 2\pi$};
	\end{scope}

    \begin{scope}[shift={(0,0)}]
    \coordinate (P1) at (-1,0);
    \coordinate (P2) at (-0.5,1);
    \coordinate (P3) at (0.2,0.6);
    \coordinate (P4) at (0.2,-0.6);
    \coordinate (P5) at (-0.5,-1);
	
   \draw[blue, line width = 1.5pt] (P1)--(P3);
   \draw[blue, line width = 1.5pt] (P1)--(P4);
   \draw[line width = 1.2pt,line join=round] (P1)--(P2)--(P3)--(P4)--(P5)--(P1);
   
   \node at (-0.4,0.5) {\scriptsize $\textcolor{blue}{\g_1}$};
   \node at (-0.4,-0.5) {\scriptsize $\textcolor{blue}{\g_2}$};
   \end{scope}
   
    \begin{scope}[shift={(1.95,0)}]
    \coordinate (P1) at (-1,0);
    \coordinate (P2) at (-0.5,1);
    \coordinate (P3) at (0.2,0.6);
    \coordinate (P4) at (0.2,-0.6);
    \coordinate (P5) at (-0.5,-1);
	
   \draw[blue, line width = 1.5pt] (P1)--(P3);
   \draw[blue, line width = 1.5pt] (P3)--(P5);
   \draw[line width = 1.2pt,line join=round] (P1)--(P2)--(P3)--(P4)--(P5)--(P1);
   
   \node at (-0.4,0.5) {\scriptsize $\textcolor{blue}{\g_1}$};
   \node at (-0.02,-0.4) {\scriptsize $\textcolor{blue}{-\g_2}$};
   \end{scope}
   
    \begin{scope}[shift={(3.9,0)}]
    \coordinate (P1) at (-1,0);
    \coordinate (P2) at (-0.5,1);
    \coordinate (P3) at (0.2,0.6);
    \coordinate (P4) at (0.2,-0.6);
    \coordinate (P5) at (-0.5,-1);
	
   \draw[blue, line width = 1.5pt] (P2)--(P5);
   \draw[blue, line width = 1.5pt] (P3)--(P5);
   \draw[line width = 1.2pt,line join=round] (P1)--(P2)--(P3)--(P4)--(P5)--(P1);
   
   \node at (-0.7,0.15) {\scriptsize $\textcolor{blue}{-\g_1}$};
   \node at (-0.02,-0.4) {\scriptsize $\textcolor{blue}{-\g_2}$};
   \end{scope}
   
    \begin{scope}[shift={(5.85,0)}]
    \coordinate (P1) at (-1,0);
    \coordinate (P2) at (-0.5,1);
    \coordinate (P3) at (0.2,0.6);
    \coordinate (P4) at (0.2,-0.6);
    \coordinate (P5) at (-0.5,-1);
	
   \draw[blue, line width = 1.5pt] (P2)--(P5);
   \draw[blue, line width = 1.5pt] (P2)--(P4);
   \draw[line width = 1.2pt,line join=round] (P1)--(P2)--(P3)--(P4)--(P5)--(P1);
   
   \node at (-0.7,0.15) {\scriptsize $\textcolor{blue}{-\g_1}$};
   \node at (0,0.15) {\scriptsize $\textcolor{blue}{\g_2}$};
   \end{scope}

    \begin{scope}[shift={(7.8,0)}]
    \coordinate (2P1) at (-1,0);
    \coordinate (2P2) at (-0.5,1);
    \coordinate (2P3) at (0.2,0.6);
    \coordinate (2P4) at (0.2,-0.6);
    \coordinate (2P5) at (-0.5,-1);
	
   \draw[blue, line width = 1.5pt] (2P1)--(2P4);
   \draw[blue, line width = 1.5pt] (2P2)--(2P4);
   \draw[line width = 1.2pt,line join=round] (2P1)--(2P2)--(2P3)--(2P4)--(2P5)--(2P1);
   
   \node at (0,0.15) {\scriptsize $\textcolor{blue}{\g_2}$};
   \node at (-0.4,-0.45) {\scriptsize $\textcolor{blue}{\g_1}$};
   \end{scope}
   
    \begin{scope}[shift={(1.3,-2.5)}]
    \coordinate (1T1) at (-1,0);
    \coordinate (1T2) at (-0.5,1);
    \coordinate (1T3) at (0.2,0.6);
    \coordinate (1T4) at (0.2,-0.6);
    \coordinate (1T5) at (-0.5,-1);
	
   \draw[line width = 1.2pt,line join=round] (1T3)--(1T5);
   \draw[line width = 1.2pt,line join=round, dashed] (1T1)--(1T4);
   \draw[line width = 1.2pt,line join=round] (1T1)--(1T3);
   \draw[orange, line width = 1.2pt,line join=round] (1T1)--(1T5)--(1T4)--(1T3);
   
  \node at (-0.4,-0.1) {\textcolor{red}{{\bf 1}}};
  \node at (0,-0.25) {\textcolor{red}{{\bf 2}}};
  \node at (-0.25,0.15) {\textcolor{red}{ $\bar{{\bf8}}$}};
  \node at (-0.5,-0.45) {\textcolor{red}{ $\bar{{\bf6}}$}};
   \end{scope}
   
    \begin{scope}[shift={(3.3,-2.5)}]
    \coordinate (2T1) at (-1,0);
    \coordinate (2T2) at (-0.5,1);
    \coordinate (2T3) at (0.2,0.6);
    \coordinate (2T4) at (0.2,-0.6);
    \coordinate (2T5) at (-0.5,-1);
	
   \draw[line width = 1.2pt,line join=round] (2T2)--(2T5);
   \draw[line width = 1.2pt,line join=round, dashed] (2T1)--(2T3);
   \draw[line width = 1.2pt,line join=round] (2T3)--(2T5);
   \draw[orange, line width = 1.2pt,line join=round] (2T5)--(2T1)--(2T2)--(2T3);
   
   \node at (-0.7,-0.2) {\textcolor{red}{{\bf 3}}};
  \node at (-0.2,0.2) {\textcolor{red}{{\bf 4}}};
  \node at (-0.4,0.45) {\textcolor{red}{ $\bar{{\bf7}}$}};
  \node at (-0.35,-0.25) {\textcolor{red}{ $\bar{{\bf1}}$}};
    \end{scope}
   
    \begin{scope}[shift={(4.8,-2.5)}]
    \coordinate (3T1) at (-1,0);
    \coordinate (3T2) at (-0.5,1);
    \coordinate (3T3) at (0.2,0.6);
    \coordinate (3T4) at (0.2,-0.6);
    \coordinate (3T5) at (-0.5,-1);
	
   \draw[line width = 1.2pt,line join=round] (3T2)--(3T4);
   \draw[line width = 1.2pt,line join=round, dashed] (3T3)--(3T5);
   \draw[line width = 1.2pt,line join=round] (3T2)--(3T5);
   \draw[orange,line width = 1.2pt,line join=round] (3T2)--(3T3)--(3T4)--(3T5);
   
   \node at (-0.3,0.1) {\textcolor{red}{{\bf 5}}};
  \node at (0.1,0.1) {\textcolor{red}{{\bf 6}}};
  \node at (-0.1,0.45) {\textcolor{red}{ $\bar{{\bf4}}$}};
  \node at (-0.1,-0.45) {\textcolor{red}{ $\bar{{\bf2}}$}};   
  \end{scope}
   
    \begin{scope}[shift={(6.8,-2.5)}]
    \coordinate (4T1) at (-1,0);
    \coordinate (4T2) at (-0.5,1);
    \coordinate (4T3) at (0.2,0.6);
    \coordinate (4T4) at (0.2,-0.6);
    \coordinate (4T5) at (-0.5,-1);
	
   \draw[line width = 1.2pt,line join=round] (4T1)--(4T4);
   \draw[line width = 1.2pt,line join=round, dashed] (4T2)--(4T5);
   \draw[line width = 1.2pt,line join=round] (4T2)--(4T4);
   \draw[orange,line width = 1.2pt,line join=round] (4T2)--(4T1)--(4T5)--(4T4);
   
   \node at (-0.35,0.2) {\textcolor{red}{{\bf 8}}};
  \node at (-0.35,-0.5) {\textcolor{red}{{\bf 7}}};
  \node at (-0.7,0.15) {\textcolor{red}{ $\bar{{\bf3}}$}};
  \node at (-0.25,-0.15) {\textcolor{red}{ $\bar{{\bf5}}$}};   
   \end{scope}
   
	\draw[dashed,<->,>={Stealth[scale=0.5]},green!90!black,line width=0.5pt] (0,-0.53) to[out=-88, in=85] (0.65,-2.65);
	\draw[dashed,<->,>={Stealth[scale=0.5]},green!90!black,line width=0.5pt] (1.6,-0.53) to[out=-140, in=110] (1.3,-2.3);
	\draw[dashed,<->,>={Stealth[scale=0.5]},green!90!black,line width=0.5pt] (1.8,0.4) to[out=-50, in=90] (2.6,-2.3);
	\draw[dashed,<->,>={Stealth[scale=0.5]},green!90!black,line width=0.5pt] (3.35,-0.3) to[out=-140, in=40] (2.83,-1.85);
	\draw[dashed,<->,>={Stealth[scale=0.5]},green!90!black,line width=0.5pt] (3.62,-0.6) to[out=-60, in=140] (4.5,-3);
	\draw[dashed,<->,>={Stealth[scale=0.5]},green!90!black,line width=0.5pt] (5.8,-0.1) to[out=-150, in=50] (4.48,-1.86);
	\draw[dashed,<->,>={Stealth[scale=0.5]},green!90!black,line width=0.5pt] (5.4,-0.8) to[out=-10, in=170] (6.27,-2.1);
	\draw[dashed,<->,>={Stealth[scale=0.5]},green!90!black,line width=0.5pt] (7.7,-0.47) to[out=-140, in=70] (6.7,-2.9);

\end{tikzpicture}
\caption{\label{fig: 5gon flip} A sequence of flips of a triangulated pentagon for $A_2$ theory, starting from the left to the right along the $\th$ direction, is depicted on top. The assigned charge $\g_i$ flips to $-\g_i$ as the internal edge flips. After the full sequence, the first and the last pentagons should be identified along edge by edge with the same charges. The four gaps are filled with four tetrahedra whose diagonal edges are identified to the flipping edges as depicted by green arrows. All the faces of the tetrahedra are labeled by red numbers where bottom faces are denoted with a bar. All the pairs of faces with same number are glued together to obtain the three-manifold $M_3^{(k=1)}$. Here we colored in orange the boundary edge $C_\infty$ which generates the first homology of the three-manifold.  }
\end{figure}
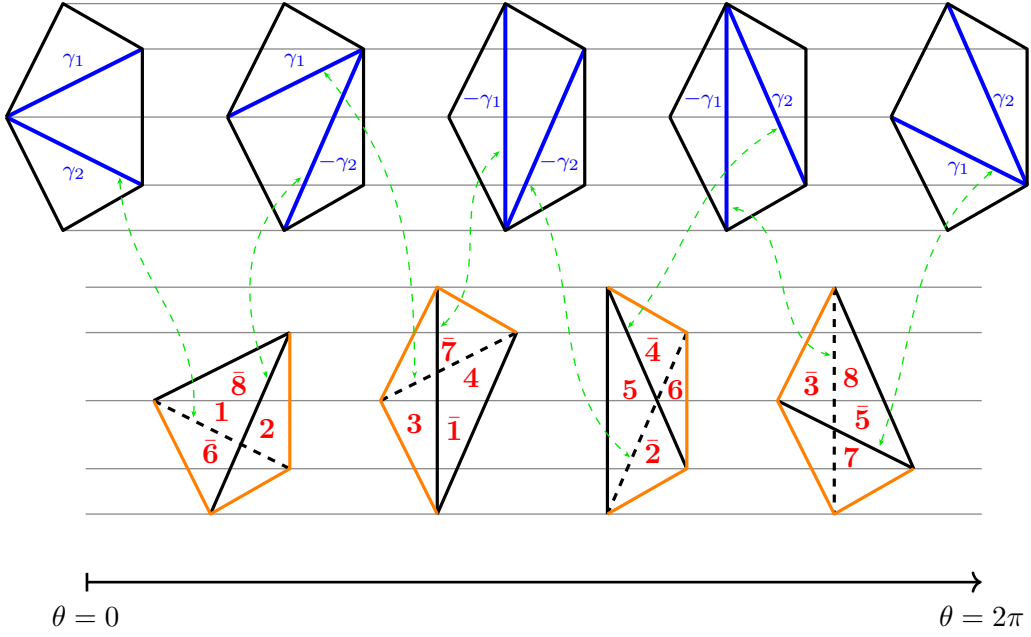

\medskip
\noindent {\bf Example: the $A_2$ triangulation.}  Figure~\ref{fig: 5gon flip} displays our construction for the $A_2$ theory, where the four transitions between the five Gaiotto curves are realised by four tetrahedra along the $\th$-direction. Here, the four pairs of diagonal edges of the tetrahedra relevant to the flips are identified with the internal edges (displayed in blue) of  ideal triangulations of $\CC$, with the identifications indicated by green arrows. Importantly, the Gaiotto curves at $\th=0$ and $\th=2\pi$ are exactly the same, up to a ${4\pi\ov 5}$ rotation, and they are thus identified in order to close up the $S^1_r$ circle with coordinate $\theta\in [0, 2\pi]$. Each face of each tetrahedron in figure~\ref{fig: 5gon flip} is labeled with a red number, with bottom faces indicated by a bar; all the 8 pairs of faces sharing the same number are glued pairwise, producing $M_3^{(k)}$ as a triangulated three-manifold.

\medskip
\noindent
{\bf General twist $\mathfrak{t}=e^{2\pi i k}$.} 
 We can similarly consider the general twist~\eqref{def t U1r} with $k$ and $N$ coprime. For $k>0$, we iterate the mutation sequence~\eqref{eq: A2n mutation} $k$ times, while for $k<0$ we should consider the sequence of left mutations
\bea
    &{\g_{1}
    \to
    \cdots
    \to
    \g_{2n-1}
    \to
    \;\;\g_{2}
    \to
    \cdots
    \to
    \g_{2n}
    \to}
\\
    &{\;\;-\g_{1}
    \to
    \cdots
    \to
    -\g_{2n-1}
    \to
    \;\;-\g_{2}
    \to
    \cdots
    \to
    -\g_{2n}
    \,,}
    \label{eq: A2n mutation bis}
\eea
 repeated $|k|$ times. In either case, the triangulated Gaiotto curve at $\theta=2\pi$ is the same as the one at $\theta=0$ rotated by $4\pi k/N$, and we obtain the three-manifold~\eqref{M3 foliated} 
triangulated by $4n|k|$ tetrahedra. Note that $M_3^{(k)}$ and $M_3^{(-k)}$ are related by an orientation reversal of the triangulated manifold. While we further discuss  the topology and geometry of this three-manifold in the next subsection, the actual triangulation --- the specific gluing rules for the $4n|k|$ tetrahedra --- is all that is needed for a direct application of the DGG machinery, which we will carry out explicitly in section~\ref{sec: AD to TQFT}.

\subsection{Towards a 3d/3d correspondence for rank-0 theories}\label{subsec:3d3drk0}
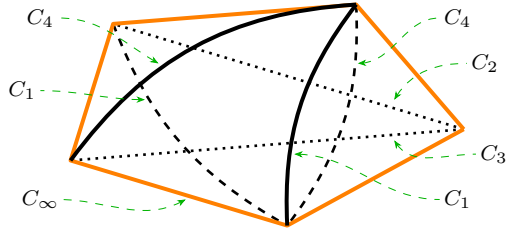
\begin{figure}[tbp]
\centering
\begin{tikzpicture}[scale=1.2, rotate=0]
\usetikzlibrary{calc}

\foreach \i in {1,...,5} {
    \coordinate (P\i) at ({2.3*cos(-10+360/5*(\i-1))},{1.3*sin(-10+360/5*(\i-1))});
}
\draw[dashed,line width=1pt,line join=round] (P3) to[bend right=20] (P5);
\draw[dashed,line width=1pt,line join=round] (P2) to[bend left=20] (P5);
\draw[dotted,line width=1pt,line join=round] (P1) -- (P3) ;
\draw[dotted,line width=1pt,line join=round] (P1) -- (P4) ;

\draw[orange,line width=1.5pt,line join=round] (P1) -- (P2) -- (P3) -- (P4) -- (P5) -- (P1) ;
\draw[line width=1.5pt,line join=round] (P2) to[bend right=25] (P4);
\draw[line width=1.5pt,line join=round] (P2) to[bend right=20] (P5);

\draw[dashed,<-,>={Stealth[scale=0.7]},green!70!black,line width=0.3pt] 
(1.1,0.5) to[out=0, in=180] (2,1);
\node at (2.2,1) {\scriptsize $C_4$};

\draw[dashed,<-,>={Stealth[scale=0.7]},green!70!black,line width=0.3pt] 
(1.5,0.1) to[out=60, in=180] (2.3,0.5);
\node at (2.5,0.5) {\scriptsize $C_2$};

\draw[dashed,<-,>={Stealth[scale=0.7]},green!70!black,line width=0.3pt] 
(1.5,-0.35) to[out=-60, in=180] (2.4,-0.5);
\node at (2.6,-0.5) {\scriptsize $C_3$};

\draw[dashed,<-,>={Stealth[scale=0.7]},green!70!black,line width=0.3pt] 
(-1.1,0.5) to[out=150, in=0] (-2.2,1);
\node at (-2.4,1) {\scriptsize $C_4$};

\draw[dashed,<-,>={Stealth[scale=0.7]},green!70!black,line width=0.3pt] 
(-1.25,0.1) to[out=180, in=0] (-2.4,0.2);
\node at (-2.6,0.2) {\scriptsize $C_1$};

\draw[dashed,<-,>={Stealth[scale=0.7]},green!70!black,line width=0.3pt] 
(0.4,-0.5) to[out=0, in=180] (2,-1);
\node at (2.2,-1) {\scriptsize $C_1$};

\draw[dashed,<-,>={Stealth[scale=0.7]},green!70!black,line width=0.3pt] 
(-0.8,-1) to[out=210, in=0] (-2.2,-1);
\node at (-2.4,-1) {\scriptsize $C_\infty$};

\end{tikzpicture}
\caption{\label{fig: L52} The lens space $L(5,2)$ obtained from the gluing of the four tetrahedra in figure~\protect\ref{fig: 5gon flip}. We draw the two top/internal/bottom edges with real/dotted/dashed lines, respectively. The two pentagons are glued along the edges $C_1$ and $C_4$ where $C_I$'s are indicated consistently with figure~\protect\ref{fig: A2ex} below. The orange-colored edges are all identified when gluing the upper and lower faces of the `lens' together, and they give us the external edge $C_\infty$.
}
\end{figure}

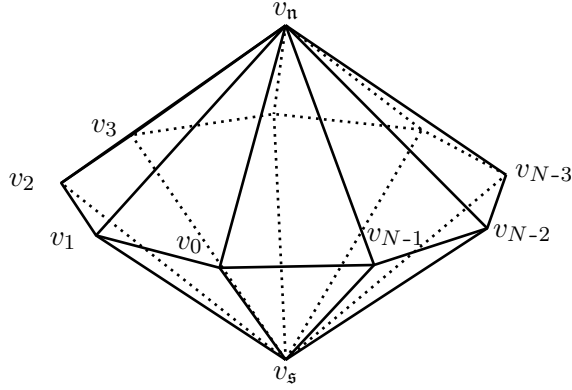
\begin{figure}[tbp]
\centering
\begin{tikzpicture}[scale=1.3, rotate=0]
\usetikzlibrary{calc}

\foreach \i in {1,...,9} {
    \coordinate (P\i) at ({2.3*cos(13+360/9*(\i-1))},{0.8*sin(13+360/9*(\i-1))});
}
\foreach \i in {1,...,9} {
    \coordinate (V\i) at ({2.7*cos(13+360/9*(\i-1))},{0.9*sin(13+360/9*(\i-1))});
}

\coordinate (Pn) at (0,1.7);
\coordinate (Ps) at (0,-1.7);
\node at (0,1.85) {$v_{\mathfrak{n}}$};
\node at (0,-1.85) {$v_{\mathfrak{s}}$};

\draw[line width=1pt,line join=round] (P4) -- (P5) -- (P6) -- (P7) -- (P8) -- (P9) -- (P1) ;
\draw[dotted,line width=1pt,line join=round] (P1) -- (P2) -- (P3) -- (P4) ;

\draw[line width=1pt,line join=round] (Pn) -- (P1) ;
\draw[dotted,line width=1pt,line join=round] (Pn) -- (P2) ;
\draw[dotted,line width=1pt,line join=round] (Pn) -- (P3) ;
\draw[line width=1pt,line join=round] (Pn) -- (P4) ;
\draw[line width=1pt,line join=round] (Pn) -- (P5) ;
\draw[line width=1pt,line join=round] (Pn) -- (P6) ;
\draw[line width=1pt,line join=round] (Pn) -- (P7) ;
\draw[line width=1pt,line join=round] (Pn) -- (P8) ;
\draw[line width=1pt,line join=round] (Pn) -- (P9) ;

\draw[dotted,line width=1pt,line join=round] (Ps) -- (P1) ;
\draw[dotted,line width=1pt,line join=round] (Ps) -- (P2) ;
\draw[dotted,line width=1pt,line join=round] (Ps) -- (P3) ;
\draw[dotted,line width=1pt,line join=round] (Ps) -- (P4) ;
\draw[dotted,line width=1pt,line join=round] (Ps) -- (P5) ;
\draw[line width=1pt,line join=round] (Ps) -- (P6) ;
\draw[line width=1pt,line join=round] (Ps) -- (P7) ;
\draw[line width=1pt,line join=round] (Ps) -- (P8) ;
\draw[line width=1pt,line join=round] (Ps) -- (P9) ;

\node at (V9) {$v_{N\texttt{-}2}$};
\node at (V1) {$v_{N\texttt{-}3}$};
\node at (V4) {$v_{3}$};
\node at (V5) {$v_{2}$};
\node at (V6) {$v_{1}$};
\node at (-0.98,-0.53) {$v_{0}$};
\node at (1.12,-0.45) {$v_{N\texttt{-}1}$};

\end{tikzpicture}
\caption{\label{fig: lens} Topologically, the lens space $L(N,q)$ is obtained by gluing the two hemispherical $N$-gons that form the boundary a three-ball with a $q$-unit rotation. In this figure, each 2-simplex $(v_\mathfrak{n} v_{[i]_N} v_{[i+1]_N})$ is identified with $(v_{\mathfrak{s}} v_{[i+q]_N} v_{[i+1+q]_N} )$, where $[m]_N:= m \, \text{mod}\,N$, for $i=0,\cdots,N-1$. For our $M_3^{(k)}$, the edges connecting $v_{i}$ to $v_{[i+1]_N}$ descend to $C_\infty$, while $C_0$ is isotopic to any path going vertically inside the solid lens from $v_{\mathfrak{s}}$ to $v_{\mathfrak{n}}$.
}
\end{figure}

Topologically, the three-manifold constructed in the previous subsection is the lens space:
\be\label{M3 as lens}
M_3^{(k)}\cong L(N, 2k)~, \qquad\quad N=2n+3~.
\ee
Indeed, the lens space $L(N,q)$ is defined as the quotient $S^3/\Z_N$:
\be\label{lens space Lpq def}
S^3 = \{ (z_1, z_2)\in \mathbb{C}^2 \, | \, |z_1|^2 +|z_2|^2= R_\infty^2\}~, \qquad\quad
(z_1, z_2) \sim (e^{2\pi i q\ov N}z_1, e^{2\pi i \ov N}z_2)~.
\ee
Let us write $z_1=z$ and $z_2= r e^{i \varphi}$ with $\varphi\in [0, {2\pi \ov N}]$. At any fixed $\varphi$, $z$ is the coordinate on a disk of radius $R_\infty$, with:
\be
|z|\leq R_\infty~, \qquad r= \sqrt{R_\infty^2-|z|^2}~.
\ee
This is a standard presentation of $L(N,q)$ as a solid `lens', which is topologically a ball where the boundary disks at $\varphi=0$ and $\varphi={2\pi \ov N}$ (and all disks at any value of $\varphi$ in-between) are glued at the equator $|z|= R_\infty$~\cite{seifert2004lehrbuch}. The upper and lower disks are identified according to~\eqref{lens space Lpq def}, as pictured in figure~\ref{fig: lens}. Introducing the coordinate $\theta= N\varphi$ to match the $S^1_r$ coordinate, this reads
\be\label{sft equiv}
(z, \theta) \sim (e^{2\pi i q\ov N} z, \theta+2\pi)~,
\ee
which is precisely the construction of $M_3^{(k)}$ discussed in the previous subsection with $q=2k$. Indeed, as we glue successive tetrahedra together, we find a set of `external' edges which are all attached to the $N$ external edges of the boundary polygon on $\CC$, which themselves are all identified together since $2k$ and $N$ are coprime --- see figures~\ref{fig: 5gon flip} and~\ref{fig: L52} for the $N=5$, $k=1$ example. We also checked the identification~\eqref{M3 as lens} directly in many examples by inputting our explicit triangulations into \textsc{Regina}~\cite{regina, em/1103749834}.

The boundary of the polygon $\CC$ is uplifted to a single vertex and a single edge in any triangulation of $M_3^{(k)}$ induced by a mutation sequence. The one-cycle $C_\infty = \{z_2=0\}$ corresponds to that edge in our lens space, and it generates the first homology:
\be
[C_\infty]\in H_1(L(N, 2k), \Z)= \Z_N~, \qquad N[C_\infty]=0~.
\ee
This is clearly seen by going to the covering space, where $N C_\infty$ is equivalent to the boundary of the disk at $|z|=R_\infty$, hence contractible. 
Another, inequivalent generator, $[C_0]$, is given by the central circle 
\be
C_0= \{z=0\} \times S^1_r \subset L(N, 2k)~.
\ee
The tubular neighbourhood of $C_0$ in $M_3^{(k)}$ is the fibered torus $T(N, 2k)$, corresponding to \eqref{sft equiv} with $(p,q)=(N, 2k)$.  Note that $[C_0]= 2k [C_\infty]$ in homology. Let us also recall that any lens space admits a Heegaard splitting $L(N, 2k)\cong \nu(C_0) \sqcup \nu(C_\infty)$  into two solid tori. For our triangulated $L(N, 2k)$, they are obtained by drilling out either $C_\infty$ or $C_0$:
\be\label{Heegaard}
\nu(C_0) \cong M_3^{(k)}- C_\infty \cong D^2\times S^1~, \qquad\qquad \nu(C_\infty) \cong M_3^{(k)}- C_0\cong D^2\times S^1~.
\ee
Removing $C_\infty$ gives us a three-manifold homeomorphic to a tubular neighbourhood of $C_0$, and vice versa. A useful topological result is that any lens space admits a unique Heegaard splitting into solid tori up to isotopy~\cite{Bonahon1983}. Here, it will be useful to make this splitting completely explicit. On the covering space $S^3$, consider the torus:
\be
\tilde{T}_\rho = \big\{|z_2|^2= \rho^2 \big\}  = \big\{|z_1|^2= R_\infty^2-\rho^2 \big\}
\ee
for some fixed radius $0<\rho<R_\infty$, which splits the  $S^3$ into two solid tori $|z_2|\leq \rho$ and $|z_1|\leq \sqrt{R_\infty^2-\rho^2}$, respectively. On the covering space, we have the deck transformations $g:(z_1, z_2)\rightarrow (e^{2\pi i q\ov N}z_1, e^{2\pi i \ov N}z_2)$.  The torus $\tilde{T}_\rho$ is parameterised by the angles $(\arg(z_1), \arg(z_2))$, and let $\mu_0=(1,0)$ and $\mu_\infty=(0,1)$ denote the corresponding basis of $H_1(\tilde{T}_\rho,\Z)$. Now, upon going to the quotient~\eqref{lens space Lpq def}, we have the same Heegaard splitting of $L(N, q)$ with the boundary torus $T_\rho \cong \tilde{T}_\rho/\Z_N$ obtained by quotienting by the deck transformation vector $d = {1\ov N}(q, 1)$. A basis of $H_1(T_\rho, \Z)$ is provided by $\{\mu_0, d\}$, and we note that $\mu_\infty = N d- q \mu_0$. Interestingly, $\mu_0$ and $\mu_\infty$ are the meridians around the cores $C_0$ and $C_\infty$, respectively, of the two solid tori~\eqref{Heegaard}. It is also useful to consider the isotopy that pulls $C_0$ and $C_\infty$ to one-cycles on $T_\rho$, which gives:
\be
C_0 \rightsquigarrow \hat C_0 \cong d~, \qquad \quad C_\infty \rightsquigarrow  \hat C_\infty \cong q^{-1} d + {1-q q^{-1}\ov N}\mu_0~.
\ee
Taking the orientation $\langle \mu_0, d\rangle=1$ for the intersection pairing on the torus, we then have the intersection numbers $\langle \mu_0, \hat C_0\rangle =1$ and $\langle \mu_0, \hat C_\infty\rangle =q^{-1}$, as well as $\langle \mu_0, \mu_\infty\rangle =N$.

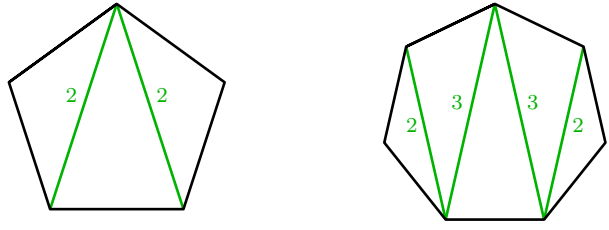
\begin{figure}[tbp]
\centering
\begin{tikzpicture}[scale=1, rotate=0]
\usetikzlibrary{calc}

\foreach \i in {1,...,5} {
    \coordinate (P\i) at ({1.5*cos(90+360/5*(\i-1))},{1.5*sin(90+360/5*(\i-1))});
}

\foreach \i in {1,...,7} {
    \coordinate (H\i) at ({5+1.5*cos(90+360/7*(\i-1))},{1.5*sin(90+360/7*(\i-1))});
}

\draw[green!70!black,line width=1pt,line join=round] (P1) -- (P3) ;
\draw[green!70!black,line width=1pt,line join=round] (P1) -- (P4) ;
\draw[line width=1pt,line join=round] (P1) -- (P2) -- (P3) -- (P4) -- (P5) -- (P1) -- (P2) ;

\draw[green!70!black,line width=1pt,line join=round] (H2) -- (H4) ;
\draw[green!70!black,line width=1pt,line join=round] (H4) -- (H1) ;
\draw[green!70!black,line width=1pt,line join=round] (H1) -- (H5) ;
\draw[green!70!black,line width=1pt,line join=round] (H5) -- (H7) ;
\draw[line width=1pt,line join=round] (H1) -- (H2) -- (H3) -- (H4) -- (H5) -- (H6) -- (H7) -- (H1) --(H2) ;

\node at (-0.6,0.3) {\scriptsize \textcolor{green!70!black}{$2$}};
\node at (0.6,0.3) {\scriptsize \textcolor{green!70!black}{$2$}};

\node at (3.9,-0.1) {\scriptsize \textcolor{green!70!black}{$2$}};
\node at (4.5,0.2) {\scriptsize \textcolor{green!70!black}{$3$}};
\node at (5.5,0.2) {\scriptsize \textcolor{green!70!black}{$3$}};
\node at (6.1,-0.1) {\scriptsize \textcolor{green!70!black}{$2$}};

\end{tikzpicture}
\caption{\label{fig: chord} Examples of chord distances $s_I$ of internal edges, shown in green, for $N=5$ and $N=7$.
}
\end{figure}

Let us now consider the one-cycle $C_I$ corresponding to an ``internal'' edge of the triangulation of $M_3^{(k)}$, which is simply one of the internal edges of one of the triangulations of $\CC$ in the mutation sequence which starts with~\eqref{eq: (2n+3)-gon}. A simple yet crucial observation is that:
\be
[C_I]  = \pm s_I [C_\infty]~,\quad\qquad s_I \in \left\{ 2, \cdots, {N-1\ov 2} \right\}~,
\ee
where $s_I$ denotes the `chord distance' of $C_I$ along the polygon --- that is, the end points of $C_I$ are separated by an angle ${2\pi s_I\ov N}$, and it is clear that $s_I \neq 1$ as that would correspond to an external edge instead --- see figure~\ref{fig: chord}.  For our purpose, we also need to determine the homology class of $\hat{C}_I$ obtained by `pulling' $C_I$ onto the Heegaard splitting torus $T_\rho$. This is:
\be\label{torus knot CIhat}
C_I \rightsquigarrow \hat{C}_I = {s_I \ov N}\mu_0 +{t_I\ov N}\mu_\infty~, \qquad q t_I= s_I \;({\rm mod}\; N)~,
\ee
since moving a vertex of the polygon by $2\pi s_I/N$ in the covering space corresponds to a deck transformation $g^{t_I}$. In the integral homology basis, this gives:
\be
\hat C_I= t_I d + w_I \mu_0 \qquad \text{with} \quad w_I = {s_I- t_I q\ov N}~,
\ee
for some choice of representative for $w_I$. 
If and only if $s_I=\pm q$, then $t_I=\pm 1$ and $\langle \mu_0, \hat{C}_I \rangle = t_I= \pm 1$. This means that $\hat C_I$ meets the meridian of $\nu(C_0)$ in~\eqref{Heegaard} exactly once, hence it is a longitude and  isotopic to the core $C_0$. Therefore, we find that $C_I$ is isotopic to $C_0$ if and only if $s_I=\pm q=\pm 2k$. This will be very important in section~\ref{sec: AD to TQFT} when we discuss the explicit 3d $\CN=2$ ACSM theories obtained from our three-manifold.%
\footnote{We can also verify this general claim on a case-by-case basis using \textsc{Regina}~\protect\cite{regina}, by checking that indeed $M_3^{(k)}- C_I$ is a solid torus if and only if $s_I=\pm 2k$; this implies the main claim by the uniqueness of the Heegaard splitting. Such explicit checks confirm our general argument in every case we tested.} We also note that the isotopy-invariant information about the torus knot (or cable) $\hat{C}_I\subset T_\rho$ is $(s_I, |t_I|)$. 

\medskip
\noindent
Incidentally, it is interesting to note that our triangulation of $M_3^{(k)}$ contains exactly 
\be\label{E euler}
E= L+1
\ee
edges, where $L=4 n|k|$ is the number of tetrahedra, which is also the number of internal edges $C_I$, while the extra edge is the `external' edge $C_\infty$. The relation~\eqref{E euler} follows from the fact that the Euler characteristic of any closed orientable three-manifold vanishes. Indeed, we have $\chi= V-E+F-L=0$ with a single vertex in our triangulation ($V=1$) while $F=2L$ because the faces are glued pairwise.

\medskip
\noindent {\bf TQFT and SCFT from $M_3^{(k)}$.}  
The 3d/3d correspondence is most well-developed when $M_3$ is a closed hyperbolic three-manifold, in which case the Bethe vacua of $T[M_3]$ correspond to irreducible $SL(2,\mathbb{C})$ flat connections on $M_3$. In the case at hand, the three-manifold~\eqref{M3 as lens} is certainly not hyperbolic. Moreover, there is a distinguished one-cycle $[C_\infty]$ which is inherited from the (regularisation of the) irregular singularity at $z=\infty$ on the curve $\CC$. We expect that the `correct' flat connections, in this case, correspond to the $U(1)_r$ fixed points on the wild character variety discussed in subsection~\ref{subsec:CB rev}, as these are also the $\Z_N$ fixed points~\cite{Dedushenko:2018bpp}. 

A precise discussion of the 3d/3d correspondence for either the rank-0 3d $\CN=4$ SCFTs $T[M_3^{(k)}]$ or for the 3d TQFTs $T_A[M_3^{(k)}]$ would likely involve a careful definition of the path integral of $SL(2,\mathbb{C})$ complex Chern--Simons theory on $M_3^{(k)}$ including, in particular, some specific boundary conditions at $C_\infty$ that would give us the 3d uplift of the Stokes phenomenon on the Gaiotto curve. 
Instead of going down this challenging route, here we present a concrete proposal on how to distinguish between the 3d TQFT $T_A[M_3^{(k)}]$ and the 3d $\CN=4$ fixed point $T[M_3^{(k)}]$ itself, as well as how to interpolate between the two:

\begin{itemize}
    \item[(a)] The 3d TQFT $T_A[M_3^{(k)}]$ is obtained by considering the lens space~\eqref{M3 as lens} itself as a closed manifold, with all internal edges `closed off' --- in the sense that the dihedral angle around them is $2\pi$, as we will review in section~\ref{subsec:DGG}. We will show in section~\ref{subsec:DGG and SCFT} that this necessarily implies the existence of a conical defect along $C_\infty$, which is the 3d remnant of the irregular puncture.

    \item[(b)] The non-trivial rank-0 3d $\CN=4$ SCFT $T[M_3^{(k)}]$ is obtained by considering the three-manifold:
    \be\label{M3 - CI}
     M_3^{(k)} - C_I \cong D^2\times S^1
    \ee
    obtained by `drilling out' (a tubular neighbourhood of) $C_I$ for any choice of an internal edge $C_I$ which is isotopic to $C_0$. As explained above, this is equivalent to the condition:
    \be\label{CI to 2k Cinf}
      s_I = \pm 2k \; {\rm mod}\; N
    \ee
    on the chord distance $s_I$ of $C_I$. As we will show in section~\ref{subsec:DGG and SCFT}, the SCFT point can be thought of as our triangulated lens space~\eqref{M3 as lens} where we moved the conical defect from $C_\infty$ to $C_I$. 

\item[(c)] Starting with the drilled three-manifold~\eqref{M3 - CI}, we can introduce a corresponding flavour holonomy $\hat{\nu}_\CA$ for some symmetry $U(1)_\CA$, which is interpreted as the axial symmetry of the 3d SCFT seen as a 3d $\CN=2$ field theory --- that is, $\hat{\nu}_\CA \neq 0$ corresponds to a soft breaking of 3d $\CN=4$ to 3d $\CN=2$ supersymmetry. The 3d $\CN=2$ $R$-symmetries at the SCFT and TQFT points are related precisely by a unit mixing with $\CA$~\cite{Closset:2016arn, Go:2025ixu}:
\be\label{rel between R and RTQFT}
R_{\rm TQFT}= R_{\rm SCFT} + \hat{\nu}_\CA \CA \qquad \text{with}\quad \hat{\nu}_\CA=-1~. 
\ee
We can continuously move from $\hat{\nu}_{\CA}=0$ to $\hat{\nu}_\CA=-1$ by tuning the dihedral angle along $C_I$ in the DGG construction, as we will explain in more detail in section~\ref{subsec:DGG and SCFT}. Conversely, if we start with the DGG construction with generic $\hat{\nu}_\CA$, the geometry predicts that $\hat{\nu}_{\CA}=0$ is dynamically obtained as a result of $F$-maximisation~\cite{Jafferis:2010un, Closset:2012vg}.

\end{itemize}

\noindent
This proposal will be further explained and corroborated in section~\ref{sec: AD to TQFT}, where we will present many explicit computations using DGG 3d $\CN=2$ field theories. The condition~\eqref{CI to 2k Cinf} on the choice of $C_I$ --- that is,  $s_I= \pm 2k$ mod $N$ --- is the most important one. It directly implies that a non-trivial SCFT is possible only if $2k \neq \pm 1$ (mod $N$). Instead, for $2k=\pm 1$ --- that is, for $k= \pm (n+1)$ mod $N$ --- there cannot exist any non-trivial SCFT, which is consistent with the previous expectation that $T_A[M_3^{(k)}]$ is a unitary TQFT precisely in this case, as explained around~\eqref{da unitaryTQFT}.

One can think of the proposal (a)--(c) above as a 3d/3d geometric reframing of the discussion of subsections~\ref{subsec:CB rev} and~\ref{subsec:twisted comp}. The solid torus with $C_\infty$ at its core,
\be\label{Mminuscinf}
\nu(C_\infty)\cong  M_3^{(k)}- C_0 \cong D^2\times S^1~,
\ee
and with the generic $U(1)_\CA$ holonomy turned on, corresponds to the circle reduction of the 4d AD theory to 3d with a generic fugacity $\mathfrak{t}$, where the $U(1)_r$ symmetry becomes the axial symmetry $\CA$ in the 3d description. The specific capping off of this solid torus into the lens space~\eqref{M3 as lens} corresponds to setting $\mathfrak{t}=e^{2\pi i k}$ to obtain either a  3d TQFT or a 3d $\CN=4$ SCFT. The two realisations differ only in subtle ways, as $T[M_3^{(k)}]$ essentially has a conical defect at $C_0$ while for $T_A[M_3^{(k)}]$ this conical angle is moved to the `external' edge $C_\infty$ of the lens space. In the DGG realisation as ACSM theories to be described in section~\ref{sec: AD to TQFT}, these two realisations differ only in the absence or presence of a particular superpotential term.

\medskip
\noindent
{\bf More general torus-knot complements.} 
We can naturally ask what happens if we drill out some internal edge $C_I$ which does not satisfy $s_I = \pm 2k$. Then, the resulting three-manifold is a knot complement
\be\label{X knot complement}
X_{(s_I, |t_I|)} \cong M_3^{(k)}- C_I~,
\ee
of the torus knot $\hat{C}_I \cong T_{(s_I, |t_I|)}$ inside the lens space $L(N, 2k)$, as discussed around~\eqref{torus knot CIhat}; more precisely, $\hat{C}_I$ is a torus knot if $\gcd(s_I, t_I)=\gcd(s_I, N)=1$, otherwise we have a cable (essentially, $\gcd(s_I, N)$ parallel copies of the minimal knot).  Of course, if $s_I=\pm 2k$, the knot complement~\eqref{X knot complement} is simply the solid torus~\eqref{Mminuscinf}. For $s_I \neq \pm 2k$, on the other hand, we expect that $T[X_{(s_I, |t_I|)}]$ gives us some 3d $\CN=2$ SCFT without supersymmetry enhancement. We hope to come back to this interesting point in future work, and to explore possible connections to related approaches in the literature --- see~{\it e.g.}~\cite{Gang:2018wek, Gukov:2019mnk, Chung:2022ypb, Gang:2024tlp, Gang:2025ykf, Chung:2026qfv}.

\medskip
\noindent
{\bf Abelian flat connections and Bethe vacua.} To conclude this section, let us discuss how the Bethe vacua of the 3d TQFT $T_A[M_3^{(k)}]$ can be naively matched to flat connections on the closed lens space~\eqref{M3 as lens} itself.
Since it has a fundamental group 
\be
\pi_1(L(N, 2k))\cong \{ g \,|\, g^N=1\}~,
\ee
the only $SL(2)$ flat connections are the $n+1$ abelian flat connections:
\begin{align}
    \r(g) = 
    \begin{pmatrix}
        e^{ \frac{ 2\pi i }{N}l} & 0 \\
        0 & e^{- \frac{2\pi i }{N}l}
    \end{pmatrix}
    \,,
    \label{eq: flat con}
\end{align}
for $l=1,\cdots,n+1$; here we exclude the $l=0$ case as it is fixed under the $\Z_2$ Weyl symmetry. On the other hand, the lens space admits a Seifert fibration with a single exceptional fiber,
\begin{align}
    L(N,2k) \cong \big[0;0;\big((-2k)^{-1},N\big)\big]
    \,.
\end{align}
Then, by treating the flat connections in \eqref{eq: flat con} {\it as if they were irreducible}, the $SU(2)$ Chern--Simons invariant and the Reidemeister torsion formula for Seifert manifolds~\cite{Rozansky:1995qft} give us:
\begin{align}
    {\rm CS}_l = \frac{2k}{N}l^2 \,\text{mod}\, 1
    \;,\;\;\qquad 
    \Tor_l = \frac{N}{4 \sin^2(2\pi \frac{2kl}{N})}
    \,.\label{CS and Tor computed}
\end{align}
By introducing a map ($\a=0,\cdots,n$):
\begin{align}
    l(\a) = 
    \Big\{
    \begin{array}{cc}
       n+1-\frac{\a}{2}  & \;\;\text{for even}\;\a~, \\
       \frac{\a+1}{2}  & \;\;\text{for odd}\;\a~,
    \end{array}
\end{align}
we precisely find that
\begin{align}
    \Tor_{l(\a)}
    =\big( S_{0\a}^{(k)} \big)^{-2}
    \;,\;\;\;\qquad
    e^{2\pi i ( {\rm CS}_{l(\a)} - {\rm CS}_{n+1} )} = T_{\a\a}^{(k)}/T_{00}^{(k)}
    \,.
    \label{eq: top inv}
\end{align}
In section~\ref{sec: AD to TQFT} below, we will provide strong consistency checks for the claim that the 3d $\CN=2$ ACSM theory $T_A[M_3^{(k)}]$ flows to a 3d TQFT in the infrared and reproduces the modular $S$ and $T$ matrices~\eqref{ST VOAk}. On the supersymmetric side of the correspondence, there is therefore a one-to-one relation between Bethe states $\ket{\alpha}$ and the topological lines of the infrared TQFT. The equalities~\eqref{eq: top inv} then strongly suggest that the Bethe states can also be understood as the abelian flat connections~\eqref{eq: flat con} on $M_3^{(k)}$ on the `topological' side of the 3d/3d correspondence. 
Note that, in~\eqref{CS and Tor computed}, we used the formula for the torsion of an irreducible flat connection, even though these lens-space flat connections are abelian and hence reducible; in particular, what we called $\Tor_l$ is not the actual torsion for these flat connections~\cite{Reidemeister:1935}. Our working hypothesis is that this discrepancy should be explained by considering~\eqref{Mminuscinf} with appropriate boundary conditions. We hope to address this question more thoroughly in future work.

Finally, we recall that one-form symmetries of $T[M_3]$ are encoded by the homology group~$H^1(M_3, \Z_2)$~\cite{Eckhard:2019jgg}, but since $N$ is odd this group is trivial in our case. Hence our 3d $\CN=2$ supersymmetric field theories have no non-trivial one-form symmetry, which we can also confirm directly by looking at the ACSM theories that we will construct explicitly in section~\ref{sec: AD to TQFT}.

\section{Probing the 3d infrared with supersymmetric partition functions}\label{sec: susy partition functions}
In this section, we review the computation of supersymmetric partition functions for 3d abelian gauge theories. In the next section, we use such computations to check that the 3d $\CN=2$ ACSM theories defined from $M_3^{(k)}$ indeed flow to the proposed 3d TQFTs $T_A[M_3^{(k)}]$ or 3d SCFTs $T[M_3^{(k)}]$ in the infrared. Here we will be particularly careful in keeping track of the phase of supersymmetric partition functions on Seifert manifolds.

\subsection{The 3d \texorpdfstring{$\CN=2$}{N=2}   \texorpdfstring{$A$}{A}-model and the topologically twisted index}
The topologically twisted index is the partition function of the theory on the three-manifold \(\CN_3 = S^1 \times \Sigma_g\) with the 3d $\CN=2$ $A$-twist~\cite{Nekrasov:2014xaa, Benini:2015noa, Benini:2016hjo, Closset:2016arn}. It can be computed using the 3d $A$-model formalism, which also allows us to compute supersymmetric partition functions on any Seifert manifold --- see~\cite{Closset:2018ghr, Closset:2019hyt} for a more detailed account and \cite{Closset:2023vos,Closset:2023jiq,Closset:2023bdr,Closset:2023izb,Closset:2024sle,Closset:2025lqt} for more recent applications. The 3d $A$-model is simply a topological $A$-model on $\Sigma_g$ obtained after integrating out Kaluza-Klein modes on the Coulomb branch. It is thus governed by the effective twisted superpotential \(\CW\) as well as by the effective dilaton \(\Omega\) which couples to the curvature of $\Sigma_g$. The 3d $\CN=2$ gauge theories of interest to us are specific ACSM theories given by an abelian gauge group $U(1)^L$ for some $L$, some $L\times L$ integer-valued matrix $K$ of CS levels, and $L$ chiral multiplets $\Phi_i$  with gauge charges $Q_{ij}= \delta_{ij}$ under $U(1)_j\subset U(1)^L$, all  with vanishing R-charges. We also have $K_{ij}^{GF}= \delta_{ij}$ for the mixed CS levels between the $U(1)_i$ gauge groups and the topological symmetries $U(1)_{T_j}$, which simply corresponds to having complexified FI parameters which we denote by $\nu_j$. We thus have:
\bea    \label{eq: ACSM WO}
 &   \CW(u,v)
   & =&\;\;{
    \frac{1}{2}\sum_{i,j=1}^L K_{ij}u_i u_j
    +\frac{1}{2}(1+2\nu_R) \sum_{i=1}^L K_{ii} u_i
    +\sum_{i=1}^L u_i\,\nu_i
    +\frac{K_g}{24}}\\
   && &\;\;{+\frac{1}{(2\pi i)^2} \sum_{i=1}^L \text{Li}_2 \big( e^{2\pi i \, u_i} \big)~,}\\
   & \Omega(u,v)
    &=&\; {\frac{1}{2\pi i} \sum_{i=1}^L \log \big( 1- e^{2\pi i \, u_i} \big)~,}
\eea
where $u_i$ and $\nu_i$ (mod $1$) are the $S^1$ holonomies for the $i$-th gauge and for the corresponding $U(1)_{T_i}$ symmetry, respectively, and $\nu_R$ is the $U(1)_R$ R-symmetry holonomy. Note also the appearance of the gravitational CS level $K_g \in \Z$ in~\eqref{eq: ACSM WO}, which we will discuss further in later sections. 
The $A$-model is governed by the massive vacua corresponding to the solutions to the Bethe equations derived from the twisted superpotential~\cite{Nekrasov:2009uh}. Let us compute the gauge flux operators for the $U(1)_i$ gauge groups and the flavour flux operators for the topological symmetries $U(1)_{T_i}$, respectively:
\bea
  &  \Pi_i(u,\nu)=
    \text{exp}\bigg( 2\pi i \frac{\partial\CW(u,\nu)}{\partial u_i} \bigg)
    = (-1)^{(1+2 \nu_R) K_{ii} } y_i \frac{ \prod_{j=1}^L x_j^{K_{ij}}  }{1-x_i}~,\\
   & \Pi_i^f(u,\nu)=
    \text{exp}\bigg( 2\pi i \frac{\partial\CW(u,\nu)}{\partial \nu_i} \bigg)= x_i~,
\eea
as well as the handle-gluing operator:
\be
    \CH(u,\nu) =
    e^{2\pi i \, \Omega(u,\nu)} \text{det}_{i,j}
    \bigg( \frac{ \partial^2 \CW(u,\nu) }{\partial u_i \partial u_j}  \bigg)
    =\prod_{i=1}^L (1-x_i) \,\times\,
    \text{det}_{i,j}  \Big( 
    K_{ij} + \delta_{i,j} \frac{x_i}{1-x_i}
    \Big)~~.
\ee
Here we use the holomorphic variables \(x_i=e^{2\pi i u_i}\) and \(y_i = e^{2\pi i \nu_i}\) for the gauge and flavour fugacities, respectively. We denote the set of Bethe vacua by:
\begin{equation}
    \CS_{\rm BE} = 
    \Big\{
    \hat{u}^{(\alpha)}\;\Big|\;
    \Pi_i\big(\hat{u}^{(\a)},\nu\big) = 1
    \; ;\;
    i=1,\cdots,L
    \Big\}\,~.
    \label{eq: Bethe eq}
\end{equation}
The topologically twisted index of the 3d $\CN=2$ ACSM theory can then be computed as a sum over Bethe vacua:
\be\label{ZSigmaggen}
Z_{S^1 \times \Sigma_g}(\nu)_\mathfrak{n} = \sum_{\hat{u}^{(\a)} \in \CS_{\rm BE}} \CH(\hat{u}^{(\alpha)},\nu)^{g-1}\, \Pi^f(\hat{u}^{(\alpha)},\nu)^\mathfrak{n}~.
\ee
Here $\mathfrak{n}_i$ denotes a flavour flux for $U(1)_{T_i}$, and we used the notation $(\Pi^f)^\mathfrak{n}\equiv \prod_i (\Pi^{f}_i)^{\mathfrak{n}_i}$ for the total flavour flux operator. 
In the supersymmetric ACSM theories of interest here, the flavour symmetries can mix with the R-symmetry as the theory flows to the infrared, as: 
\begin{align}\label{R mixing gen}
 R=R_0 + \sum_{i=1}^L \m_i\,T_i
    \,,
\end{align}
where $T_i$ is the charge of $U(1)_{T_i}$ topological symmetry, and $\m_i$ is a mixing parameter. Here, we define the reference $R$-symmetry $R_0$ in the UV to be the one that assigns vanishing $R$-charges to the chiral multiplets.  Equivalently, the conserved currents mix as $j_R \to j_R + \sum_i \m_i\, j_{T_i}$, which can be equivalently described as a shift of the background gauge fields for the flavour symmetries~\cite{Closset:2013vra, Closset:2014uda}: 
\be\label{flav mix gen}
    \nu_i \to \nu_i + \m_i\, \nu_R~,
    \qquad\qquad
    \mathfrak{n}_i \to \mathfrak{n}_i + \m_i \, \mathfrak{n}_R~,
\ee
where $\mathfrak{n}_R=g-1$ is the R-symmetry flux. 
Once we turn on monopole terms in the 3d superpotential $W$, the corresponding topological symmetries are explicitly broken and the mixing parameters are fixed by the fact that $R[W]=2$. When considering the theories $T_A[M_3^{(k)}]$ expected to flow to 3d TQFTs, we will see that the topological symmetries are completely fixed, amounting to a specific choice:
\be\label{flav mix R TQFT}
    \nu_i \to \m_i\, \nu_R~,\qquad\quad
    \mathfrak{n}_i \to \m_i\, \mathfrak{n}^R~,
\ee
with  $\mathfrak{n}^R=g-1$ on $\Sigma_g$. 
In that case, we can write the twisted index as:  
\be 
    Z_{S^1\times \Sigma_g}\big(\mu,\nu_R\big)
   =\sum_{\hat{u}^{(\a)} \in \CS_{\rm BE}} \hat{\CH}_\a (\m,\nu_R)^{g-1}
    \,,
    \label{eq: shifted ptf}
\ee
where we defined the shifted on-shell handle-gluing operator 
\begin{equation}
\begin{aligned}
    &\hat{\CH}_\a (\m,\nu_R):=
    \CH\big(\hat{u}^{(\a)},\m \nu_R\big)
    \prod_{i=1}^L
    \Pi_i^f\big(\hat{u}^{(\a)},\m \nu_R\big)^{\m_i}~.
    \label{Halpha def}
\end{aligned}    
\end{equation}
taking into account this $R$-symmetry mixing.

\medskip
\noindent 
{\bf Probing RG flows to infrared TQFTs.} In this work, our main interest is in constructing 3d $\CN=2$ gauge theories which flow to non-trival 3d TQFTs in the infrared. Due to the RG-invariance of supersymmetric partition functions, we can directly compare the supersymmetric computations to the expected answer in the conjectured infrared TQFT. In any semi-simple 3d TQFT,%
\footnote{That is, essentially, a 3d TQFT with a finite number of lines and with a modular tensor category structure. In this work, we are not exploring in full details the difference between bosonic and spin TQFTs; we hope to address the necessary refinements in future work.} 
 the $\Sigma_g\times S^1$ partition functions can be computed as:
\begin{equation}
    Z_{S^1\times\Sigma_g} = \mathrm{dim}\, \mathscr{H}(\Sigma_g) = \sum_{\alpha}(S_{0\alpha})^{2-2g}~,
\end{equation}
where \(\mathscr{H}(\Sigma_g)\) is the Hilbert space of the TQFT quantised on \(\Sigma_g\), and \(S_{\alpha\beta}\) denotes the matrix elements of the modular \(S\)-matrix in the basis of simple objects of the TQFT. By comparison with~\eqref{eq: shifted ptf}, we see the relation 
\begin{equation}
    \hat\CH_\a (\m,\nu_R) = (S_{0\alpha})^{-2}
    \label{eq: HtoS2}
\end{equation}
should hold whenever the UV theory flows to this TQFT, given some one-to-one identification between the Bethe vacua~\eqref{eq: Bethe eq} and the simple lines --- we refer to~\cite{Closset:2025lqt} for a more detailed explanation of this important point.

\subsection{Seifert fibering operators from supersymmetry to TQFT}

While the relation~\eqref{eq: HtoS2} between the $S$-matrix elements and the on-shell handle-gluing operators gives us a first robust test of any conjectured RG flow from an ACSM theory to an infrared TQFT, we can perform much more refined tests by considering partition functions on general Seifert fibrations over a $\Sigma_g$ orbifold~\cite{Closset:2018ghr}. A generic compact Seifert three-manifold is a circle bundle over an orbifold Riemann surface, indexed by the following data: 
\begin{equation}\label{N3 def}
    \CN_3 \cong \left[g\,;\,\mathbf{d}\,;\, (q_1,p_1),...,(q_\mathtt{N},p_\mathtt{N})  \right]~.
\end{equation}
Here the integer \(\mathbf{d}\) is the degree of the circle fibration and the Seifert invariants \((q_a,p_a)\) parametrise the local trivialisation around the orbifold points; see \cite{Closset:2018ghr} for more details and for a full account of our conventions. Starting from $\Sigma\times S^1$ (that is, $\mathbf{d}=0$ and $\mathtt{N}=0$), one builds up the supersymmetric partition function on any Seifert $\CN_3$ by the introduction of Seifert fibering operators. 
In particular, the degree $\mathbf{d}$ can be increased by the insertion of the (ordinary) \textit{fibering operator}~\cite{Closset:2017zgf}:
\bea
 &  {\CF(u,\nu) =
    \text{exp}
    \bigg[2\pi i
    \bigg(
    \CW(u,\nu)
    -
    \sum_{i=1}^L u_i \frac{\partial \CW(u,\nu)}{\partial u_i}
    -
    \sum_{i=1}^L \nu_i \frac{\partial \CW(u,\nu)}{\partial \nu_i}
    \bigg)\bigg]} \\
&
  \quad= {e^{\pi i K_g\ov 12} \prod_{i=1}^L x_i^{-\nu_i}} \\
  &\qquad{\times\text{exp}
    \bigg[
    \frac{1}{2\pi i}
    \bigg(
    \sum_{i=1}^L \Big(\text{Li}_2(x_i)
    +\log(x_i) \log(1\!-\!x_i)\Big)
    \!-\!\frac{1}{2}\sum_{i,j}^L K_{ij} \log(x_i)\log(x_j)
    \bigg)
    \bigg]~.}
\eea
Similarly, the insertion of a \((q,p)\) Seifert fiber is achieved via the \textit{Seifert fibering operator} \(\CG_{q,p}\), with $q>0$ and gcd$(q,p)=1$; it reduces to the ordinary fibering operator for $(q,p)=(1,1)$. For a more general Seifert invariant $(q,p)$, the Seifert fibering operator does not have a simple expression in terms of \(\CW\), but it is still explicitly computed in terms of the UV gauge theory data~\cite{Closset:2018ghr}. For the ACSM theory considered here, it reads:
\begin{equation}\label{SeifertOp}
    \CG_{q,p}(u,\nu)_{\mathfrak n} = \frac{1}{q^{\frac{L}{2}}}\sum_{\mathfrak{m} \in \Lambda(q)} \left(\CG^{(0)}\right)^{K_g} \CG^{GG}_{q,p}(u)_{\mathfrak m}\, \CG^{GF}_{q,p}(u,\nu)_{\mathfrak m,\mathfrak n}\, \CG_{q,p}^{\Phi}(u)_{\mathfrak m}~,
\end{equation}
where gravitational, gauge and mixed gauge-flavour Chern--Simons terms contribute as:\footnote{Here $s(p,q)$ is the Dedekind sum:
\be\nn
s(p,q)= {1\ov 4 q}\sum_{l=1}^{q-1}\cot{\left({\pi l\ov q}\right)}\cot{\left({\pi l p\ov q}\right)}~,
\ee
and \(l^R\) relates to the parametrisation of the R-symmetry line bundle, and for the \(\mathtt N\) Seifert fibers must obey \(\sum_{a=0}^\mathtt N l_a^R = 0\). When considering individual fibering operators we can take \(l^R=0\).}
\bea\nn
   & {\CG^{(0)}_{q,p} = \exp\left(\pi i \left({p\ov 12 q}- s(p,q)\right)\right)~,}\\
       & {\CG^{GG}_{q,p}(u)_{\mathfrak m} = (-1)^{K^{ii}\mathfrak{m}_i(1+t+l^Rt+2\nu_R s)}\exp\left(-\frac{i \pi}{q}\sum_{i,j=1}^L K_{ij}(pu_iu_j - 2\mathfrak{m}_i u_j + t \mathfrak{m}_i \mathfrak{m}_j)\right)~, }\\
      &{  \CG^{GF}_{q,p}(u,\nu)_{\mathfrak m,\mathfrak n} =\exp\left(-\frac{2\pi i}{q}\sum_{i=1}^L(pu_i \nu_i - \mathfrak{n}_i u_i - \mathfrak{m}_i \nu_i+ t \mathfrak{m}_i \mathfrak{n}_i)\right)~,}
\eea
while the chiral multiplets contribute the more complicated-looking terms:
\bea
 & \CG_{q,p}^{\Phi}(u)_{\mathfrak m} =\\
 &\quad { \left(e^{\frac{2\pi i u}{q}}; e^{\frac{2\pi i t}{q}}\right)_{-\mathfrak m} \exp{\sum^{q-1}_{l=0}\left[ \frac{p}{2\pi i}\mathrm{Li}_2(e^{2\pi i\frac{u+t l}{q}})+\frac{pu + l}{q} \log\left( 1-e^{2\pi i\frac{u+t l}{q}}\right)\right]}~.}
\eea
Here \(t\) and \(s\) are the modular inverses of \(p\) and \(q\), respectively, and we refer to~\cite[eq.(4.67)]{Closset:2018ghr} for the definition of the Pochhammer symbol prefactor. The sum in \eqref{SeifertOp} is over the \(\mathbb{Z}_q\) reduction of the magnetic flux lattice of the theory, denoted by \(\Lambda(q)\); in the case at hand, we simply have \(\Lambda(q)\cong \Z_q^L\). Finally, note the factor depending on $K_g$ in~\eqref{SeifertOp}; since this is a $u$-independent factor, it contributes an overall phase to the supersymmetric partition function on $\CN_3$. We will further discuss the choice of such gravitational (and $U(1)_R$) CS contact terms in the next section. 

\medskip
\noindent {\bf Supersymmetric partition function on $\CN_3$.} Given these fibering operators, the partition function on the Seifert manifold~\eqref{N3 def} is again given as a sum over Bethe vacua,
\be
  Z_{\CN_3}(\mu, \nu_R)_\mathfrak{n}=
   \sum_{\hat{u}^{(\a)} \in \CS_{\rm BE}}
    \CH\big(\hat{u}^{(\alpha)},\nu\big)^{g-1} \,\Pi^f(\hat{u}^{(\alpha)},\nu)^\mathfrak{n}\,
    \CF\big(\hat{u}^{(\alpha)},\nu\big)^\mathbf{d} \prod_{a=1}^\mathtt{N} \CG_{q_a,p_a}(\hat{u}^{(\alpha)},\nu)_\mathfrak{n}~,
    \label{eq: twisted ptf}
\ee
generalising~\eqref{ZSigmaggen}. 
Taking into account the flavour mixing corresponding to an infrared TQFT, we have the same shifts in the flavour holonomies and fluxes as in~\eqref{flav mix R TQFT}, except that the R-symmetry fluxes are now fixed as 
\begin{align}
    \mathfrak{n}_0^R = g-1 + \nu_R\,\mathbf{d} +\frac{l_0^R}{2} &
    \;\;,\;\;\;\quad
    \mathfrak{n}^{R}_a = \frac{q_a-1}{2} +  \nu_R\, p_a +\frac{l_a^R q_a}{2}
    \,~, 
\end{align}
where \(\mathfrak{n}_0^R\) refers to a unit of regular flux and \(\mathfrak{n}_a^R\) are the fractional $U(1)_R$ flux contributions that mix with the fractional fluxes $\mathfrak{n}_a\in \Z_{q_a}$ associated to each special Seifert fiber $(q_a, p_a)$. The Seifert partition function then reads:
\begin{equation}
\begin{aligned}
    Z_{\CN_{3}}\big(\m,\nu_R\big)
    =
    \sum_{\hat{u}^{(\a)} \in \CS_{\rm BE}}
    \hat{\CH}_\a (\m,\nu_R)^{g-1}\,
    \hat{\CF}_\a (\m,\nu_R)^{\mathbf d}\,\prod_{a=1}^\mathtt{N} \hat{\CG}^{\,\alpha}_{q_a,p_a}(\mu,\nu_R)
    \,,
    \label{eq: shifted ptf bis}
\end{aligned}
\end{equation}
where $\hat{\CH}_\a$  was defined in~\eqref{Halpha def}, and we similarly define:%
\begin{align}
    &\hat{\CF}_\a (\m,\nu_R):=
    \CF\big(\hat{u}^{(\a)},\m \nu_R\big)
    \prod_{i=1}^L
    \Pi_i^f\big(\hat{u}^{(\a)},\m \nu_R\big)^{\nu_R\m_i}~,\\
    &\hat{\CG}^{\,\a}_{q_a,p_a}(\mu,\nu_R):= \CG_{q_a,p_a}\big(\hat{u}^{(\a)},\mu \nu_R\big)_{\mu \left(\frac{q_a-1}{2} +\nu_R\, p_a\right)}~.
    \label{eq: modified handle fibering}
\end{align}
Here we have set $l_a^R =0$ for simplicity of presentation.

\medskip
\noindent
{\bf Matching to TQFT modular matrices.} The precise connection between the geometry-changing line operators in the 3d \(A\)-model and the modular matrix elements from Seifert surgery in the infrared 3d TQFT~\cite{Witten:1988hf} was studied in~\cite{Closset:2025lqt} in the case of 3d $\CN=2$ supersymmetric Chern--Simons theory. In particular, the relationships 
\begin{equation}
\begin{gathered}
T_{\a\a} = \hat{\CF}_\a(\m,\nu_R)^{-1}~, \qquad \qquad\frac{\CU^{(q,p)}_{\alpha0}}{S_{0\alpha}} = \,\,  e^{\frac{i \pi t}{12q}c_{\text{2d}}}\, \hat{\CG}^{\,\a}_{q,p}(\mu,\nu_R)~.
    \label{eq: HF to ST} 
\end{gathered}
\end{equation}
%
were shown to hold, where \(T_{\alpha\alpha}\) and \(\CU^{(q,p)}_{\alpha\beta}\) are the matrix elements of the modular \(T\)-matrix and of a general Seifert surgery modular matrix \(U^{(q,p)}\), which depends on both $(q,p)$ and on some choice of integers $(s,t)$ such that $sq+tp=1$. Note here the appearance of the central charge \(c_{\text{2d}}\) of the 3d TQFT, which is also the central charge of the 2d CFT living on the holomorphic boundary --- in the case of the Chern--Simons theory $G_k$, this is the corresponding WZW model. The phase proportional to \(c_{\text{2d}}\) acts as a counterterm moving us between the supersymmetric scheme, which mildly depends on a choice of metric on $\CN_3$, and the TQFT scheme, which does not but introduces a framing anomaly~\cite{Witten:1988hf}. 

The theories considered in this paper are ACSM theories which do not obviously flow to 3d TQFTs in the IR, unlike the pure 3d $\CN=2$ CS theories studied in~\cite{Closset:2025lqt}. Nonetheless, we conjecture that, whenever a theory $T_A[M_3^{(k)}]$ does flow to a topological phase, the same relations~\eqref{eq: HtoS2} and~\eqref{eq: HF to ST} between supersymmetric and TQFT quantities still hold true. In particular, this provides some fine checks of the central charge $c_{\rm 2d}$ of any candidate boundary VOA. 
At the level of supersymmetric partition functions on closed Seifert manifolds, we thus have the conjectural relation between the supersymmetric and TQFT computations:
\begin{equation}\label{Ztqft to susy}
    Z_{\rm TQFT} = e^{S_{\rm ct}}Z_{\rm SUSY}~,\qquad \quad     S_{\rm ct} = \frac{i\pi}{12}c_{\text{2d}}\sum_{i=1}^\mathtt{N} \frac{t_i}{q_i}~,
\end{equation}
with the explicit counterterm depending on $c_{\rm 2d}$. Note that shifting $K_g$ by an integer is equivalent to shifting $c_{\rm 2d}$ by the same integer, as discussed previously around~\eqref{c and K shift} and as is also apparent from the first relation in~\eqref{eq: HF to ST}. Once we fix $K_g$ on the supersymmetric side, the relations~\eqref{Ztqft to susy} then determine $c_{\rm 2d}$ fully. More precisely, while the first relation in~\eqref{eq: HF to ST} only determines $c_{\rm 2d}$ mod $24$, the second relation with arbitrary $(q,p)$ determines it to any required accuracy --- in particular, from the $(q,1)$ fibering operator one can determine $c_{\rm 2d}$ mod $24q$.

\subsection{The 3d \texorpdfstring{$\CN=2$}{N=2} superconformal index}
Another important observable with 3d $\CN=2$ gauge theory is the superconformal index, also known as the untwisted $S^2\times S^1$ partition function~\cite{Kim:2009wb,Imamura:2011su}. Whenever the 3d $\CN=2$ gauge theory flows to an SCFT, the index counts gauge invariant local BPS operators of the latter:
\begin{align}
    \CI_{S^2 \times S^1}(y;\q) = 
    \Tr_{\mathscr{H}}
    (-1)^R
    \q^{\frac{R}{2}+j_3}
    y^{f}~.
\end{align}
Here the trace is over the radially quantized Hilbert space $\mathscr{H}$ on $S^2$ of the IR fixed point, $R$ is the superconformal R-charge, $j_3$ is the Cartan of the $SO(3)$ isometry of $S^2$, and $y$ denotes the fugacities for possible flavour symmetries with flavour charges $f$. In particular, whenever our ACSM flows to a semi-simple TQFT in the infrared, we must find a trivial result,
\be
 \CI_{S^2 \times S^1}(y;\q) = 1~, \qquad \qquad
 Z_{S^1\times \Sigma_0}=1~,
\ee
for both the untwisted and twisted index on $S^2$, 
reflecting the absence of non-trivial local operators in this low-energy phase; equivalently, this is the statement that a 3d TQFT has a unique state on the sphere~\cite{Witten:1988hf}.

The superconformal index can be computed for any 3d $\CN=2$ supersymmetric gauge theory as a supersymmetric partition function on $S^2\times S^1$. For our ACSM theory, the localisation formula takes the form of an integral over the gauge fugacities $x_i$~\cite{Imamura:2011su, Kapustin:2011jm}:
\begin{align}
    \CI_{S^2\times S^1} (y;\q) = 
    \sum_{\mathfrak{m}\in \mathbb{Z}^L}
    \oint \prod_{i=1}^{L} \frac{ dx_i}{ 2\pi i \, x_i }
    x_i^{ \sum_{j=1}^L K_{ij} \mathfrak{m}_j }
    \,y_i^{\mathfrak{m}_i}\,
    \CI_\D (\mathfrak{m}_i,x_i;\q)~,
    \label{eq: index int form}
\end{align}
where the chiral multiplet contribution $\CI_\D(\mathfrak{m},x;\q)$ is called the tetrahedron index~\cite{Dimofte:2011py}:
\begin{align}
    \CI_\D(\mathfrak{m},x;\q) = 
    \prod_{l=0}^\infty
    \frac{ 1 - \q^{l - \frac{\mathfrak{m}}{2} + 1} x^{-1} }{1- \q^{l-\frac{\mathfrak{m}}{2} } x} = \sum_{e \in \mathbb{Z}} \CJ_{\q}(\mathfrak{m},e) \, x^e~,
\end{align}
where in the second equality we expanded in the charge basis with:
\begin{align}
 \CJ_{\q}(\mathfrak{m},e)= 
    \sum_{l= |e_-| }^\infty
    \frac{
    (-1)^l \q^{\frac{l(l+1)}{2} -(l + \frac{e}{2}) \mathfrak{m} }
    }{ (\q)_l (\q)_{l+e} }~,
    \label{eq: CJ}
\end{align}
where $ e_- \equiv \frac{e-|e|}{2}$, and $(\q)_l \equiv \prod_{i=1}^{l}(1-\q^i) $ is the $\q$-Pochhammer symbol. By substituting this expansion into \eqref{eq: index int form}, we obtain a useful expression for the index as:
\begin{align}
    \CI_{S^2\times S^1} (y;\q)
    =
    \sum_{\mathfrak{m} \in \mathbb{Z}^L}
    \prod_{i=1}^{L}
    \bigg[
\,    y_i^{\mathfrak{m}_i}\,
    \CJ_{\q}
    \Big(
    \mathfrak{m}_i
    \,,\,
    -\sum_{j=1}^{L}
    K_{ij} \mathfrak{m}_j
    \Big)
    \bigg]~.
    \label{eq: SCI}
\end{align}
These formulas for the index are at the reference $R$-charge $R_0$ that assigns vanishing $R$-charge to the chiral multiplets. When considering a general mixing with the $U(1)^L$ topological symmetry as in~\eqref{R mixing gen}, we should shift the flavour parameters according to:
\be
y_i \to \big(-\q^{\frac{1}{2}} \big)^{\m_i} y_i~.
\ee
Interestingly, whenever $\CI_{S^2 \times S^1} = 1$ holds because the ACSM theory flows to a TQFT, this implies a very non-trivial identity such that the expansion~\eqref{eq: SCI} vanishes at every positive order in $\q$. 

\subsection{The 3d \texorpdfstring{$\CN=2$}{N=2} half-index}
Similarly, one can also consider the half-index~\cite{Gadde:2013sca,Gadde:2013wq,Yoshida:2014ssa,Dimofte:2017tpi} which essentially counts local operators at the torus boundary of $D^2\times S^1$ that satisfy a 2d $\CN=(0,2)$ boundary condition  $\CB$,
\begin{align}
    \CI_{D^2\times S^1}(y;\q)
    =
    \Tr_{{\rm Ops}_\CB} (-1)^R \q^{\frac{R}{2} + j_3} y^f
    \,.
\end{align}
In this paper, we only consider the simple boundary condition  $\CB = (\CD,D_c)$, wherein we choose Dirichlet boundary conditions $\CD$ for the $\CN=2$ vector multiplets and deformed-Dirichlet boundary conditions $D_c$ for the chiral multiplets for some constant $c$~\cite{Dimofte:2017tpi}:
\begin{align}
    &\mathcal{D}
    \;:\;\;
    A_\pm |_\partial = D |_\partial = \l_- |_\partial = 0\,,
    \nonumber\\
    & D_c
    \;:\;\;
    \phi |_\partial = c 
    \;\;,\;\;\;
    \psi_+ |_\partial = 0\,.
\end{align}
We can then use the following formula~\cite{Chung:2023qth,Gang:2023ggt,Gang:2024loa} for the half-index of our ACSM theory:
\begin{align}
    \CI_{D^2\times S^1}
    (y;\q)
    =
    \sum_{\mathfrak{m} \in \mathbb{Z}_{\geq 0}^{L}}
    \frac{
    \q^{ \frac{1}{2} \mathfrak{m}^T \cdot K \cdot \mathfrak{m}}
    \prod_{i=1}^L
    y_i^{-\mathfrak{m}_i}
    }
    {
    \prod_{i=1}^L
    (\q)_{\mathfrak{m}_i}
    }~,
    \label{eq: half index}
\end{align}
here again written for the reference $R$-charge $R_0$. 
Whenever the 3d infrared is a TQFT, the half-index should only receive contributions from boundary operators. In favourable circumstances, it therefore computes the vacuum character $\chi_0$ of the holomorphic boundary VOA, up to an overall normalisation which captures the 2d central charge:
\be
  \CI_{D^2\times S^1}
    (y;\q)=\q^{c_{\rm 2d}\ov 24}  \chi_0({\rm VOA})~.
\ee
In general, however, it is not  clear {\it a priori} which precise boundary conditions we should impose in the UV to reproduce the vacuum character of the boundary theory. We leave these questions as an important challenge for future work.

\section{From abelian Chern--Simons-matter theories to TQFTs} 
\label{sec: AD to TQFT}

In this section, we explicitly construct the ACSM theories flowing to $T[M_3^{(k)}]$ and $T_A[M_3^{(k)}]$ using the DGG construction~\cite{Dimofte:2011ju}. We first spell out the general DGG formalism and how to see the TQFT and SCFT points from the geometry. We then turn to working out various examples. For $k=1$, we recover the expected 3d TQFT that mediates the SCFT/VOA correspondence for $\CT_{A_{2n}}$. For $k=-1$, we explicitly verify that the 3d TQFT supports the affine $osp(1|2n)_1$ VOA on its holomorphic boundary. We also check that the theory $T_A[M_3^{(-2)}]$ for $n=1$ is a 3d TQFT with the affine $(G_2)_1$ VOA on its holomorphic boundary, in agreement with~\cite{Kim:2024dxu}. Explicit computations of supersymmetric partition functions become somewhat prohibitive for higher values of $n$ and $|k|$, but we nevertheless discuss general aspects of the ACSM for $T_A[M_3^{(k)}]$ for general $n$ and $k$.

\subsection{The Dimofte--Gaiotto--Gukov construction}\label{subsec:DGG}

Let us first review the relevant aspects of the DGG construction~\cite{Dimofte:2011ju}, which provides us with a 3d $\CN=2$ abelian gauge theory that works as a 3d UV completion for the compactification of the 6d $\CN=(2,0)$ $A_1$-type theory on a three-manifold $M_3$ topologically twisted by the $SO(3)_R \subset SO(5)_R$ R-symmetry.
 Consider an explicit triangulation of the three-manifold into $L$ ideal tetrahedra,
\begin{align}
    M_3 = \bigcup_{i=1}^L \D_i~,
\end{align}
where $\D_i$ denotes the $i$-th ideal tetrahedron. Each ideal tetrahedron, as depicted in figure \ref{fig: tetra}, has its vertices on the boundary at infinity of the hyperbolic 3-space $\mathbb{H}^3$, and its hyperbolic structure is determined by a complex cross-ratio of the vertex locations captured by variables $z = e^{Z}$, where:
\begin{align}\label{Z tor angle def}
    Z = (\text{torsion}) + i\, (\text{dihedral angle})
    \,.
\end{align}
The cross-ratio can be defined for the other edges $z'=e^{Z'}$, $z''=e^{Z''}$ as well, defining the boundary phase space for each tetrahedron \cite{Dimofte:2011gm}:
\begin{align}
    \CP_{\partial \D_i} = 
    \big\{
    (Z_i,Z'_i,Z''_i) \in \mathbb{C} \backslash 2\pi i\, \mathbb{Z}
    \;
    \big|
    \;
    Z_i+Z'_i+Z''_i = \pi i
    \big\}~,
\end{align}
which is an affine linear space with a symplectic structure given by the Poisson bracket
\begin{align}
    \{Z_i,Z'_j\} = \{Z'_i,Z''_j\} = \{Z''_i,Z_j\} = \d_{ij}
    \,.
\end{align}
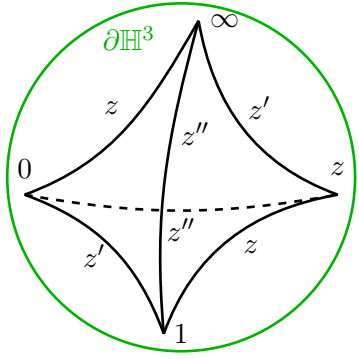
\begin{figure}[tbp]
\centering
\vspace{-10pt}
\begin{tikzpicture}[scale=2.3, rotate=0]
\usetikzlibrary{calc}

    \coordinate (P1) at (0.1,0.9);
    \coordinate (P2) at (-0.1,-0.9);
    \coordinate (P3) at (-0.9,-0.1);
    \coordinate (P4) at (0.9,-0.1);

    \draw[green!70!black, line width=1pt] (0,0) circle (1);

    \draw[dashed, line width=1pt] (P3) to[out=-10, in=190] (P4);
    \draw[line width=1pt] (P1) to[out=-105, in=95] (P2);
    \draw[line width=1pt] (P1) to[out=-120, in=20] (P3);
    \draw[line width=1pt] (P1) to[out=-80, in=160] (P4);
    \draw[line width=1pt] (P3) to[out=-20, in=110] (P2);
    \draw[line width=1pt] (P4) to[out=-170, in=70] (P2);

    \node at (-0.3,0.8) {\textcolor{green!70!black}{$\partial \mathbb{H}^3$}};
    \node at (0.25,0.9) {$\infty$}; 
    \node at (-0.9,0.05) {$0$};
    \node at (0,-0.9) {$1$};
    \node at (0.9,0.05) {$z$};
    \node at (0.4,-0.4) {$z$};
    \node at (-0.4,0.4) {$z$};
    \node at (0.45,0.4) {$z'$};
    \node at (-0.5,-0.45) {$z'$};
    \node at (0.08,0.25) {$z''$};
    \node at (0,-0.3) {$z''$};

\end{tikzpicture}
\caption{\label{fig: tetra} An ideal tetrahedron. The vertices are placed on $\partial \mathbb{H}^3$. }
\end{figure}

\medskip
\noindent
{\bf Choice of polarisation.} 
To characterise the 3d $\CN=2$ gauge theory $T[M_3]$, we specify a {\it polarisation} of each tetrahedron, labeled by $\s_i \in \{0,1,2\}$:
\begin{align}
    (X_i,P_i)
    =
    \Bigg\{
    \begin{array}{cc}
        (Z_i,Z''_i) & \;\;\text{if}\;\; \s_i = 0~, \\
        (Z''_i,Z'_i) & \;\;\text{if}\;\; \s_i = 1~, \\
        (Z'_i,Z_i) & \;\;\text{if}\;\; \s_i = 2~,
    \end{array}
\end{align}
which chooses canonically-conjugate position and momentum coordinates on the phase space $\CP_{\partial \D_i}$. Let us take $\s_i=0$ for the moment. 

As the faces of the tetrahedra are glued pairwise, the total dihedral angle around every merged internal edge should equal $2\pi$ with vanishing torsion in~\eqref{Z tor angle def} in order to obtain a smooth hyperbolic metric on $M_3$. This condition reads:%
\footnote{By a slight abuse of notation, we will use the notation $C_I$ for the variable associated with the internal edge $C_I$ itself. This should cause no confusion.}
\begin{align}
    C_I
    :=
    \sum_{j=1}^L
    \Big(
    g_{Ij}^{(0)} Z_j
    +
    g_{Ij}^{(1)} Z'_j
    +
    g_{Ij}^{(2)} Z''_j
    \Big)
    =
    2\pi i
    \,,
    \label{eq: CI=ZZZ}
\end{align}
where the index $I$ labels the internal edges. Here, the coefficients $g_{Ij}^{(s)}$ count the number of $j$-th dihedral-angle variables $(Z_j,Z'_j,Z''_j)$ meeting at the $I$-th internal edge.  
Note that we will consider relaxing somewhat the condition $C_I=2\pi i$ in~\eqref{eq: CI=ZZZ} momentarily --- see subsection~\ref{subsec:DGG and SCFT}.  The $C_I$ variables mutually commute, $\{ C_I,C_J \} = 0$. In general, there are other mutually commuting edges $\{E_J\}$ that describe the boundary phase space of $M_3$ \cite{NeumannZagier1985},
\begin{align}
    \CP_{\partial M_3}
    =
    \bigg(
    \prod_{i=1}^L
    \CP_{\partial \D_i}
    \bigg)
    \Big/
    \!\!\!
    \Big/
    \{C_I = 2\pi i\}
    \,,
\end{align}
which is a symplectic quotient of the product of the phase spaces for the individual tetrahedra. 
 Altogether, the union $\{C_I\} \cup \{E_J\}$ parametrises $L$ independent positions of the $L$ tetrahedra $\{\D_i\}$, and one can find their conjugate momenta $\{\G_J\} \cup \{\L_J\}$ satisfying
\begin{align}
    \{ C_I,\G_J \} = -\d_{I,J}
    \;,\;\;\quad
    \{ E_I, \L_J \} = - \d_{I,J}
    \;,\;\;\quad
    \{ \G_I,\L_J \} = 0
    \,.
\end{align}
These can be organised into an element $h \in Sp(2L,\mathbb{Z})$ as:
\begin{align}
    \begin{pmatrix}
        C_1 \\
        \vdots \\
        E_1 \\
        \vdots \\
        \hline
        \G_1 \\
        \vdots \\
        \L_1 \\
        \vdots
    \end{pmatrix}
    \;=\;
    h\cdot
    \begin{pmatrix}
        Z_1 \\
        \vdots \\
        \vdots \\
        Z_L \\
        \hline
        Z''_1 \\
        \vdots\\
        \vdots \\
        Z''_L
    \end{pmatrix}
    \,,
    \label{eq: sp element}
\end{align}
which, by setting $h=\left(
    \begin{array}{c|c}
       A  & B  \\
       \hline
        D & F
    \end{array}
    \right)$, can be decomposed into three generators
\begin{align}
    h_S 
    = 
    \left(
    \begin{array}{c|c}
       {\bf 1}-V  & -V \\
       \hline
        V & {\bf 1} -V
    \end{array}
    \right)
    \;,\;\quad
    h_T
    =
    \left(
    \begin{array}{c|c}
       {\bf 1}  & {\bf 0} \\
       \hline
        K & {\bf 1}
    \end{array}
    \right)
    \;,\;\quad
    h_U
    =
    \left(
    \begin{array}{c|c}
       U  & {\bf 0} \\
       \hline
        {\bf 0} & U^{-1 t}
    \end{array}
    \right)
    \,,
\end{align}
where $V = \text{diag}(v_1,\cdots,v_L)$ with $v_j \in \{0,1\}$, and $U\in GL(L,\mathbb{Z})$. We note that, in the case of interest in this section, there will not be any `boundary' edges $E_J$.

\medskip
\noindent
{\bf Gluing rules.} In the DGG construction, each tetrahedron $\D_i$ corresponds to a 3d $\CN=2$ theory $T_{\D_i}$ of a chiral multiplet coupled to a background $U(1)$ symmetry with effective CS level $-\half$ \cite{Dimofte:2011ju}. Then, the $Sp(2L,\mathbb{Z})$ action $h$ on the $L$ chirals $\{T_{\D_i}\}$ implements a particular gauging resulting in a specific 3d $\CN=2$ abelian gauge theory. Namely, the generator $h_U$ relabels the $U(1)^L$ background gauge fields by $U$, $h_T$ adds mixed CS terms with level matrix $K$, and $h_S$ makes the $j$-th $U(1)$ background gauge field into a dynamical gauge field if $v_j = 1$. 

The 3d $\CN=2$ superpotential terms $O_I$ of half-BPS chiral primaries are given by `easy' internal edges $C_I$ defined by the condition that {\it at most one of} $(g_{Ij}^{(0)},g_{Ij}^{(1)},g_{Ij}^{(2)})$ {\it is nonzero for each} $j$. We then consider:
\begin{align}
    W = \sum_{C_I\in \text{easy}}
    O_I
    \,.
    \label{eq: easy superpotential}
\end{align}
Consequently, we formally write the resulting 3d $\CN=2$ gauge theory as:
\begin{align}
    T\big[M_3,\{\s_i=0\}\big]
    \;=\;
    \bigg[\;
    h \circ 
    \Big( \bigotimes_{i=1}^{L} T_{\D_i} \Big)
    \;\;
    \text{with superpotential}
    \;\eqref{eq: easy superpotential}
    \;
    \bigg]~.
\end{align}
For a generic polarisation choice $\s=\{\s_i\}$, the construction follows in a parallel way, resulting in a different gauge-theory description $T[M_3,\s]$. Nevertheless, the distinct gauge theories for a given $M_3$ flow to the same IR fixed point, which we generally denote by $T[M_3]$ regardless of choices of polarisation and triangulation, since the infrared theory should only depend on the topology of $M_3$.

\medskip
\noindent {\bf The simple ACSM description for $|\det(B)|=1$.} 
For a generic polarisation choice $\s$, the coefficients in \eqref{eq: CI=ZZZ} reshuffle as
\begin{align}
    g_{Ij} := g_{Ij}^{([\s_i])}
    \;,\;\;
    g'_{Ij} := g_{Ij}^{([\s_i+1])}
    \;,\;\;
    g''_{Ij} := g_{Ij}^{([\s_i+2])}
    \,,
    \label{eq: s shuffle}
\end{align}
where $[s] := s \, \text{mod}\, 3$. Then, the two $L\times L$ matrices $(A,B)$ in the $Sp(2L,\mathbb{Z})$ element $h$, called the {\it Neumann-Zagier} (NZ) matrices \cite{Dimofte:2012qj,NeumannZagier1985}, become
\begin{align}
    A = g-g'
    \;,\;\;\qquad
    B= g''-g'
    \,.
\end{align}
Whenever $|\text{det}(B)|=1$, the gauge theory description simplifies considerably.\footnote{For a recent resolution for the gauge theory description with $\det(B) \neq 1$ cases, see \cite{Gang:2025ykf}.} Indeed, we then have a 3d $\CN=2$ $U(1)^L$ gauge theory with a bare CS level matrix
\begin{align}
    K = B^{-1} A
    \,,
    \label{eq: K}
\end{align}
and $L$ chiral multiplets $\Phi_i$ ($i=1,\cdots,L$) of charge $+\d_{ij}$ under the $j$-th $U(1)$, and with vanishing R-charge \cite{Dimofte:2012qj}. The $L$ `internal' edges $C_I$ have linearly-independent angle variables since the matrix $(A|B)$ in~\protect\eqref{eq: sp element} has rank $L$. They are also all easy edges, and the corresponding superpotential terms $O_I$ are:
\begin{align}
    O_I
    =
    \Big(
    \prod_{j=1}^{L} \phi_j^{g_{Ij}}
    \Big)
    V_{\mathfrak{m}^{(I)}}
    \,,
    \label{eq: half BPS op}
\end{align}
where $\phi_j$ is the scalar component of the chiral multiplet $\Phi_j$, while $V_{\mathfrak{m}^{(I)}}$ is a bare monopole operator with $j$-th magnetic flux $\mathfrak{m}_{j}^{(I)} = -B_{Ij}$. Here the reference R-charge of $O_I$ is given by (see {\it e.g.}~\cite{Bashkirov:2010kz,Benini:2011cma,Gang:2024loa}):
\begin{align}
    R_0[O_I] = \sum_{j=1}^L \frac{\big| \mathfrak{m}_j^{(I)} \big| + \mathfrak{m}_j^{(I)}}{2}
    \,,
\end{align}
so that the infrared R-charge with topological symmetry mixing is given by
\begin{align}
    R = R_0 + \sum_{j=1}^L \m_j\, T_j
    \,,
\end{align}
where $T_j$ and $\m_j$ are the $U(1)_{T_j}$ topological symmetry charge and its corresponding mixing parameter. Finally, since the superpotential has $R$-charge $2$, we have a constraint:
\begin{align}
    R[O_I]
    =
    R_0[O_I] + \sum_{j=1}^L \m_j\, \mathfrak{m}_j^{(I)}
    =2
    \label{eq: mu R}
\end{align}
for each superpotential term $O_I$. This fully determines $\mu$ at the TQFT point, as we explain momentarily. 

\subsection{DGG theories on lens spaces and SCFT points}
\label{subsec:DGG and SCFT}

A crucial fact, worth emphasising again, is that our triangulation of the lens space $L(N, 2k)$ gives us directly the 3d TQFT that we denoted by $T_A[M_3^{(k)}]$, for reasons that we already sketched in section~\ref{subsec:3d3drk0} and which we will further explore here. In the following, we will clarify how to directly construct the 3d $\CN=2$ ACSM theory corresponding to the 3d $\CN=4$ SCFT $T[M_3^{(k)}]$ itself, whenever it exists, as anticipated in section~\ref{subsec:3d3drk0}.

Let us start this discussion with a simple observation. As recalled above, for each tetrahedron $\Delta_i$ we have 
\be\label{2Z2pii}
2(Z_i + Z_i' + Z_i'')= 2\pi i~,
\ee
where the left-hand-side is the sum over the variables slotted into the six edges of $\Delta_i$. Then, in the full triangulated $M_3^{(k)}$, we have $L$ linearly independent edges $C_I$ defined as in~\eqref{eq: CI=ZZZ}, but where we might not necessarily impose the condition $C_I= 2\pi i$ yet. 
The triangulation also has an external `hard' edge $C_\infty$, as discussed around~\eqref{E euler}. We then always have a linear relation between edge variables:
\be\label{CI lin rel}
C_\infty + \sum_{I=1}^L C_I = 2\pi i L~.
\ee
This holds identically as a relation amongst the dihedral angle variables, since summing over all $L+1$ edges is equivalent to exhausting all $6L$ edge slots using the $3L$ variables $(Z_i, Z_i', Z_i'')$, by the definition~\eqref{eq: CI=ZZZ} applied to both the `internal' edge $C_I$ and to the `external' edge $C_\infty$ in the triangulation. Therefore~\eqref{CI lin rel} simply follows from~\eqref{2Z2pii}. 

\medskip
\noindent
{\bf The TQFT point.} 
First, note that the standard DGG construction on $M_3^{(k)}$ instructs us to turn on all monopole superpotential terms~\eqref{eq: half BPS op}, namely:
\be\label{W TQFT}
W= \sum_{I=1}^L O_I~.
\ee
The $R$-charge condition~\eqref{eq: mu R} then uniquely determines the mixing parameters $\mu=\mu^\ast$ at which we have the TQFT point:
\be\label{muTQFT gen form}
\mu_{\rm TQFT}^\ast =B^{-1}(R_0- 2\rho) = -\half \rho  + \half B^{-1} |B| \rho- 2 B^{-1}\rho~,
\ee
where $\rho=(1, 1, \cdots, 1)$ and $|B|$ denotes the $L\times L$ matrix with entries $|B_{ij}|$. Note that this TQFT point is then automatically reached once we impose the condition $C_I= 2\pi i$ on every easy edge, which is the geometric counterpart of the field-theory constraint~\eqref{eq: mu R}. It then follows from the linear relation~\eqref{CI lin rel} that
\be
C_\infty=0
\ee
for the dihedral angle of the `external' edge of the lens space.%
\footnote{This means that the meridian holonomy around $C_\infty$ is $-\mathbf{1} \in  SL(2,\mathbb{C})$, hence trivial in $PSL(2,\mathbb{C})$.} 
The external edge $C_\infty$ is therefore completely degenerate, which is naturally interpreted as the 3d uplift of the irregular singularity. 

Note that the presence of such a degenerate edge is forced upon us by the topology of the triangulation. Indeed, the link of the single vertex is a two-sphere, and one can show using the Gauss--Bonnet theorem that the total angle deficit is then
\be
\sum_{e\in {\rm edges}} (2\pi i - C_e)= \pi i \, \chi(S^2)= 2\pi i
\ee
for any one-vertex triangulation of a closed three-manifold. This gives us a second derivation of~\eqref{CI lin rel}. That linear relation then localises the unavoidable deficit angle at $C_\infty$ once all the internal edges are closed off.

\medskip
\noindent
{\bf Flowing to SCFT points.} To obtain the 3d $\CN=4$ rank-0 fixed point, when it exists, we need to turn on some additional fugacity parameters $\nu$ for some combination of the topological symmetries of the ACSM theory without superpotential. The minimal prescription, which we consider here, consists in removing a single term in the superpotential~\eqref{W TQFT}. That is, we choose an edge $C_I$ and consider the superpotential with $L-1$ terms:
\be\label{W OI removed}
W= \sum_{J\neq I} O_J~.
\ee
This frees up one symmetry $U(1)_{\CA^{(I)}}$ such that
\be
\CA^{(I)}[O_J]= -\delta_{IJ}~.
\ee
It is given explicitly by:
\be\label{CAI def}
\CA^{(I)} = \sum_{j=1}^L (\mu_{\CA^{(I)}})_j T_j~, \qquad \qquad \mu_{\CA^{(I)}} = B^{-1} e_I \in \Z^L~,
\ee
where we defined the index vector $(e_I)_j=\delta_{Ij}$. 
Turning on a 3d $\CN=2$ (complexified) real mass $\nu_{\CA^{(I)}}\in \mathbb{C}$ for $\CA^{(I)}$ corresponds to
\be
\nu = \nu_{\CA^{(I)}} \mu_{\CA^{(I)}} \qquad \leftrightarrow\qquad  \nu_{\CA^{(I)}} = (B\nu)_I~,
\ee
where $\nu$ denotes the vector of FI parameters as defined around~\eqref{eq: ACSM WO}. 
The real mass for $U(1)_{\CA^{(I)}}$ is realised geometrically as a non-trivial holonomy around the internal edge $C_I$~\cite{Dimofte:2010tz, Terashima:2011qi, Dimofte:2011ju, Dimofte:2014ija, Gang:2014ema}. It follows from~\eqref{CI lin rel} that we have $C_I+C_\infty=2\pi i$, and hence we also have a non-trivial holonomy around the external edge:
\be
C_I - 2\pi i = - 2\pi i \nu_{\CA^{(I)}}~, \qquad \quad
C_\infty  =2\pi i \nu_{\CA^{(I)}}~.
\ee
In general, the 3d $\CN=2$ RG flow for the ACSM theory with superpotential~\eqref{W OI removed} could lead us to some non-trivial 3d $\CN=2$ fixed point in the infrared, whose properties can be probed by supersymmetric partition functions upon finding the critical value $\nu_{\CA^{(I)}}^\ast$ that determines the superconformal $R$-charge by $F$-maximisation. If the IR fixed point has enhanced 3d $\CN=4$ supersymmetry such that $U(1)_{\CA^{(I)}}$ is an axial symmetry, the larger supersymmetry fixes the mixing parameter entirely to $\nu_{\CA^{(I)}}=1$ at the SCFT point. That corresponds to the relation~\eqref{rel between R and RTQFT} between $R$-charges or, equivalently, to the relation
\be\label{mu SCFT and TQFT}
\mu_{\rm SCFT}^\ast = \mu_{\rm TQFT}^\ast + \mu_{\CA^{(I)}}
\ee
between the values of $\mu$ at the SCFT and TQFT points~\cite{Closset:2016arn, Go:2025ixu}.
Here, an ``axial symmetry'' means a $U(1)$ flavour symmetry $\CA$ in the UV that becomes part of the $SU(2)_C\times SU(2)_H$ $R$-symmetry of the 3d $\CN=4$ supersymmetric fixed point with $\CA= C-H$. Note also that the mixing parameter $\hat{\nu}_\CA$ in~\eqref{rel between R and RTQFT} and our parameter $\nu_\CA$ are related as
\be\label{mu vs mutilde}
\hat{\nu}_{\CA^{(I)}} = \nu_{\CA^{(I)}}-1~,
\ee
the TQFT point being at $\hat{\nu}_{\CA^{(I)}}=-1$ and the SCFT point at $\hat{\nu}_{\CA^{(I)}}=0$. In summary:
\bea
&{\rm TQFT:}\quad &&\nu_{\CA^{(I)}}=0 \quad &&\longleftrightarrow\quad && (C_I~,\, C_\infty) =(2\pi i, 0)~,\\
&{\rm SCFT:}\quad &&\nu_{\CA^{(I)}}=1 \quad &&\longleftrightarrow\quad && (C_I, C_\infty) =(0~,\, 2\pi i)~.
\eea
The relation~\eqref{mu SCFT and TQFT} then gives us a {\it prediction} for the result of $F$-maximisation in our 3d $\CN=2$ ACSM theory with superpotential~\eqref{W OI removed} whenever it flows to a 3d $\CN=4$ fixed point: $F(\nu_{\CA^{(I)}})$ should have a maximum at $\nu_{\CA^{(I)}}=1$. We then have the integral vector:
\be
\mu_{\rm SCFT}^\ast = -\half \rho  +  B^{-1}\left(\half |B| \rho- 2\rho+ e_I \right) \in \Z^L~.
\ee
Now, as anticipated in section~\ref{subsec:3d3drk0}, not every choice of $C_I$ leads to a 3d $\CN=4$ SCFT in the IR. Instead, we must choose the `drilled out' edge $C_I$ such that it is isotopic to $C_0$, the $S^1_r$ circle itself, which gives us the condition~\eqref{CI to 2k Cinf} on the chord length $s_I$ of $C_I$ inside the polygon describing the Gaiotto curve.

If we choose an edge $C_I$ such that $s_I \neq \pm 2k$ mod $N$ instead, we are really considering a different class of DGG construction on a non-trivial knot complement~\eqref{X knot complement}. Then we do not expect any supersymmetry enhancement, and $F$-maximisation on the parameter $\nu_{\CA^{(I)}}$ should generally lead to a non-trivial 3d $\CN=2$ fixed point at irrational values of the $R$-charges. Another `experimental' confirmation that such a theory is unrelated to the circle compactification of $\CT_{A_{2n}}$ is that there are typically more than $n+1$ Bethe vacua in this case --- this we find out by numerically solving the Bethe equations in such cases. We will discuss one example of this phenomenon for $n=1$, $k=-2$ below, but leave a more complete investigation for future studies. See also~\cite{Gang:2018wek} for some closely related work.

\medskip
\noindent
{\bf Dropping multiple edges.} We could similarly consider dropping several edges $C_I$ for $I\in S$ some subset $S\subseteq \{1, 2, \cdots, L\}$. We then consider the superpotential:
\be\label{W S removed}
W= \sum_{J\notin S} O_J~.
\ee
Each $C_I$ we drop corresponds to re-introducing a topological symmetry $\CA^{(I)}$ defined as in~\eqref{CAI def}. In the geometry $M_3^{(k)}$, we have the holonomies:
\be
C_I - 2\pi i = - 2\pi i \nu_{\CA^{(I)}} \quad \forall I\in S~, \qquad \quad
C_\infty  =2\pi i \sum_{I\in S}\nu_{\CA^{(I)}}~.
\ee
In general, the 3d RG flow from this ACSM theory could reach a variety of 3d $\CN=2$ fixed points, which one can investigate using $F$-maximisation. Instead, let us focus on the case where $C_I$ is isotopic to $C_0$ for every $I\in S$. Then, we expect that the theory still flows to the 3d $\CN=4$ SCFT $T[M_3^{(k)}]$, with the axial symmetry
\be\label{CA sum IinS}
\CA\equiv \sum_{I\in S}\CA^{(I)}~,
\ee
while all other symmetries orthogonal to $\CA$, namely $\tilde{\CA}^{(IJ)}= \CA^{(I)}-\CA^{(J)}$ for $I, J\in S$ with $I\neq J$, are `flat directions' corresponding to symmetries that decouple in the IR. The prediction is then that all 3d $\CN=2$ supersymmetric partition functions on closed three-manifolds $\CN_3$ will only depend on the parameter $\nu_{\CA}$ for~\eqref{CA sum IinS} and be independent of the parameters $\tilde{\nu}_{\tilde{\CA}^{(IJ)}}$ for $\tilde{\CA}^{(IJ)}$. Geometrically, this is because, since $C_I \cong C_0$ for all $I$, holonomies can effectively cancel out between $C_I$ and $C_J$ in $M_3^{(k)}$. The physically important fact remains that we will have the SCFT point at:
\be
\nu_{\CA} =1 \qquad \leftrightarrow\qquad   C_\infty =  2\pi i~.
\ee
This is again a geometric prediction for the result of $F$-maximisation in such a theory. In the rest of this section, we will see how our general formalism plays out in explicit examples.

\subsection{The \texorpdfstring{$(A_1, A_2)$}{(A1,A2)} Argyres--Douglas theory}

In this subsection, we consider the $(A_1, A_2)$ theory, which is the original AD 4d $\CN=2$ fixed point~\cite{Argyres:1995jj}, dimensionally reduced on $S^1_r$.

\subsubsection{The case \texorpdfstring{$k=1$}{k=1}: \texorpdfstring{$M(2,5)$}{M(2,5)} VOA}
\begin{figure}[tbp]
\centering
\begin{tikzpicture}[scale=1.6, rotate=0]
\usetikzlibrary{calc}
	
   \draw[gray!90!white,line width=0.5pt] (-1,0)--(-1+7.8,0);
   \draw[gray!90!white,line width=0.5pt] (-0.5,1)--(-0.5+7.8,1);
   \draw[gray!90!white,line width=0.5pt] (0.2,0.6)--(0.2+7.8,0.6);
   \draw[gray!90!white,line width=0.5pt] (0.2,-0.6)--(0.2+7.8,-0.6);
   \draw[gray!90!white,line width=0.5pt] (-0.5,-1)--(-0.5+7.8,-1);
   
	\draw[|->,line width=1pt] (-0.3,-1.6)--(-0.2+7.8,-1.6);
	\node at (-0.3,-1.9) {$\th = 0$};
	\node at (-0.2+7.8,-1.9) {$\th = 2\pi$};

    \begin{scope}[shift={(0,0)}]
    \coordinate (P1) at (-1,0);
    \coordinate (P2) at (-0.5,1);
    \coordinate (P3) at (0.2,0.6);
    \coordinate (P4) at (0.2,-0.6);
    \coordinate (P5) at (-0.5,-1);
	
   \draw[blue, line width = 1.5pt] (P1)--(P3);
   \draw[blue, line width = 1.5pt] (P1)--(P4);
   \draw[line width = 1.2pt,line join=round] (P1)--(P2)--(P3)--(P4)--(P5)--(P1);
   
   \node at (-0.4,0.5) {\scriptsize $\textcolor{blue}{\g_1}$};
   \node at (-0.4,-0.5) {\scriptsize $\textcolor{blue}{\g_2}$};
   
   \end{scope}
   
    \begin{scope}[shift={(1.7,0)}]
    \coordinate (1T1) at (-1,0);
    \coordinate (1T2) at (-0.5,1);
    \coordinate (1T3) at (0.2,0.6);
    \coordinate (1T4) at (0.2,-0.6);
    \coordinate (1T5) at (-0.5,-1);
	
   \draw[line width = 1.2pt,line join=round] (1T1)--(1T5)--(1T4)--(1T3);
   \draw[line width = 1.2pt,line join=round] (1T3)--(1T5);
   \draw[line width = 1.2pt,line join=round, dashed] (1T1)--(1T4);
   \draw[green!70!black,line width = 1.7pt,line join=round] (1T1)--(1T3);
   
  \node at (-0.3,-0.25) {\tiny $Z_1$};
  \node at (-0.6,0.35) {\tiny $Z_1'$};
  \node at (0,-0.85) {\tiny $Z_1'$};
  \node at (-0.9,-0.55) {\tiny $Z_1''$};
  \node at (0.35,-0.2) {\tiny $Z_1''$};
  \node at (-0.1,0.7) {\scriptsize \textcolor{green!70!black}{$C_1$}};
   \end{scope}
   
    \begin{scope}[shift={(3.4,0)}]
    \coordinate (2T1) at (-1,0);
    \coordinate (2T2) at (-0.5,1);
    \coordinate (2T3) at (0.2,0.6);
    \coordinate (2T4) at (0.2,-0.6);
    \coordinate (2T5) at (-0.5,-1);
	
   \draw[line width = 1.2pt,line join=round] (2T5)--(2T1)--(2T2)--(2T3);
   \draw[line width = 1.2pt,line join=round] (2T2)--(2T5);
   \draw[line width = 1.2pt,line join=round, dashed] (2T1)--(2T3);
   \draw[green!70!black,line width = 1.7pt,line join=round] (2T3)--(2T5);
   
   \node at (-0.35,0.1) {\tiny $Z_2$};
  \node at (-0.9,0.5) {\tiny $Z_2'$};
  \node at (0,-0.2) {\tiny $Z_2'$};
  \node at (-0.95,-0.45) {\tiny $Z_2''$};
  \node at (0,0.85) {\tiny $Z_2''$};
  \node at (-0.2,-0.7) {\scriptsize \textcolor{green!70!black}{$C_2$}};
   \end{scope}
   
    \begin{scope}[shift={(4.6,0)}]
    \coordinate (3T1) at (-1,0);
    \coordinate (3T2) at (-0.5,1);
    \coordinate (3T3) at (0.2,0.6);
    \coordinate (3T4) at (0.2,-0.6);
    \coordinate (3T5) at (-0.5,-1);
	
   \draw[line width = 1.2pt,line join=round] (3T2)--(3T3)--(3T4)--(3T5);
   \draw[line width = 1.2pt,line join=round] (3T2)--(3T4);
   \draw[line width = 1.2pt,line join=round, dashed] (3T3)--(3T5);
   \draw[green!70!black,line width = 1.7pt,line join=round] (3T2)--(3T5);
   
   \node at (-0.23,0.1) {\tiny $Z_3$};
  \node at (-0.6,0.2) {\tiny $Z_3'$};
  \node at (0.33,-0.1) {\tiny $Z_3'$};
  \node at (0,0.85) {\tiny $Z_3''$};
  \node at (0,-0.85) {\tiny $Z_3''$};
  \node at (-0.65,-0.2) {\scriptsize \textcolor{green!70!black}{$C_3$}};
   \end{scope}
   
    \begin{scope}[shift={(6.3,0)}]
    \coordinate (4T1) at (-1,0);
    \coordinate (4T2) at (-0.5,1);
    \coordinate (4T3) at (0.2,0.6);
    \coordinate (4T4) at (0.2,-0.6);
    \coordinate (4T5) at (-0.5,-1);
	
   \draw[line width = 1.2pt,line join=round] (4T2)--(4T1)--(4T5)--(4T4);
   \draw[line width = 1.2pt,line join=round] (4T1)--(4T4);
   \draw[line width = 1.2pt,line join=round, dashed] (4T2)--(4T5);
   \draw[green!70!black,line width = 1.7pt,line join=round] (4T2)--(4T4);
   
   \node at (-0.6,-0.3) {\tiny $Z_4$};
  \node at (0,0.2) {\tiny $Z_4'$};
  \node at (-0.9,-0.5) {\tiny $Z_4'$};
  \node at (-0.9,0.5) {\tiny $Z_4''$};
  \node at (0,-0.85) {\tiny $Z_4''$};
  \node at (0.2,-0.2) {\scriptsize \textcolor{green!70!black}{$C_4$}};
   \end{scope}
	
    \begin{scope}[shift={(7.8,0)}]
    \coordinate (2P1) at (-1,0);
    \coordinate (2P2) at (-0.5,1);
    \coordinate (2P3) at (0.2,0.6);
    \coordinate (2P4) at (0.2,-0.6);
    \coordinate (2P5) at (-0.5,-1);
	
   \draw[blue, line width = 1.5pt] (2P1)--(2P4);
   \draw[blue, line width = 1.5pt] (2P2)--(2P4);
   \draw[line width = 1.2pt,line join=round] (2P1)--(2P2)--(2P3)--(2P4)--(2P5)--(2P1);
   
   \node at (0,0.3) {\scriptsize $\textcolor{blue}{\g_2}$};
   \node at (-0.4,-0.5) {\scriptsize $\textcolor{blue}{\g_1}$};
   \end{scope}

\end{tikzpicture}
\caption{\label{fig: A2ex} An ideal triangulation of $M_3^{(1)}$ into four tetrahedra. Internal edges $C_I$ are shown in green. The two triangulations of the Gaiotto curves at $\th = 0$ and $\th = 2\pi$ are identified.}
\end{figure}
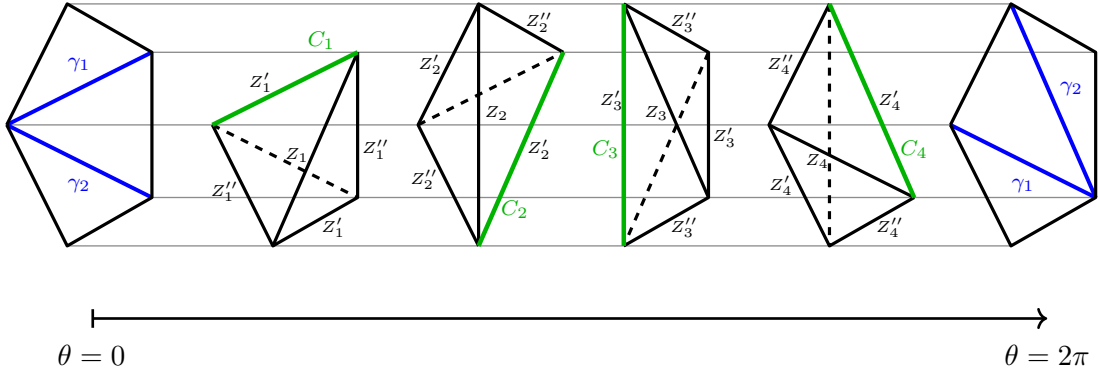
We already explained how to construct 
\be
M_{3}^{(1)}\cong L(5,2)
\ee
for $\CT_{A_2}$ in section~\ref{subsec:M3 from Delta}; see figure~\ref{fig: 5gon flip}. 
This three-manifold can be triangulated into four ideal tetrahedra as shown in figure \ref{fig: A2ex}, where green edges denote the four internal edges
\begin{align}
    C_1 & = Z_1' + Z_2 + Z_4
    \;,\;\;\qquad
    C_2  = Z_1 + Z_2' + Z_3\;,
    \nonumber\\
    C_3 & = Z_2 + Z_3' + Z_4
    \;,\;\;\qquad
    C_4 = Z_1 + Z_3 + Z_4'
    \,,
\end{align}
from which the coefficient matrices $g^{(i)}$ are organised as
\begin{align}
    g^{(0)} =
    \begin{pmatrix}
        0 & 1 & 0 & 1 \\
        1 & 0 & 1 & 0 \\
        0 & 1 & 0 & 1 \\
        1 & 0 & 1 & 0
    \end{pmatrix}
    \;\;,\;\;\;\quad
    g^{(1)} = 
    \begin{pmatrix}
        1 & 0 & 0 & 0 \\
        0 & 1 & 0 & 0 \\
        0 & 0 & 1 & 0 \\
        0 & 0 & 0 & 1
    \end{pmatrix}
    \;\;,\;\;\;\quad
    g^{(2)} = 
    \begin{pmatrix}
        0 & 0 & 0 & 0 \\
        0 & 0 & 0 & 0 \\
        0 & 0 & 0 & 0 \\
        0 & 0 & 0 & 0
    \end{pmatrix}\,.
    \label{eq: A2 ggg}
\end{align}
This is obtained for the polarisation $\s=(0,0,0,0)$, to which we will come back below. For now, let us consider another polarisation, $\s = (1,1,0,0)$, which generalises more naturally to $k\neq 1$. Then the shuffle \eqref{eq: s shuffle} gives us:
\begin{align}
    g =
    \begin{pmatrix}
        1 & 0 & 0 & 1 \\
        0 & 1 & 1 & 0 \\
        0 & 0 & 0 & 1 \\
        0 & 0 & 1 & 0
    \end{pmatrix}
    \;\;,\;\;\;\quad
    g' = 
    \begin{pmatrix}
        0 & 0 & 0 & 0 \\
        0 & 0 & 0 & 0 \\
        0 & 0 & 1 & 0 \\
        0 & 0 & 0 & 1
    \end{pmatrix}
    \;\;,\;\;\;\quad
    g'' = 
    \begin{pmatrix}
        0 & 1 & 0 & 0 \\
        1 & 0 & 0 & 0 \\
        0 & 1 & 0 & 0 \\
        1 & 0 & 0 & 0
    \end{pmatrix}
    \,,
\end{align}
so that the NZ matrices $(A,B)$ become
\begin{align}
    A = g - g' = 
    \begin{pmatrix}
        1 & 0 & 0 & 1 \\
        0 & 1 & 1 & 0 \\
        0 & 0 & -1 & 1 \\
        0 & 0 & 1 & -1
    \end{pmatrix}
    \;\;,\;\;\;\quad
     B = g'' - g' = 
     \begin{pmatrix}
        0 & 1 & 0 & 0 \\
        1 & 0 & 0 & 0 \\
        0 & 1 & -1 & 0 \\
        1 & 0 & 0 & -1
    \end{pmatrix}
    \,.
\end{align}
Since $\text{det}(B)=-1$, we find the CS level matrix $K$ of the DGG theory $T[M_3^{(1)},\s]$ as
\begin{align}
    K = B^{-1}A
     = 
     \begin{pmatrix}
        0 & 1 & 1 & 0 \\
        1 & 0 & 0 & 1 \\
        1 & 0 & 1 & 0 \\
        0 & 1 & 0 & 1
    \end{pmatrix}
    \,,
    \label{eq: A2 K}
\end{align}
with superpotential terms from the four easy internal edges \eqref{eq: half BPS op}:
\begin{align}
  W =  \phi_1 \phi_4 V_{(0,-1,0,0)} \;+\;
    \phi_2 \phi_3 V_{(-1,0,0,0)}\;+\;
    \phi_4 V_{(0,-1,1,0)}\;+\;
    \phi_3 V_{(-1,0,0,1)}~,
    \label{eq: A2 k=1 monopoles}
\end{align}
corresponding to $O_1, \cdots, O_4$, respectively. 
This maximal superpotential gives us the TQFT $T_A[M_3^{(1)}]$, fixing the mixing vector $\mu$ to:
\be
\mu_{\rm TQFT}^\ast= (-2,-2,-1,-1)~,
\ee
in agreement with~\eqref{muTQFT gen form}.

\medskip
\noindent
{\bf SCFT loci.} As is clear from the left-hand-side of figure~\ref{fig: chord}, all internal edges $C_I$ have chord distance $s_I=2$ in this example, thus satisfying the condition~\eqref{CI to 2k Cinf}. Therefore, every $C_I$ is isotopic to $C_0$ and dropping a single term $O_I$ from~\eqref{eq: A2 k=1 monopoles} allows us to reach the SCFT point most easily. Indeed, `drilling out' any one of the four edges $C_I$, we get the axial symmetry and corresponding 3d $\CN=4$ SCFT points:
\bea\label{CI n1k1 expls}
&C_1\; : \; &&\CA^{(1)}= T_2+T_3~, \quad &&\mu_{\CA^{(1)}}= (0,1,1,0)~, \quad &&\mu_{\rm SCFT}^\ast= (-2,-1,0,-1)~,\\
&C_2\; : \; &&\CA^{(2)}= T_1+T_4~, \quad &&\mu_{\CA^{(2)}}= (1,0,0,1)~, \quad &&\mu_{\rm SCFT}^\ast= (-1,-2,-1,0)~,\\
&C_3\; : \; &&\CA^{(3)}= -T_3~, \quad &&\mu_{\CA^{(3)}}= (0,0,-1,0)~, \quad &&\mu_{\rm SCFT}^\ast= (-2,-2,-2,-1)~,\\
&C_4\; : \; &&\CA^{(4)}= -T_4~, \quad &&\mu_{\CA^{(4)}}= (0,0,0,-1)~, \quad &&\mu_{\rm SCFT}^\ast= (-2,-2,-1,-2)~.\\
\eea
We have checked explicitly, using $F$-maximisation and the superconformal index, that we indeed have the same SCFT in each case, with:
\begin{align}\label{F GY again}
    F     \approx 0.642965~,
\end{align}
matching the known value of the Gang--Yamazaki SCFT~\cite{Gang:2021hrd}. Indeed, by dropping out several superpotential terms we can interpolate between the four points~\eqref{CI n1k1 expls} by tuning symmetry parameters that decouple in the IR. We will display an example of this below.

\medskip
\noindent
{\bf The Gaiotto-Kim theory from the $\s=0$ polarisation.} Coming back to the polarisation $\s=(0,0,0,0)$, by a similar computation as above, we find another ACSM theory with a CS level matrix 
\begin{align}
    K      = 
  \begin{pmatrix}
1 & -1 & 0 & -1 \\
-1 & 1 & -1 & 0 \\
0 & -1 & 1 & -1 \\
-1 & 0 & -1 & 1
\end{pmatrix}~,
    \label{eq: A2 K for GK}
\end{align}
and with the four diagonally-charged chiral multiplets. We find four half-BPS monopole operators  from the easy internal edges,
\begin{align}
    \phi_2 \phi_4 V_{(1,0,0,0)}
    \;,\;\;
    \phi_1 \phi_3 V_{(0,1,0,0)}
    \;,\;\;     
    \phi_2 \phi_4 V_{(0,0,1,0)}\;,\;\;
    \phi_1 \phi_3 V_{(0,0,0,1)}
    \,.
\end{align}
The discussion of the SCFT loci is the same as above. For definiteness, consider dropping $O_1$ from the superpotential, so that we retain a flavour symmetry:
\be
\CA^{(1)}= - T_1~,
\ee
which becomes an axial symmetry for the 3d $\CN=4$ enhanced supersymmetry at the SCFT point. Here the TQFT and SCFT points are:
\be
\mu_{\rm TQFT}^\ast=(1,1,1,1)~, \qquad \mu_{\rm SCFT}^\ast=(0,1,1,1)~,
\ee
as we can verify by $F$-maximisation. By construction, this theory is infrared dual to the one constructed above, as one can check explicitly by computing supersymmetric partition functions. In fact, the ACSM theory $T_A[M_3^{(1)}, \sigma=(0,0,0,0)]$ is exactly the one constructed by Gaiotto and Kim from the same minimal BPS chamber that determined our triangulation~\cite{Gaiotto:2024ioj}.

\medskip
\noindent {\bf Superconformal index.}
Going back to the first $U(1)^4$ theory above for concreteness, we can compute the superconformal index \eqref{eq: SCI} from the CS level matrix \eqref{eq: A2 K} and for any mixing parameter $\mu$. For definiteness, consider the ACSM with the superpotential $W=O_1+O_2$. Then we have the symmetries
\be\label{def CA and CAt A2k1 expl}
\CA\equiv \CA^{(3)}~, \qquad \tilde{\CA} \equiv \CA^{(4)}-\CA^{(3)}~,
\ee
where $\CA$ becomes the axial symmetry and $\tilde{\CA}$ is expected to decouple. Indeed, using the notation:
\be
    a \equiv \mu_{\CA^{(3)}} = (0,0,-1,0)~,
\qquad\quad
    \Wa\equiv \mu_{\CA^{(4)}}- \mu_{\CA^{(3)}}  = (0,0,1,-1)
\ee
for the mixing vectors, we find:
\begin{align}
    \CI_{S^2 \times S^1}^{T[M_{3}^{(1)}]}
    (y_i=\eta^{a_i} \widetilde{\eta}^{\Wa_i},\q)
    &=
    \!\!
    \sum_{\mathfrak{m} \in \mathbb{Z}^4}
    (-\q^{\frac{1}{2}})^{-2\mathfrak{m}_1 -2\mathfrak{m}_2 -2\mathfrak{m}_3-\mathfrak{m}_4}
    \eta^{-\mathfrak{m}_3}
    \widetilde{\eta}^{\mathfrak{m}_3-\mathfrak{m}_4}
    \CJ_{\q} (\mathfrak{m}_1, \!-\mathfrak{m}_2 - \!\mathfrak{m}_3)
    \nonumber\\
    &\qquad\times
    \CJ_{\q} (\mathfrak{m}_2, - \mathfrak{m}_1 - \mathfrak{m}_4)
    \CJ_{\q} (\mathfrak{m}_3, - \mathfrak{m}_1 - \mathfrak{m}_3)
    \CJ_{\q} (\mathfrak{m}_4, - \mathfrak{m}_2 - \mathfrak{m}_4)
    \nonumber\\
    &=
    1 - q 
    - \Big(\eta+\frac{1}{\eta}\Big) q^{\frac{3}{2}}
    - 2 q^2
    -\Big(\eta+\frac{1}{\eta}\Big) q^{\frac{5}{2}}
    - 2 q^3 + \cdots~,
\end{align}
with the fugacities $\eta$ and $\widetilde{\eta}$ for $\CA$ and $\tilde\CA$, respectively. Note that the superconformal index does not depend on $\widetilde{\eta}$, as expected. This index exactly matches with that of the Gang--Yamazaki minimal rank-0 SCFT \cite{Gang:2018huc}. The topological $A$-twist is here implemented by the substitution $\eta \to (- \q^{1/2})^{-1}$, in which case one can check by explicit computation that:%
\footnote{Here we did an explicit computation at order $\q^4$, for simplicity; of course conjecturally the full index trivialises.}
\begin{align}
    \CI_{S^2 \times S^1}^{T_A[M_{3}^{(1)}]}
    (\q)
    =
    \CI_{S^2 \times S^1}^{T[M_{3}^{(1)}]}
    (y_i=\eta^{a_i} \widetilde{\eta}^{\Wa_i},\q)
    \big|_{\eta \to (-\q^{1/2})^{-1}}
    =
    1 + \CO(\q^5)
    \,.
\end{align}
This strongly supports the claim that the $T_A[M_3^{(1)}]$ theory becomes a 3d TQFT without any local operators.

\medskip
\noindent {\bf Modular data.}
The modular data at the TQFT point can be also extracted from the 3d A-model results by computing the handle gluing and the fibering operators as in \eqref{eq: HtoS2} and \eqref{eq: HF to ST}. We find the ordered sets:
\begin{align}
    \{ |S_{0\a}| \} = 
    \Big\{
    \frac{2}{\sqrt{5}} \sin\Big( \frac{2\pi}{5} \Big)
    ,
    \frac{2}{\sqrt{5}} \sin\Big( \frac{\pi}{5} \Big)
    \Big\}
    \;,\;\;\quad
    \{ T_{\a\a}  \} = 
    \Big\{ e^{2\pi i ({11\ov 60})} , e^{2\pi i (\frac{59}{60})} \Big\}
    \,,
    \label{S and T matrix A2}
\end{align}
where the two Bethe vacua are ordered to match the expected TQFT answer. These are indeed compatible with the modular data of the $M(2,5)$ Virasoro minimal model. Setting $n=1$ in~\eqref{ST VOAk}, we indeed have: 
\begin{align}
    S =
    \frac{2}{\sqrt{5}}
    \begin{pmatrix}
        -\sin \big( \frac{2\pi}{5} \big) & \sin \big( \frac{\pi}{5} \big) \\
        \sin \big( \frac{\pi}{5} \big) & \sin \big( \frac{2\pi}{5} \big)
    \end{pmatrix}
    \;,\;\;\quad
    T= e^{-{2\pi i\ov 24} (-\frac{22}{5})} 
    \text{diag}
    \,
    \Big(
    e^{2\pi i (0)} , e^{2\pi i (\frac{4}{5})}
    \Big)\;.
\end{align}
More generally, we have checked the second relation in~\eqref{eq: HF to ST} numerically for various values of $(q,p)$, which gives us more refined checks of the infrared modular data. Such exact matching of the phases of supersymmetric and TQFT partition functions depends on taking the `correct' choice for the gravitational CS contact term $K_g$, as we discuss next.

\medskip\noindent
{\bf Central charge $c_{\rm 2d}$ and the choice of $K_g$.} To match the fibering operator with the $T$ matrix as in~\eqref{S and T matrix A2}, we have set the gravitational CS level to be:
\be
K_g= 4~.
\ee
Similarly, for the GK theory~\eqref{eq: A2 K for GK}, we would need to fix $K_g=26$ to reproduce the correct infrared data. From the point of view of the 3d $\CN=2$ ACSM theory, the choice of $K_g$ is a Chern--Simons contact term which is part of the UV definition of the theory in 3d~\cite{Closset:2012vp}, and it is a choice we can make at will. In principle, the `correct' $K_g$ should be fully determined by the compactification of the $A_1$ 6d $\CN=(2,0)$ SCFT on the triangulated manifold $M_3$ with polarisation $\sigma$. However, since we are not aware of any explicit derivation of such gravitational contact terms from 6d, we simply computed the needed $K_g$ for each choice of $\sigma$ such that~\eqref{eq: HF to ST} holds true for
\be\label{c2d 225 expl}
c_{2d}=-\frac{22}{5}
\ee
and for any Seifert fibering operator. Indeed, we computed $K_g$ for 43 distinct polarisations leading to `easy' ACSM theories using~\eqref{eq: HF to ST}, which also provides a check that all Seifert partition functions agree for these 43 distinct ACSM theories, providing strong evidence for the claim that they are all infrared dual. Note that the fact that we find $K_g$ to be an integer in order to match to~\eqref{c2d 225 expl}, for any ACSM theory determined by a fixed polarisation, is itself a strong check of our formalism, as indeed the bare CS level $K_g$ should be an integer~\cite{Closset:2012vp}.

\medskip
\noindent {\bf Half-index.}
Turning on the same symmetries as in~\eqref{def CA and CAt A2k1 expl}, computing the half-index of $T_A[M_3^{(1)}]$ with the $(\CD,D_c)$ boundary condition gives%
\footnote{Note that we need to set $\widetilde{\eta}=1$ here. Indeed, while the symmetry $\tilde\CA$ decouples in the bulk, boundary conditions are UV data that need to be set appropriately.}
\begin{align}
    \CI_{D^2 \times S^1}^{T_A[M_3^{(1)}]}(\q)
    &:=
    \CI_{D^2 \times S^1}^{T[M_3^{(1)}]} (y_i=\eta^{a_i},\,\q)
    \big|_{\eta \to (-\q^{1/2})^{-1}}
    \nonumber\\
    &=
    \sum_{\mathfrak{m}\in \mathbb{Z}_{\geq 0}^4}
    \frac{(-1)^{\mathfrak{m}_3+\mathfrak{m}_4} 
    \q^{\mathfrak{m}_1 \mathfrak{m}_2 + \mathfrak{m}_1 \mathfrak{m}_3 + \mathfrak{m}_2 \mathfrak{m}_4 + \mathfrak{m}_1 + \mathfrak{m}_2 
    + \frac{1}{2} \mathfrak{m}_3 (\mathfrak{m}_3 + 1)
    + \frac{1}{2} \mathfrak{m}_4 (\mathfrak{m}_4 + 1) 
    } 
    }{(\q)_{\mathfrak{m}_1}(\q)_{\mathfrak{m}_2}(\q)_{\mathfrak{m}_3}(\q)_{\mathfrak{m}_4}}
    \nonumber\\
    &=
    1 + 0 \q + \q^2 + \q^3 + \q^4 + \q^5 + 2 \q^6 + 2 \q^7 + 3 \q^8 + 3 \q^9  + \cdots
    \,,
    \label{eq: A2 half ind}
\end{align}
which coincides with the vacuum character of $M(2,5)$ up to the overall $\q^{c_{\rm 2d}\ov 24}$ factor
\begin{align}
    \q^{-\frac{11}{60}}
    \chi_0^{M(2,5)}(\q)
    &=1 + 0 \q + \q^2 + \q^3 + \q^4 + \q^5 + 2 \q^6 + 2 \q^7 + 3 \q^8 + 3 \q^9
    +\cdots\,.
\end{align}
Indeed, this can be calculated from the Schur index of the $A_2$ theory by the SCFT/VOA correspondence \cite{Cordova:2015nma}
\begin{align}
    \q^{-\frac{11}{60}}\CI_{\text{Schur}}^{A_2}(\q)
    &=
    (\q)_\infty^2
    \sum_{l_1,l_2=0}^\infty
    \frac{\q^{l_1 + l_2 + l_1 l_2}}{(\q)_{l_1}^2 (\q)_{l_2}^2 }
    \nonumber\\
    &=\sum_{l_1,l_2,k_1,k_2=0}^{\infty}
    \frac{
    (-1)^{k_1+k_2}
    \q^{\frac{k_1(k_1+1)}{2}+\frac{k_2(k_2+1)}{2} +l_1 k_1 + l_2 k_2 + l_1 l_2 + l_1 + l_2 }
    }{(\q)_{l_1}(\q)_{l_2}(\q)_{k_1}(\q)_{k_2}}
    \label{eq: A2 Schur}
\end{align}
where we used a $\q$-Pochhammer identity in the second line
\begin{align}
    \frac{1}{(\q)_n} = \frac{1}{(\q)_{\infty}} \sum_{k=0}^{\infty}
    \frac{ (-1)^k \q^{\frac{k(k-1)}{2}} }{(\q)_k} \q^{(n+1)k}
    \,.
    \label{eq: qpoch identity}
\end{align}
Upon relabeling the summation variables $(l_1,l_2,k_1,k_2)$ $\to$ $(\mathfrak{m}_1,\mathfrak{m}_2,\mathfrak{m}_3,\mathfrak{m}_4)$ the Schur index expression \eqref{eq: A2 Schur} perfectly matches with the half-index \eqref{eq: A2 half ind} up to turning off the decoupling fugacity $\widetilde{\eta} \to 1$. Therefore, we verify
\begin{align}
    \CI_{D^2 \times S^1}^{T_A[M_{3}^{(1)}]}(\q)
    =
    \q^{-\frac{11}{60}}
    \chi_0^{M(2,5)}(\q)
    \,,
\end{align}
in agreement with the SCFT/VOA correspondence and the expectation that the 3d TQFT $T_A[ M_{3}^{(1)}]$ can support the $M(2,5)$ VOA on its holomorphic boundary~\cite{Dedushenko:2018bpp, Dedushenko:2023cvd}.

\subsubsection{The case \texorpdfstring{$k=-1$}{k=-1}: \texorpdfstring{$osp(1|2)_1$}{osp(1|2)1} }
The triangulation of $M_3^{(-1)}$ is simply obtained by the orientation reversal of the ideal tetrahedra in figure \ref{fig: A2ex}, {\it i.e.} by exchanging the variables
\begin{align}
    Z_i'
    \;\;\longleftrightarrow \;\;
    Z_i''
    \,,
\end{align}
so that the four internal edges now become
\begin{align}
    C_1 & = Z_1'' + Z_2 + Z_4
   \;,\;\;\qquad
    C_2 = Z_1 + Z_2'' + Z_3
    \nonumber\\
    C_3 & = Z_2 + Z_3'' + Z_4
    \;,\;\;\qquad
    C_4 = Z_1 + Z_3 + Z_4''
    \,.
    \label{eq: k=-1 A2 internal edge}
\end{align}
With a polarisation choice $\s=(1,1,2,2)$, we get the coefficient matrices
\begin{align}
    g =
    \begin{pmatrix}
        0 & 0 & 0 & 0 \\
        0 & 0 & 0 & 0 \\
        0 & 0 & 1 & 0 \\
        0 & 0 & 0 & 1
    \end{pmatrix}
    \;\;,\;\;\;\quad
    g' = 
    \begin{pmatrix}
        1 & 0 & 0 & 1 \\
        0 & 1 & 1 & 0 \\
        0 & 0 & 0 & 1 \\
        0 & 0 & 1 & 0
    \end{pmatrix}
    \;\;,\;\;\;\quad
    g'' = 
    \begin{pmatrix}
        0 & 1 & 0 & 0 \\
        1 & 0 & 0 & 0 \\
        0 & 1 & 0 & 0 \\
        1 & 0 & 0 & 0
    \end{pmatrix}
    \,,
\end{align}
and the NZ matrices as
\begin{align}
    A = g-g' =
    \begin{pmatrix}
        -1 & 0 & 0 & -1 \\
        0 & -1 & -1 & 0 \\
        0 & 0 & 1 & -1 \\
        0 & 0 & -1 & 1
    \end{pmatrix}
    \;\;,\;\;\;\quad
    B = g''-g' =
    \begin{pmatrix}
        -1 & 1 & 0 & -1 \\
        1 & -1 & -1 & 0 \\
        0 & 1 & 0 & -1 \\
        1 & 0 & -1 & 0
    \end{pmatrix}
    \,,
\end{align}
where $\text{det}(B)=-1$, thereby the DGG theory has the CS level matrix
\begin{align}
    K = B^{-1}A
    =
    \begin{pmatrix}
        1 & 0 & 1 & 0 \\
        0 & 1 & 0 & 1 \\
        1 & 0 & 2 & -1 \\
        0 & 1 & -1 & 2
    \end{pmatrix}
    \,.
    \label{eq: A2 k=-1 K}
\end{align}
The superpotential that allows us to flow directly to the TQFT point comes from the four easy internal edges:
\begin{align}
   W= V_{(1,-1,0,1)}
    \;+\;
    V_{(-1,1,1,0)}
  \;+\;
    \phi_3 V_{(0,-1,0,1)}
  \;+\;
    \phi_4 V_{(-1,0,1,0)}
    \,,
\end{align}
which corresponds to $\mu_{\rm TQFT}^\ast = (-1,-1,0,0)$. 

\medskip
\noindent
{\bf SCFT loci.} As in the previous example, all internal edges $C_I$ have $s_I=2= \pm 2k$, therefore we can remove any one of the terms $O_I$ from the superpotential to reach the SCFT point most easily. There are thus four possibilities, giving us:
\bea\label{CI n1kminus1 expls}
&C_1\; : \; &&\CA^{(1)}= -T_1-T_3~, \quad  &&\mu_{\rm SCFT}^\ast= (-2,-1,-1,0)~,\\
&C_2\; : \; &&\CA^{(2)}= -T_2-T_4~, \quad &&\mu_{\rm SCFT}^\ast= (-1,-2,0,-1)~,\\
&C_3\; : \; &&\CA^{(3)}= T_1+T_3-T_4~,  \quad &&\mu_{\rm SCFT}^\ast= (0,-1,1,-1)~,\\
&C_4\; : \; &&\CA^{(4)}= T_2-T_3+T_4~, \quad &&\mu_{\rm SCFT}^\ast= (-1,0,-1,1)~.\\
\eea
This can again be checked by $F$-maximisation, and one finds that the $F$-function at the 3d $\CN=4$ fixed point takes the same value~\eqref{F GY again} as in the $k=1$ case. Indeed, in all examples we checked, we find that the $F$-function of $T[M_3^{(k)}]$ at fixed $n$ is given by~\eqref{Fscft}, independently of $k$. 

\medskip
\noindent {\bf Superconformal index.}
Let us focus on removing $C_1$. Then, by denoting the mixing vector as $a \equiv \m_{\CA^{(1)}} = (-1,0,-1,0)$ together with the SCFT point $\mu_{\rm SCFT}^\ast= (-2,-1,-1,0)$, the superconformal index reads
\begin{align}
    \CI_{S^2 \times S^1}^{T[M_{3}^{(-1)}]}
    (y_i=\eta^{a_i},\q)
    &=
    \sum_{\mathfrak{m}\in\mathbb{Z}^4}
    \big(-\q^{\frac{1}{2}}\big)^{-2\mathfrak{m}_1-\mathfrak{m}_2-\mathfrak{m}_3}
    \eta^{-\mathfrak{m}_1-\mathfrak{m}_3}
    \nonumber\\
    &\qquad\qquad\times
    \CJ_{\q} ( \mathfrak{m}_1,-\mathfrak{m}_1-\mathfrak{m}_3 )
    \CJ_{\q} ( \mathfrak{m}_2,-\mathfrak{m}_2-\mathfrak{m}_4 )
    \nonumber\\
    &\qquad\qquad\times
    \CJ_{\q} ( \mathfrak{m}_3,-\mathfrak{m}_1-2\mathfrak{m}_3+\mathfrak{m}_4 )
    \CJ_{\q} ( \mathfrak{m}_4,-\mathfrak{m}_2+\mathfrak{m}_3-2\mathfrak{m}_4 )
    \nonumber\\
    &=
    1 - q 
    - \Big(\eta+\frac{1}{\eta}\Big) q^{\frac{3}{2}}
    - 2 q^2
    -\Big(\eta+\frac{1}{\eta}\Big) q^{\frac{5}{2}}
    - 2 q^3 + \cdots
    \,.
\end{align}
This again coincides with the superconformal index of the GY theory as expected, since the orientation reversal $M_3^{(1)} \to M_3^{(-1)}$ implements the parity conjugation of the GY theory, under which the theory is invariant. Upon the topological A-twist, the index again becomes trivial:
\begin{align}
    \CI_{S^2 \times S^1}^{T_A[M_{3}^{(-1)}]}
    (\q)
    =
    \CI_{S^2 \times S^1}^{T[M_{3}^{(-1)}]}
    (y_i=\eta^{a_i},\q)
    \big|_{\eta \to (-\q^{1/2})^{-1}}
    =
    1~.
\end{align}

\medskip
\noindent {\bf Modular data and choice of $K_g$.}
The modular data of the TQFT can be extracted similarly to the $k=1$ example, as
\begin{align}
    \{ |S_{0\a}| \} = 
    \Big\{
    \frac{2}{\sqrt{5}} \sin\Big( \frac{2\pi}{5} \Big)
    ,
    \frac{2}{\sqrt{5}} \sin\Big( \frac{\pi}{5} \Big)
    \Big\}
    \;,\;\;
    \{ T_{\a\a} \} = 
    e^{-\frac{2\pi i}{24} (\frac{2}{5})} 
    \Big\{ e^{2\pi i (0)} , e^{2\pi i (\frac{1}{5})} \Big\}
    \,,
\end{align}
which is compatible with the modular data of the two irreducible modules in affine $osp(1|2)_1$ VOA
\begin{align}
    S =
    \frac{2}{\sqrt{5}}
    \begin{pmatrix}
        \sin \big( \frac{2\pi}{5} \big) & -\sin \big( \frac{\pi}{5} \big) \\
        -\sin \big( \frac{\pi}{5} \big) & -\sin \big( \frac{2\pi}{5} \big)
    \end{pmatrix}
    \;,\;\;
    T= e^{2\pi i (-\frac{1}{60})} 
    \text{diag}\,
    \Big(
    e^{2\pi i (0)} , e^{2\pi i (\frac{1}{5})}
    \Big)\,.
\end{align}
This ACSM theory with $K_g=4$ recovers the central charge $c_{\text{2d}}={2\ov 5}$ exactly, as one can check using~\eqref{eq: HF to ST}. 

\medskip
\noindent {\bf Half-index.}
The half-index of $T_A[M_3^{(-1)}]$ is calculated as follows:
\begin{align}
    \CI_{D^2 \times S^1}^{T_A[M_3^{(-1)}]}(\q)
    &:=
    \CI_{D^2 \times S^1}^{T[M_3^{(-1)}]} (y_i=\eta^{a_i} ,\q)
    \big|_{\eta \to (-\q^{1/2})^{-1}}
    \nonumber\\
    &=
    \sum_{\mathfrak{m}\in\mathbb{Z}_{\geq 0}^4}
    \frac{
    (-1)^{\mathfrak{m}_1+\mathfrak{m}_2}
    \q^{
    \frac{1}{2} \mathfrak{m}_1(\mathfrak{m}_1+1) +\frac{1}{2}\mathfrak{m}_2(\mathfrak{m}_2+1) + \mathfrak{m}_3^2 + \mathfrak{m}_4^2
    +\mathfrak{m}_1 \mathfrak{m}_3 + \mathfrak{m}_2 \mathfrak{m}_4 - \mathfrak{m}_3 \mathfrak{m}_4 
    }    }
    {(\q)_{\mathfrak{m}_1}(\q)_{\mathfrak{m}_2}(\q)_{\mathfrak{m}_3}(\q)_{\mathfrak{m}_4}}
    \nonumber\\
    &=1 + \q + \q^2 + \q^3 + 2 \q^4 + 2 \q^5 + 3 \q^6 + 3 \q^7 + 4 \q^8 + 5 \q^9 + \cdots
    \,.
    \label{eq: A2 k=-1 half-index}
\end{align}
This coincides with the vacuum character of the affine $osp(1|2)_1$ VOA computed from the trace of the monodromy operator~\cite{Cecotti:2010fi,Cecotti:2015lab,Kim:2024dxu}:
\begin{align}
    \tilde\chi_0^{osp(1|2)_1}(\q)
    &=
    (\q)_\infty^2
    \sum_{l_1,l_2=0}^{\infty}
    \frac{
    \q^{l_1^2 + l_2^2 - l_1 - l_2}
    }{(\q)_{l_1}^2 (\q)_{l_2}^2 }
    \nonumber\\
    &=1 + \q + \q^2 + \q^3 + 2 \q^4 + 2 \q^5 + 3 \q^6 + 3 \q^7 + 4 \q^8 + 5 \q^9  + \cdots
    \,,
\end{align}
where we defined $\tilde\chi_0 \equiv \q^{c_{\rm 2d}\ov 24}\chi_0$. This can be massaged using the $\q$-Pochhammer identity \eqref{eq: qpoch identity} into
\begin{align}
   \tilde\chi_0^{osp(1|2)_1}(\q)
    =
    \sum_{l_1,l_2,k_1,k_2=0}^{\infty}
    \frac{
    (-1)^{k_1 + k_2}
    \q^{
    \frac{1}{2}k_1(k_1+1) + \frac{1}{2}k_2(k_2+1)
    + l_1^2 + l_2^2
    + l_1 k_1 + l_2 k_2 - l_1 l_2
    }
    }
    {(\q)_{l_1} (\q)_{l_2} (\q)_{k_1} (\q)_{k_2} }
    \,,
\end{align}
which matches the half-index expression \eqref{eq: A2 k=-1 half-index} upon relabeling the summation variables $(k_1,k_2,l_1,l_2)$ $\to$ $(\mathfrak{m}_1,\mathfrak{m}_2,\mathfrak{m}_3,\mathfrak{m}_4)$. Thus, these computations support the claim that the DGG theory $T_A[M_3^{(-1)}]$ is a 3d TQFT that supports the affine $osp(1|2)_1$ VOA on its holomorphic boundary.

\subsubsection{The case \texorpdfstring{$k=-2$}{k=-2} : \texorpdfstring{$(G_2)_1$}{(G2)1} }
\begin{figure}[tbp]
\centering
\begin{tikzpicture}[scale=1.19, rotate=0]
\usetikzlibrary{calc}
	
   \draw[gray!90!white,line width=0.3pt] (-1,0)--(-1+11.4,0);
   \draw[gray!90!white,line width=0.3pt] (-0.5,1)--(-0.5+11.4,1);
   \draw[gray!90!white,line width=0.3pt] (0.2,0.6)--(0.2+11.4,0.6);
   \draw[gray!90!white,line width=0.3pt] (0.2,-0.6)--(0.2+11.4,-0.6);
   \draw[gray!90!white,line width=0.3pt] (-0.5,-1)--(-0.5+11.4,-1);
   
	\draw[|->,line width=1pt] (-0.3,-1.6)--(-0.2+11.4,-1.6);
	\node at (-0.3,-1.9) {\scriptsize $\th = 0$};
	\node at (-0.2+11.4,-1.9) {\scriptsize $\th = 2\pi$};

    \begin{scope}[shift={(0,0)}]
    \coordinate (P1) at (-1,0);
    \coordinate (P2) at (-0.5,1);
    \coordinate (P3) at (0.2,0.6);
    \coordinate (P4) at (0.2,-0.6);
    \coordinate (P5) at (-0.5,-1);
	
   \draw[blue, line width = 1.2pt] (P1)--(P3);
   \draw[blue, line width = 1.2pt] (P1)--(P4);
   \draw[line width = 1pt,line join=round] (P1)--(P2)--(P3)--(P4)--(P5)--(P1);
   
   \node at (-0.4,0.5) {\scriptsize $\textcolor{blue}{\g_1}$};
   \node at (-0.4,-0.5) {\scriptsize $\textcolor{blue}{\g_2}$};
   
   \end{scope}
   
    \begin{scope}[shift={(1.4,0)}]
    \coordinate (1T1) at (-1,0);
    \coordinate (1T2) at (-0.5,-1);
    \coordinate (1T3) at (0.2,-0.6);
    \coordinate (1T4) at (0.2,0.6);
    \coordinate (1T5) at (-0.5,1);
	
   \draw[line width = 1pt,line join=round] (1T1)--(1T5)--(1T4)--(1T3);
   \draw[line width = 1pt,line join=round] (1T3)--(1T5);
   \draw[line width = 1pt,line join=round, dashed] (1T1)--(1T4);
   \draw[green!70!black,line width = 1.2pt,line join=round] (1T1)--(1T3);
   
  \node at (-0.3,0.2) {\tiny $Z_1$};
  \node at (-0.4,-0.1) {\tiny $Z_1'$};
  \node at (-0.17,0.65) {\tiny $Z_1'$};
  \node at (-0.6,0.45) {\tiny $Z_1''$};
  \node at (0.06,0.2) {\tiny $Z_1''$};
  \node at (-0.5,-0.4) {\scriptsize \textcolor{green!70!black}{$C_1$}};
   \end{scope}
   
    \begin{scope}[shift={(2.8,0)}]
    \coordinate (2T1) at (-1,0);
    \coordinate (2T2) at (-0.5,-1);
    \coordinate (2T3) at (0.2,-0.6);
    \coordinate (2T4) at (0.2,0.6);
    \coordinate (2T5) at (-0.5,1);
	
   \draw[line width = 1pt,line join=round] (2T5)--(2T1)--(2T2)--(2T3);
   \draw[line width = 1pt,line join=round] (2T2)--(2T5);
   \draw[line width = 1pt,line join=round, dashed] (2T1)--(2T3);
   \draw[green!70!black,line width = 1.2pt,line join=round] (2T3)--(2T5);
   
   \node at (-0.35,-0.1) {\tiny $Z_2$};
  \node at (-0.63,-0.4) {\tiny $Z_2'$};
  \node at (-0.3,0.2) {\tiny $Z_2'$};
  \node at (-0.67,0.33) {\tiny $Z_2''$};
  \node at (-0.2,-0.65) {\tiny $Z_2''$};
  \node at (0,0.3) {\scriptsize \textcolor{green!70!black}{$C_2$}};
   \end{scope}
   
    \begin{scope}[shift={(3.7,0)}]
    \coordinate (3T1) at (-1,0);
    \coordinate (3T2) at (-0.5,-1);
    \coordinate (3T3) at (0.2,-0.6);
    \coordinate (3T4) at (0.2,0.6);
    \coordinate (3T5) at (-0.5,1);
	
   \draw[line width = 1pt,line join=round] (3T2)--(3T3)--(3T4)--(3T5);
   \draw[line width = 1pt,line join=round] (3T2)--(3T4);
   \draw[line width = 1pt,line join=round, dashed] (3T3)--(3T5);
   \draw[green!70!black,line width = 1.2pt,line join=round] (3T2)--(3T5);
   
   \node at (-0.23,0.08) {\tiny $Z_3$};
  \node at (-0.35,-0.2) {\tiny $Z_3'$};
  \node at (0.09,0) {\tiny $Z_3'$};
  \node at (-0.1,-0.57) {\tiny $Z_3''$};
  \node at (-0.1,0.57) {\tiny $Z_3''$};
  \node at (-0.65,-0.2) {\scriptsize \textcolor{green!70!black}{$C_3$}};
   \end{scope}
   
    \begin{scope}[shift={(5.1,0)}]
    \coordinate (4T1) at (-1,0);
    \coordinate (4T2) at (-0.5,-1);
    \coordinate (4T3) at (0.2,-0.6);
    \coordinate (4T4) at (0.2,0.6);
    \coordinate (4T5) at (-0.5,1);
	
   \draw[line width = 1pt,line join=round] (4T2)--(4T1)--(4T5)--(4T4);
   \draw[line width = 1pt,line join=round] (4T1)--(4T4);
   \draw[line width = 1pt,line join=round, dashed] (4T2)--(4T5);
   \draw[green!70!black,line width = 1.2pt,line join=round] (4T2)--(4T4);
   
   \node at (-0.35,0.15) {\tiny $Z_4$};
  \node at (-0.3,-0.15) {\tiny $Z_4'$};
  \node at (-0.65,0.45) {\tiny $Z_4'$};
  \node at (-0.68,-0.3) {\tiny $Z_4''$};
  \node at (-0.22,0.65) {\tiny $Z_4''$};
  \node at (0.03,-0.2) {\scriptsize \textcolor{green!70!black}{$C_4$}};
   \end{scope}
   
    \begin{scope}[shift={(6.3,0)}]
    \coordinate (1T1) at (-1,0);
    \coordinate (1T2) at (-0.5,1);
    \coordinate (1T3) at (0.2,0.6);
    \coordinate (1T4) at (0.2,-0.6);
    \coordinate (1T5) at (-0.5,-1);
	
   \draw[line width = 1pt,line join=round] (1T1)--(1T5)--(1T4)--(1T3);
   \draw[line width = 1pt,line join=round,dashed] (1T3)--(1T5);
   \draw[line width = 1pt,line join=round] (1T1)--(1T4);
   \draw[green!70!black,line width = 1.2pt,line join=round] (1T1)--(1T3);
   
  \node at (-0.3,-0.18) {\tiny $Z_5$};
  \node at (-0.4,0.15) {\tiny $Z_5'$};
  \node at (-0.17,-0.62) {\tiny $Z_5'$};
  \node at (-0.6,-0.45) {\tiny $Z_5''$};
  \node at (0.07,-0.2) {\tiny $Z_5''$};
  \node at (-0.5,0.43) {\scriptsize \textcolor{green!70!black}{$C_5$}};
   \end{scope}
   
    \begin{scope}[shift={(7.7,0)}]
    \coordinate (1T1) at (-1,0);
    \coordinate (1T2) at (-0.5,-1);
    \coordinate (1T3) at (0.2,-0.6);
    \coordinate (1T4) at (0.2,0.6);
    \coordinate (1T5) at (-0.5,1);
	
   \draw[line width = 1pt,line join=round] (1T1)--(1T5)--(1T4)--(1T3);
   \draw[line width = 1pt,line join=round] (1T3)--(1T5);
   \draw[line width = 1pt,line join=round, dashed] (1T1)--(1T4);
   \draw[green!70!black,line width = 1.2pt,line join=round] (1T1)--(1T3);
   
  \node at (-0.3,0.2) {\tiny $Z_6$};
  \node at (-0.4,-0.1) {\tiny $Z_6'$};
  \node at (-0.17,0.65) {\tiny $Z_6'$};
  \node at (-0.6,0.45) {\tiny $Z_6''$};
  \node at (0.06,0.2) {\tiny $Z_6''$};
  \node at (-0.5,-0.4) {\scriptsize \textcolor{green!70!black}{$C_6$}};
   \end{scope}
   
    \begin{scope}[shift={(9.1,0)}]
    \coordinate (2T1) at (-1,0);
    \coordinate (2T2) at (-0.5,-1);
    \coordinate (2T3) at (0.2,-0.6);
    \coordinate (2T4) at (0.2,0.6);
    \coordinate (2T5) at (-0.5,1);
	
   \draw[line width = 1pt,line join=round] (2T5)--(2T1)--(2T2)--(2T3);
   \draw[line width = 1pt,line join=round] (2T2)--(2T5);
   \draw[line width = 1pt,line join=round, dashed] (2T1)--(2T3);
   \draw[green!70!black,line width = 1.2pt,line join=round] (2T3)--(2T5);
   
   \node at (-0.35,-0.1) {\tiny $Z_7$};
  \node at (-0.63,-0.4) {\tiny $Z_7'$};
  \node at (-0.3,0.2) {\tiny $Z_7'$};
  \node at (-0.67,0.33) {\tiny $Z_7''$};
  \node at (-0.2,-0.65) {\tiny $Z_7''$};
  \node at (0,0.3) {\scriptsize \textcolor{green!70!black}{$C_7$}};
   \end{scope}
   
    \begin{scope}[shift={(10,0)}]
    \coordinate (3T1) at (-1,0);
    \coordinate (3T2) at (-0.5,-1);
    \coordinate (3T3) at (0.2,-0.6);
    \coordinate (3T4) at (0.2,0.6);
    \coordinate (3T5) at (-0.5,1);
	
   \draw[line width = 1pt,line join=round] (3T2)--(3T3)--(3T4)--(3T5);
   \draw[line width = 1pt,line join=round] (3T2)--(3T4);
   \draw[line width = 1pt,line join=round, dashed] (3T3)--(3T5);
   \draw[green!70!black,line width = 1.2pt,line join=round] (3T2)--(3T5);
   
   \node at (-0.23,0.08) {\tiny $Z_8$};
  \node at (-0.35,-0.2) {\tiny $Z_8'$};
  \node at (0.09,0) {\tiny $Z_8'$};
  \node at (-0.1,-0.57) {\tiny $Z_8''$};
  \node at (-0.1,0.57) {\tiny $Z_8''$};
  \node at (-0.65,-0.2) {\scriptsize \textcolor{green!70!black}{$C_8$}};
   \end{scope}
   
    \begin{scope}[shift={(11.4,0)}]
    \coordinate (2P1) at (-1,0);
    \coordinate (2P2) at (-0.5,1);
    \coordinate (2P3) at (0.2,0.6);
    \coordinate (2P4) at (0.2,-0.6);
    \coordinate (2P5) at (-0.5,-1);
	
   \draw[blue, line width = 1.2pt] (2P2)--(2P5);
   \draw[blue, line width = 1.2pt] (2P3)--(2P5);
   \draw[line width = 1pt,line join=round] (2P1)--(2P2)--(2P3)--(2P4)--(2P5)--(2P1);
   
   \node at (-0.65,0) {\scriptsize $\textcolor{blue}{\g_1}$};
   \node at (0,-0.3) {\scriptsize $\textcolor{blue}{\g_2}$};
   \end{scope}

\end{tikzpicture}
\vspace{-20pt}
\caption{\label{fig: A2k2} An ideal triangulation of $M_{3}^{(-2)}$ into eight ideal tetrahedra. Internal edges $C_I$ for gluing data are denoted by green. The two ideal triangulations of the Gaiotto curves at $\th = 0$ and $\th = 2\pi$ are identified.}
\end{figure}
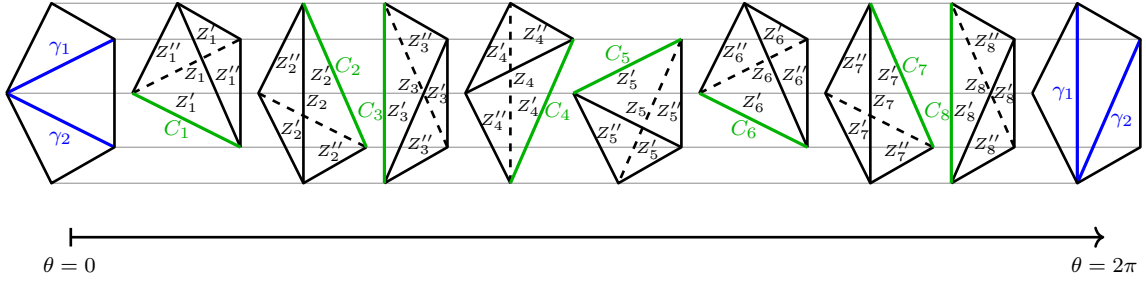
Our next example is for $k=-2$, in which case we should consider an ideal triangulation of $M_{3}^{(-2)}$ into eight ideal tetrahedra as detailed in figure \ref{fig: A2k2}, where the eight green internal edges read
\begin{align}
    C_1 & = Z_1'' + Z_2 + Z_8 \;,\;\;
    C_2 = Z_1 + Z_2'' + Z_3 \;,\;\;
    C_3 = Z_2 + Z_3'' + Z_4 \;,\;\;
    C_4 = Z_3 + Z_4'' + Z_5 \;,\;\;
    \nonumber\\
    C_5 & = Z_4 + Z_5'' + Z_6 \;,\;\;
    C_6  = Z_5 + Z_6'' + Z_7 \;,\;\;
    C_7  = Z_6 + Z_7'' + Z_8 \;,\;\;
    C_8  = Z_1 + Z_7 + Z_8''\; .\nonumber
\end{align}
For a polarisation $\s = (1,1,2,2,2,2,2,2)$, the coefficients are organised as
\begin{align}
    g=
    \begin{pmatrix}
 0 & 0 & 0 & 0 & 0 & 0 & 0 & 0 \\
 0 & 0 & 0 & 0 & 0 & 0 & 0 & 0 \\
 0 & 0 & 1 & 0 & 0 & 0 & 0 & 0 \\
 0 & 0 & 0 & 1 & 0 & 0 & 0 & 0 \\
 0 & 0 & 0 & 0 & 1 & 0 & 0 & 0 \\
 0 & 0 & 0 & 0 & 0 & 1 & 0 & 0 \\
 0 & 0 & 0 & 0 & 0 & 0 & 1 & 0 \\
 0 & 0 & 0 & 0 & 0 & 0 & 0 & 1 \\
    \end{pmatrix}
    \;,\;
    g'=
    \begin{pmatrix}
 1 & 0 & 0 & 0 & 0 & 0 & 0 & 1 \\
 0 & 1 & 1 & 0 & 0 & 0 & 0 & 0 \\
 0 & 0 & 0 & 1 & 0 & 0 & 0 & 0 \\
 0 & 0 & 1 & 0 & 1 & 0 & 0 & 0 \\
 0 & 0 & 0 & 1 & 0 & 1 & 0 & 0 \\
 0 & 0 & 0 & 0 & 1 & 0 & 1 & 0 \\
 0 & 0 & 0 & 0 & 0 & 1 & 0 & 1 \\
 0 & 0 & 0 & 0 & 0 & 0 & 1 & 0 \\
    \end{pmatrix}
    \;,\;
    g''=
    \begin{pmatrix}
 0 & 1 & 0 & 0 & 0 & 0 & 0 & 0 \\
 1 & 0 & 0 & 0 & 0 & 0 & 0 & 0 \\
 0 & 1 & 0 & 0 & 0 & 0 & 0 & 0 \\
 0 & 0 & 0 & 0 & 0 & 0 & 0 & 0 \\
 0 & 0 & 0 & 0 & 0 & 0 & 0 & 0 \\
 0 & 0 & 0 & 0 & 0 & 0 & 0 & 0 \\
 0 & 0 & 0 & 0 & 0 & 0 & 0 & 0 \\
 1 & 0 & 0 & 0 & 0 & 0 & 0 & 0 \\
    \end{pmatrix}
    \,,
\end{align}
which gives rise to the NZ matrices:
\begin{align}
    A =
    \begin{pmatrix}
 -1 & 0 & 0 & 0 & 0 & 0 & 0 & -1 \\
 0 & -1 & -1 & 0 & 0 & 0 & 0 & 0 \\
 0 & 0 & 1 & -1 & 0 & 0 & 0 & 0 \\
 0 & 0 & -1 & 1 & -1 & 0 & 0 & 0 \\
 0 & 0 & 0 & -1 & 1 & -1 & 0 & 0 \\
 0 & 0 & 0 & 0 & -1 & 1 & -1 & 0 \\
 0 & 0 & 0 & 0 & 0 & -1 & 1 & -1 \\
 0 & 0 & 0 & 0 & 0 & 0 & -1 & 1 \\
    \end{pmatrix}
    \;,\;\;
    B =
    \begin{pmatrix}
 -1 & 1 & 0 & 0 & 0 & 0 & 0 & -1 \\
 1 & -1 & -1 & 0 & 0 & 0 & 0 & 0 \\
 0 & 1 & 0 & -1 & 0 & 0 & 0 & 0 \\
 0 & 0 & -1 & 0 & -1 & 0 & 0 & 0 \\
 0 & 0 & 0 & -1 & 0 & -1 & 0 & 0 \\
 0 & 0 & 0 & 0 & -1 & 0 & -1 & 0 \\
 0 & 0 & 0 & 0 & 0 & -1 & 0 & -1 \\
 1 & 0 & 0 & 0 & 0 & 0 & -1 & 0 \\
    \end{pmatrix}
\,.
\end{align}
We thus obtain the CS level matrix $K$ of the DGG theory $T[M_3^{(-2)},\s]$ as:
\begin{align}
    K = B^{-1}A = 
    \begin{pmatrix}
 1 & 0 & 1 & 0 & -1 & 0 & 1 & 0 \\
 0 & 1 & 0 & 1 & 0 & -1 & 0 & 1 \\
 1 & 0 & 2 & -1 & -1 & 1 & 1 & -1 \\
 0 & 1 & -1 & 2 & 0 & -1 & 0 & 1 \\
 -1 & 0 & -1 & 0 & 2 & -1 & -1 & 1 \\
 0 & -1 & 1 & -1 & -1 & 2 & 0 & -1 \\
 1 & 0 & 1 & 0 & -1 & 0 & 2 & -1 \\
 0 & 1 & -1 & 1 & 1 & -1 & -1 & 2 \\
    \end{pmatrix}
    \,.
    \label{eq: A2k2 K}
\end{align}
with eight superpotential terms from the eight easy internal edges
\begin{align}
    &V_{(1,-1,0,0,0,0,0,1)}\;,\;\;
    V_{(-1,1,1,0,0,0,0,0)}\;,\;\;
    \phi_3 V_{(0,-1,0,1,0,0,0,0)}\;,\;\;
    \phi_4 V_{(0,0,1,0,1,0,0,0)}\;,
    \nonumber\\
    &
    \phi_5 V_{(0,0,0,1,0,1,0,0)}\;,\;\;
    \phi_6 V_{(0,0,0,0,1,0,1,0)}\;,\;\;
    \phi_7 V_{(0,0,0,0,0,1,0,1)}\;,\;\;
    \phi_8 V_{(-1,0,0,0,0,0,1,0)}\;,
\end{align}
that completely break the topological symmetries, fixing the mixing parameters to
\begin{align}
    \m_{\rm TQFT}^\ast = (-1,-1,0,0,0,0,0,0)\,,
    \label{eq: A2k2 m}
\end{align}
so that the $F$-function is a constant of the same value as in the $k=\pm 1$ cases. Note that this is a case where $T[M_3]=T_A[M_3]$: the RG flow from the 3d $\CN=2$ ACSM theory directly lands us on some unitary TQFT without the need for performing the $A$-twist, and indeed we do not find any `axial' flavour symmetry in this case. This is in agreement with the discussion around~\eqref{da unitaryTQFT} and with the fact that the edges $C_I$ with $s_I=2$ do not satisfy the condition $s_I= \pm 2k$ here, hence we cannot find any non-trivial 3d $\CN=4$ fixed point.

\medskip
\noindent {\bf Superconformal index.}
The superconformal index of the $T[ M_{3}^{(-2)} ]$ theory with CS level matrix \eqref{eq: A2k2 K} and mixing \eqref{eq: A2k2 m} is computed as
\begin{align}
    \CI_{S^2\times S^1}^{T[M_{3}^{(-2)}]}
    (y_i=1,\q)
    &=
    \sum_{\mathfrak{m}\in \mathbb{Z}^8}
    \big( -\q^{\frac{1}{2}} \big)^{-\mathfrak{m}_1 - \mathfrak{m}_2}
    \CJ_{\q}\left(\mathfrak{m}_1,-\mathfrak{m}_1-\mathfrak{m}_3+\mathfrak{m}_5-\mathfrak{m}_7\right) 
    \nonumber\\
    &\times
    \CJ_{\q}\left(\mathfrak{m}_2,-\mathfrak{m}_2-\mathfrak{m}_4+\mathfrak{m}_6-\mathfrak{m}_8\right) 
    \nonumber\\
    &\times
    \CJ_{\q}\left(\mathfrak{m}_3,-\mathfrak{m}_1-2\mathfrak{m}_3+\mathfrak{m}_4+\mathfrak{m}_5-\mathfrak{m}_6-\mathfrak{m}_7+\mathfrak{m}_8\right) 
    \nonumber\\
    &\times
    \CJ_{\q}\left(\mathfrak{m}_4,-\mathfrak{m}_2+\mathfrak{m}_3-2 \mathfrak{m}_4 + \mathfrak{m}_6 -\mathfrak{m}_8\right) 
    \nonumber\\
    &\times
    \CJ_{\q}\left(\mathfrak{m}_5,\mathfrak{m}_1 + \mathfrak{m}_3 -2 \mathfrak{m}_5 + \mathfrak{m}_6 + \mathfrak{m}_7 - \mathfrak{m}_8\right) 
    \nonumber\\
    &\times
    \CJ_{\q}\left(\mathfrak{m}_6,\mathfrak{m}_2 - \mathfrak{m}_3 + \mathfrak{m}_4 + \mathfrak{m}_5 - 2 \mathfrak{m}_6 + \mathfrak{m}_8\right) 
    \nonumber\\
    &\times
    \CJ_{\q}\left(\mathfrak{m}_7,-\mathfrak{m}_1 - \mathfrak{m}_3 + \mathfrak{m}_5 - 2 \mathfrak{m}_7 + \mathfrak{m}_8\right) 
    \nonumber\\
    &\times
    \CJ_{\q}\left(\mathfrak{m}_8,-\mathfrak{m}_2 + \mathfrak{m}_3 - \mathfrak{m}_4 - \mathfrak{m}_5 + \mathfrak{m}_6 + \mathfrak{m}_7 - 2\mathfrak{m}_8\right)
    \nonumber\\
    &=1+\CO(\q^{5})
    \,,
    \label{eq: TQFT SCI}
\end{align}
which strongly corroborates the claim that $T[ M_{3}^{(-2)} ,\s]$ directly flows to a 3d TQFT without local operators.

\medskip
\noindent {\bf Modular data and choice of $K_g$.}
For this ACSM theory, we choose $K_g=8$. Then, the modular data of the TQFT can be extracted as
\begin{align}
    \{ |S_{0\a}| \} = 
    \Big\{
    \frac{2}{\sqrt{5}} \sin\Big( \frac{\pi}{5} \Big)
    ,
    \frac{2}{\sqrt{5}} \sin\Big( \frac{2\pi}{5} \Big)
    \Big\}
    \;,\;\;\quad
    \{ T_{\a\a} \} = 
    e^{-\frac{2\pi i}{24} (\frac{14}{5})} 
    \Big\{ e^{2\pi i (0)} , e^{2\pi i (\frac{2}{5})} \Big\}
    \,,
\end{align}
with central charge $c_{\text{2d}}={14 \ov 5}$. These results are compatible with the modular data of the affine Lie algebra $G_2$ at level $1$:
\begin{align}
    S =
    \frac{2}{\sqrt{5}}
    \begin{pmatrix}
        \sin \big( \frac{\pi}{5} \big) & \sin \big( \frac{2\pi}{5} \big) \\
        \sin \big( \frac{2\pi}{5} \big) & -\sin \big( \frac{\pi}{5} \big)
    \end{pmatrix}
    \;,\;\;\quad
    T= e^{2\pi i (\frac{53}{60})} 
    \text{diag}\,
    \Big(
    e^{2\pi i (0)} , e^{2\pi i (\frac{2}{5})}
    \Big)\,,
\end{align}
which is also known as the modular representation of the Fibonacci MTC.

\medskip
\noindent {\bf Half-index.}
The half-index of $T[ M_{3}^{(-2)} ]$ is computed as
\begin{align}
    \CI_{D^2\times S^1}^{T[M_{3}^{(-2)}]}(y_i=1,\q)
    &=
    \sum_{\mathfrak{m}\in \mathbb{Z}_{\geq 0}^8 }
    \frac{\q^{\frac{1}{2} \mathfrak{m}^T\cdot K \cdot \mathfrak{m} } (-\q^{1/2})^{-\mathfrak{m}_1-\mathfrak{m}_2} }
    {\prod_{i=1}^{8} (\q)_{\mathfrak{m}_i} }
    \nonumber\\
    &=
    1 + 14 \q + 42 \q^2 + 140 \q^3 + 350 \q^4 + 840 \q^5 + 1827 \q^6 + \cdots
    \,,
\end{align}
which coincides, up to the overall $\q^{c_{\rm 2d}\ov 24}$ prefactor, with the vacuum character of the $(G_2)_1$ VOA as a series expansion~\cite{Kim:2024dxu}. We therefore claim that the DGG theory $T[ M_{3}^{(-2)} ]$ is a 3d unitary TQFT which can support the $(G_2)_1$ VOA on its holomorphic boundary.

\medskip
\noindent
{\bf New 3d $\CN=2$ SCFT upon dropping an edge $C_I$.} Consider the same ACSM as above but with a superpotential where we drop one term $O_I$, corresponding to `drilling out' the edge $C_I$. Let us take $C_I= C_1$ for definiteness. Then, according to our general discussion in subsection~\ref{subsec:DGG and SCFT}, the theory we obtain in the IR is a non-trivial 3d $\CN=2$ SCFT unrelated to the 4d Argyres--Douglas theory $\CT_{A_2}$. Indeed, by direct computation, we find {\it three Bethe vacua} in this case instead of two, and the SCFT point obtained by $F$-maximisation sits at
\be
\n_{\CA^{(1)}}
\approx 1.0612~,
\ee
at which point we have $F\approx 1.16108$. 
The superconformal index at this fixed point reads
\begin{align}\nn
    \CI_{S^2 \times S^1}(\xi;\q)
    =
    1 
    \!-\!
    \frac{1}{\xi}\q^{\frac{1}{2}}
    \!+\! \Big(\frac{1}{\xi^2}\!-\!1\Big)\q
    -\frac{1}{\xi^3} \q^{\frac{3}{2}}
    + \CO(\q^2)\,,
\end{align}
where $\eta$ corresponds to the symmetry $U(1)_{\CA^{(1)}}$ and we need to specify $\xi = (-\q^{1/2})^{\hat\n_{\!\CA^{(1)}}} \eta $ to take into account the mixing $\nu_{\CA^{(1)}}=1+\hat{\nu}_{\CA^{(1)}}$. Filling in $C_1$ back corresponds to turning on $O_1$. This fixes $\n_{\CA^{(1)}} = 0$ and $\eta =1$, giving us
\begin{align}
    \CI_{S^2 \times S^1}\big(\xi = (-\q^{1/2})^{-1};\q\big)
    =
    1+O(\q^4)\,,
\end{align}
consistently with the unitary TQFT result \eqref{eq: TQFT SCI}.

\subsection{The \texorpdfstring{$(A_1,A_{2n})$}{(A1,A2n)} Argyres--Douglas theory}

Let us now consider the case of general $n$. We will first consider the cases $k=1$ and $k=-1$, which are best understood, before briefly discussing the general case.

\subsubsection{ \texorpdfstring{$k=1$}{k=1} : \texorpdfstring{$M(2,2n+3)$}{M(2,2n+3)} }
For the $\CT_{A_{2n}}$ theory at general $n$, as discussed in section~\ref{subsec:M3 from Delta}, the three-manifold $M_{3}^{(1)}$ can be triangulated into $4n$ ideal tetrahedra with $4n$ internal edges. These are:
\begin{align}
    C_i =
    Z_i'
    +
    (1-\d_{i,\k(i)}) Z'_{i + \l(i)}
    +
    Z_{1+[i+n-1]_{4n}}
    +
    Z_{1+[i-n-1]_{4n}}
    \;,\;\;\;
    \text{for}
    \;\;
    i=1,\cdots,4n
    \,,
    \label{eq: A2n internal edge}
\end{align}
where $[x]_p := x\, \text{mod}\, p$, and
\begin{align}
    \l(i) &= (-1)^{\lceil i/n \rceil}
    \;\;,\;\;\;
    \k(i) = \Big\lceil \frac{i}{n} \Big\rceil \, n 
    - \Big[
    \frac{\l(i)-1}{2}
    \Big]_n 
    \,.
\end{align}
Then, the coefficient matrices read
\begin{align}
    g_{ij}^{(0)}
    =
    \d_{j,1+[i+n-1]_{4n}}
    +
    \d_{j,1+[i-n-1]_{4n}}
    \;,\;\;
    g_{ij}^{(1)} 
    = 
    \d_{j,i} + (1-\d_{i,\k(i)}) \d_{j,i+\l(i)}
    \;,\;\;
    g_{ij}^{(2)} 
    =
    0\,,
\end{align}
and for a polarisation choice $\s = (\overbrace{1,\cdots,1}^{2n},\overbrace{ 0,\cdots,0}^{2n})$, we find
\begin{align}
    g =
    \left(
    \begin{array}{c|c|c|c}
        \Xi_+ & {\bf 0} & {\bf 0} & {\bf 1} \\
        \hline
        {\bf 0}  & \Xi_- & {\bf 1} & {\bf 0} \\
        \hline
        {\bf 0} & {\bf 0} & {\bf 0} & {\bf 1} \\
        \hline
        {\bf 0} & {\bf 0} & {\bf 1} & {\bf 0}
    \end{array}
    \right)
    \;,\;\;
    g' =
    \left(
    \begin{array}{c|c|c|c}
        {\bf 0} & {\bf 0} & {\bf 0} & {\bf 0} \\
        \hline
        {\bf 0}  & {\bf 0} & {\bf 0} & {\bf 0} \\
        \hline
        {\bf 0} & {\bf 0} & \Xi_+ & {\bf 0} \\
        \hline
        {\bf 0} & {\bf 0} & {\bf 0} & \Xi_-
    \end{array}
    \right)
    \;,\;\;
    g'' =
    \left(
    \begin{array}{c|c|c|c}
        {\bf 0} & {\bf 1} & {\bf 0} & {\bf 0} \\
        \hline
        {\bf 1}  & {\bf 0} & {\bf 0} & {\bf 0} \\
        \hline
        {\bf 0} & {\bf 1} & {\bf 0} & {\bf 0} \\
        \hline
        {\bf 1} & {\bf 0} & {\bf 0} & {\bf 0}
    \end{array}
    \right)\,,
\end{align}
with $n\times n$ block matrices ($i,j=1,\cdots,n$)
\begin{align}
    (\Xi_\pm)_{ij} 
    \equiv \d_{i,j} + \d_{i,j\pm 1}
    \;\;,\;\;\;
    {\bf 1}_{ij} = \d_{i,j}
    \;\;,\;\;\;
    {\bf 0}_{ij} = 0
    \,,
\end{align}
so that we get the NZ matrices
\begin{align}
    A =
    \left(
    \begin{array}{c|c|c|c}
        \Xi_+ & {\bf 0} & {\bf 0} & {\bf 1} \\
        \hline
        {\bf 0}  & \Xi_- & {\bf 1} & {\bf 0} \\
        \hline
        {\bf 0} & {\bf 0} & -\Xi_+ & {\bf 1} \\
        \hline
        {\bf 0} & {\bf 0} & {\bf 1} & -\Xi_-
    \end{array}
    \right)
    \;\;,\;\;\;
    B = 
    \left(
    \begin{array}{c|c|c|c}
        {\bf 0} & {\bf 1} & {\bf 0} & {\bf 0} \\
        \hline
        {\bf 1}  & {\bf 0} & {\bf 0} & {\bf 0} \\
        \hline
        {\bf 0} & {\bf 1} & -\Xi_+ & {\bf 0} \\
        \hline
        {\bf 1} & {\bf 0} & {\bf 0} & -\Xi_-
    \end{array}
    \right)\,.
\end{align}
Note that $|\text{det}(B)|=1$, thus, given the inverse of $\Xi^\pm$
\begin{align}
    (\Xi_\pm)^{-1}_{ij} = \sum_{k=0}^{n-1} (-1)^k \d_{i,j\pm k}
    \,,
\end{align}
the CS level matrix $K = B^{-1}A$ for the $T[M_{3}^{(1)}]$ theory reads:\footnote{Note that this $K$ coincides with the adjacency matrix of the CPT-double of the $A_{2n}$ BPS quiver~\protect\cite{Kucharski:2025lcr}.}
\begin{align}
    K
    =
    \left(
    \begin{array}{c|c|c|c}
        {\bf 0} & {\bf 1} & {\bf 0} & {\bf 0} \\
        \hline
        {\bf 1}  & {\bf 0} & {\bf 0} & {\bf 0} \\
        \hline
        \Xi_+^{-1} & {\bf 0} & -\Xi_+^{-1} & {\bf 0} \\
        \hline
        {\bf 0} & \Xi_-^{-1} & {\bf 0} & -\Xi_-^{-1}
    \end{array}
    \right)
    \left(
    \begin{array}{c|c|c|c}
        \Xi_+ & {\bf 0} & {\bf 0} & {\bf 1} \\
        \hline
        {\bf 0}  & \Xi_- & {\bf 1} & {\bf 0} \\
        \hline
        {\bf 0} & {\bf 0} & -\Xi_+ & {\bf 1} \\
        \hline
        {\bf 0} & {\bf 0} & {\bf 1} & -\Xi_-
    \end{array}
    \right)
    =
    \left(
    \begin{array}{c|c|c|c}
        {\bf 0} & \Xi_- & {\bf 1} & {\bf 0} \\
        \hline
        \Xi_+  & {\bf 0} & {\bf 0} & {\bf 1} \\
        \hline
        {\bf 1} & {\bf 0} & {\bf 1} & {\bf 0} \\
        \hline
        {\bf 0} & {\bf 1} & {\bf 0} & {\bf 1}
    \end{array}
    \right)\,,
    \label{eq: A2n K}
\end{align}
with superpotential terms from $4n$ easy internal edges
\begin{align}
    W=\sum_{I=1}^{4n}
    \bigg(\prod_{j=1}^{4n} \phi_j^{g_{Ij}} \bigg)
    V_{\mathfrak{m}^{(I)}}
    \,,
\end{align}
where the magnetic fluxes are $\mathfrak{m}_j^{(I)} = -B_{Ij}$. This fixes the TQFT point as
\begin{align}
    \m_{\rm TQFT}^\ast = (\overbrace{-2,\cdots ,-2}^{2n} , \overbrace{-1 , \cdots, -1}^{2n} )\,.
    \label{eq: A2n m}
\end{align}
For general $n$ with $k=1$, there are four internal edges satisfying $s_I=2$:
\begin{align}
    &C_1:\;\qquad
    \CA^{(1)} = T_{n+1} - \sum_{j=1}^{n} (-1)^j T_{2n+j}
    ,\;
    \nonumber\\
    & \qquad \to\qquad \mu_{\rm SCFT}^\ast= (\overbrace{-2,\cdots,-2}^{n},-1,\overbrace{-2,\cdots,-2}^{n-1},\overbrace{0,-2,0,-2,\cdots}^{n},\overbrace{-1,\cdots,-1}^{n})
    \,,
    \nonumber\\
    &C_{2n}:\;\qquad
    \CA^{(2n)} = T_n + \sum_{j=0}^{n-1}(-1)^j T_{4n-j}
    ,\;
    \nonumber\\
    & \qquad \to\qquad \mu_{\rm SCFT}^\ast= (\overbrace{-2,\cdots,-2}^{n-1},-1,\overbrace{-2,\cdots,-2}^{n},\overbrace{-1,\cdots,-1}^{n},\overbrace{\cdots,-2,0,-2,0}^{n})
    \,,
    \nonumber\\
    &C_{2n+1}:\;\qquad
    \CA^{(2n+1)}= \sum_{j=1}^{n}(-1)^j T_{2n+j}
    ,\;
    \nonumber\\
    & \qquad \to\qquad \mu_{\rm SCFT}^\ast
    = (\overbrace{-2,\cdots,-2}^{n},\overbrace{-2,\cdots,-2}^{n},\overbrace{-2,0,-2,0,\cdots}^{n},\overbrace{-1,\cdots,-1}^{n})
    \,,
    \nonumber\\
    &C_{4n}:\;\qquad
    \CA^{(4n)}= -\sum_{j=0}^{n-1}(-1)^j T_{4n-j}
    ,\;
    \nonumber\\
    & \qquad \to\qquad \mu_{\rm SCFT}^\ast
    = (\overbrace{-2,\cdots,-2}^{n},\overbrace{-2,\cdots,-2}^{n},\overbrace{-1,\cdots,-1}^{n},\overbrace{\cdots,0,-2,0,-2}^{n})
    \,.
    \label{CI genericnk1 expls}
\end{align}
to get to the rank-0 SCFT point. Note that setting $n=1$ recovers~\eqref{CI n1k1 expls}.

\medskip
\noindent {\bf Superconformal index.}
The superconformal index of the $T_A[M_3^{(1)}]$ theory reads
\begin{align}
    \CI_{S^2\times S^1}^{T_A[M_{3}^{(1)}]}
    (\q)
    &=
    \sum_{m \in \mathbb{Z}^{4n}}
    \prod_{i=1}^{4n}
    \bigg[
    \Big(
    \big( - \q ^{\frac{1}{2}} \big)^{\m_i}
    \Big)^{\mathfrak{m}_i}
    \CJ_{\q}
    \Big(
    \mathfrak{m}_i
    \,,\,
    -\sum_{j=1}^{4n}
    K_{ij} \mathfrak{m}_j
    \Big)
    \bigg]\,.
    \label{eq: A2n SCI before}
\end{align}
where we set $\m \equiv \m_{\rm TQFT}^\ast$. For later convenience, we relabel the summation variables
\begin{align}
    &l_1 = \mathfrak{m}_{n+1},\; 
    l_2 = \mathfrak{m}_1,\;
    l_3 = \mathfrak{m}_{n+2},\;
    l_4 = \mathfrak{m}_2,\;
    \cdots,
    l_{2n-1} = \mathfrak{m}_{2n},\;
    l_{2n} = \mathfrak{m}_n,
    \nonumber\\
    &k_1 = \mathfrak{m}_{3n+1},\; 
    k_2 = \mathfrak{m}_{2n+1},\;
    k_3 = \mathfrak{m}_{3n+2},\;
    k_4 = \mathfrak{m}_{2n+2},\;
    \cdots,
    k_{2n-1} = \mathfrak{m}_{4n},\;
    k_{2n} = \mathfrak{m}_{3n}.
    \label{eq: A2n index relabel}
\end{align}
Then, we have
\begin{align}
    \CI_{S^2\times S^1}^{T_A[M_{3}^{(1)}]} (\q) 
    &=
    \sum_{l,k \in \mathbb{Z}^{2n}}
    \prod_{i=1}^{2n}
    \CJ_{\q}( l_{i-1} - l_i + l_{i+1} + k_i , l_i )
    \CJ_{\q}( -l_i - k_i , l_i)
    \,,
    \label{eq: A2n SCI}
\end{align}
where we set $l_0 = l_{2n+1} \equiv 0$, and used the triality identity for $\CJ_{\q}(\mathfrak{m},e)$,
\begin{align}
    \CJ_{\q} (\mathfrak{m},e)
    =
    \big(-\q^{1/2}\big)^{-e}
    \CJ_{\q} ( e,-e-\mathfrak{m} )
    =
    \big(-\q^{1/2}\big)^{\mathfrak{m}}
    \CJ_{\q} ( -e-\mathfrak{m} , \mathfrak{m} )
    \,.
    \label{eq: triality}
\end{align}
We numerically checked that the index becomes 1 for several small $n$ at the TQFT point \eqref{eq: A2n m}, and we therefore expect that the superconformal index \eqref{eq: A2n SCI} is identically equal to 1 at that point, for any $n$, indicating that we have indeed correctly identified the 3d TQFT $T_A[M_3^{(1)}]$.

\medskip
\noindent {\bf Half-index.}
We also compute the half-index
\begin{align}
    \CI_{D^2\times S^1}^{T_A[M_{3}^{(1)}]} (\q)
    &=
    \sum_{\mathfrak{m} \in \mathbb{Z}_{\geq 0}^{4n}}
    \frac{\q^{\frac{1}{2} \mathfrak{m}^T \cdot K \cdot \mathfrak{m} } (-\q^{1/2})^{-\m \cdot \mathfrak{m}} }{\prod_{i=1}^{4n} (\q)_{\mathfrak{m}_i} }
    \nonumber\\
    &=
    \sum_{l,k\in \mathbb{Z}_{\geq 0}^{2n}}
    \frac{\q^{\frac{1}{2} \big(
    2 \sum_{i=1}^{2n-1} l_i l_{i+1} + \sum_{i=1}^{2n} k_i^2 + \sum_{i=1}^{2n} l_i k_i
    \big)    } } 
    {\prod_{i=1}^{2n} (\q)_{l_i} (\q)_{k_i} } 
    (-\q^{1/2})^{-\sum_{i=1}^{2n} (-2 l_i - k_i) }
    \,,
    \label{eq: A2n half index}
\end{align}
where we used the relabeled summation variables \eqref{eq: A2n index relabel} in the second line. Indeed, the Schur index of the $A_{2n}$ theory can be written as~\cite{Cordova:2015nma}:
\begin{align}
    \CI_{\text{Schur}}^{A_{2n}}(\q)
    &=
    \sum_{l,k\in \mathbb{Z}_{\geq 0}^{2n}}
    \frac{\q^{\frac{1}{2} \big(
    2 \sum_{i=1}^{2n-1} l_i l_{i+1} + \sum_{i=1}^{2n} k_i^2 + \sum_{i=1}^{2n} l_i k_i
    \big)    } } 
    {\prod_{i=1}^{2n} (\q)_{l_i} (\q)_{k_i} } 
    (-\q^{1/2})^{-\sum_{i=1}^{2n} (-2 l_i - k_i) }
    \,,
    \label{eq: A2n Schur}
\end{align}
which computes the vacuum character of the $M(2,2n+3)$ minimal model via the SCFT/VOA correspondence. This perfectly matches with the half-index \eqref{eq: A2n half index}:
\begin{align}
    \CI_{D^2\times S^1}^{T_A[M_3^{(1)}]} (\q)
    =
   \q^{\frac{c_{\text{2d}}}{24}} \chi_0^{M(2,2n+3)}(\q)
    \,,
\end{align}
corroborating the claim that $T_A[M_{3}^{(1)}]$ is a 3d TQFT that can support the $M(2,2n+3)$ minimal model on its holomorphic boundary.

\subsubsection{ \texorpdfstring{$k=-1$}{k=-1} : \texorpdfstring{$osp(1|2n)_1$}{osp(1|2n)1} }
By reversing the orientations of the tetrahedra $Z_i \leftrightarrow Z_i''$, the ideal triangulation of $M_3^{(-1)}$ is easily obtained. Its gluing rules are organised by the internal edges
\begin{align}
    C_i =
    Z_i''
    +
    (1-\d_{i,\k(i)}) Z''_{i + \l(i)}
    +
    Z_{1+[i+n-1]_{4n}}
    +
    Z_{1+[i-n-1]_{4n}}
    \;,\;\;\;
    \text{for}
    \;\;
    i=1,\cdots,4n
    \,.
    \label{eq: A2n k=-1 internal edge}
\end{align}
For a polarisation choice $\s = (\overbrace{1,\cdots,1}^{2n},\overbrace{ 2,\cdots,2}^{2n})$, the coefficient matrices read
\begin{align}
    g =
    \left(
    \begin{array}{c|c|c|c}
        {\bf 0} & {\bf 0} & {\bf 0} & {\bf 0} \\
        \hline
        {\bf 0}  & {\bf 0} & {\bf 0} & {\bf 0} \\
        \hline
        {\bf 0} & {\bf 0} & \Xi_+ & {\bf 0} \\
        \hline
        {\bf 0} & {\bf 0} & {\bf 0} & \Xi_-
    \end{array}
    \right)
    \;,\;\;\quad
    g' =
    \left(
    \begin{array}{c|c|c|c}
        \Xi_+ & {\bf 0} & {\bf 0} & {\bf 1} \\
        \hline
        {\bf 0}  & \Xi_- & {\bf 1} & {\bf 0} \\
        \hline
        {\bf 0} & {\bf 0} & {\bf 0} & {\bf 1} \\
        \hline
        {\bf 0} & {\bf 0} & {\bf 1} & {\bf 0}
    \end{array}
    \right)
    \;,\;\;\quad
    g'' =
    \left(
    \begin{array}{c|c|c|c}
        {\bf 0} & {\bf 1} & {\bf 0} & {\bf 0} \\
        \hline
        {\bf 1}  & {\bf 0} & {\bf 0} & {\bf 0} \\
        \hline
        {\bf 0} & {\bf 1} & {\bf 0} & {\bf 0} \\
        \hline
        {\bf 1} & {\bf 0} & {\bf 0} & {\bf 0}
    \end{array}
    \right)\,,
\end{align}
from which, we find the NZ matrices
\begin{align}
    A = 
    \left(
    \begin{array}{c|c|c|c}
        -\Xi_+ & {\bf 0} & {\bf 0} & -{\bf 1} \\
        \hline
        {\bf 0} &-\Xi_- & -{\bf 1} & {\bf 0} \\
        \hline
        {\bf 0} & {\bf 0} & \Xi_+ & -{\bf 1} \\
        \hline
        {\bf 0} & {\bf 0} & -{\bf 1} & \Xi_-
    \end{array}
    \right)
    \;,\;\;\quad
    B=
    \left(
    \begin{array}{c|c|c|c}
        -\Xi_+ & {\bf 1} & {\bf 0} & -{\bf 1} \\
        \hline
        {\bf 1}  & -\Xi_- & -{\bf 1} & {\bf 0} \\
        \hline
        {\bf 0} & {\bf 1} & {\bf 0} & -{\bf 1} \\
        \hline
        {\bf 1} & {\bf 0} & -{\bf 1} & {\bf 0}
    \end{array}
    \right)\,,
\end{align}
where $|\text{det}(B)|=1$ and the inverse of $B$ is given by
\begin{align}
    B^{-1}=
    \left(
    \begin{array}{c|c|c|c}
        -(\Xi_+)^{-1} & {\bf 0} & (\Xi_+)^{-1} & {\bf 0} \\
        \hline
        {\bf 0}  & -(\Xi_-)^{-1} & {\bf 0} & (\Xi_-)^{-1} \\
        \hline
        -(\Xi_+)^{-1} & {\bf 0} & (\Xi_+)^{-1} & -{\bf 1} \\
        \hline
        {\bf 0} & -(\Xi_-)^{-1} & -{\bf 1} & (\Xi_-)^{-1}
    \end{array}
    \right)
    \,.
\end{align}
Hence, we obtain the CS level matrix $K$ of $T[ M_{3}^{(-1)} ]$ theory:
\begin{align}
    K = B^{-1}A
    =
    \left(
    \begin{array}{c|c|c|c}
        {\bf 1} & {\bf 0} & {\bf 1} & {\bf 0} \\
        \hline
        {\bf 0}  & {\bf 1} & {\bf 0} & {\bf 1} \\
        \hline
        {\bf 1} & {\bf 0} & 2 \cdot {\bf 1} & -\Xi_- \\
        \hline
        {\bf 0} & {\bf 1} & -\Xi_+ & 2\cdot {\bf 1}
    \end{array}
    \right)\,,
    \label{eq: A2n k=-1 K}
\end{align}
as well as $4n$ superpotential terms from the $4n$ easy internal edges
\begin{align}
    W = \sum_{I=1}^{4n}
    \bigg(\prod_{j=1}^{4n} \phi_j^{g_{Ij}} \bigg)
    V_{\mathfrak{m}^{(I)}}
    \,,
\end{align}
with magnetic fluxes $\mathfrak{m}_j^{(i)} = -B_{ij}$. It brings us to the TQFT point:
\begin{align}
    \m_{\rm TQFT}^\ast = (\overbrace{-1,\cdots,-1}^{2n},\overbrace{ 0,\cdots,0}^{2n})\,.
    \label{eq: A2n k=-1 m}
\end{align}
Similarly to the $k=1$ case, there are four internal edges with $s_I=2$, and one can reach the rank-0 SCFT point upon removing the corresponding superpotential term:
\begin{align}
    &C_1:\;\qquad
    \CA^{(1)} = \sum_{j=1}^{n} (-1)^j (T_j + T_{2n+j})
    ,\;
    \nonumber\\
    & \quad \to\quad \mu_{\rm SCFT}^\ast= (\overbrace{-2,0,-2,0,\cdots}^{n},\overbrace{-1,\cdots,-1}^{n},\overbrace{-1,1,-1,1,\cdots}^{n},\overbrace{0,\cdots,0}^{n})
    \,,
    \nonumber\\
    &C_{2n}:\;\qquad
    \CA^{(2n)} = - \sum_{j=0}^{n-1}(-1)^j ( T_{2n-j} + T_{4n-j} )
    ,\;
    \nonumber\\
    & \quad \to\quad \mu_{\rm SCFT}^\ast= (\overbrace{-1,\cdots,-1}^{n},\overbrace{\cdots,0,-2,0,-2}^{n},\overbrace{0,\cdots,0}^{n},\overbrace{\cdots,1,-1,1,-1}^{n})
    \,,
    \nonumber\\
    &C_{2n+1}:\;\qquad
    \CA^{(2n+1)}= -\sum_{j=1}^{n}(-1)^j ( T_j + T_{2n+j} ) - T_{3n+1}
    ,\;
    \nonumber\\
    & \quad \to\quad \mu_{\rm SCFT}^\ast
    = (\overbrace{0,-2,0,-2,\cdots}^{n},\overbrace{-1,\cdots,-1}^{n},\overbrace{1,-1,1,-1,\cdots}^{n},-1,\overbrace{0,\cdots,0}^{n-1})
    \,,
    \nonumber\\
    &C_{4n}:\;\qquad
    \CA^{(4n)}= \sum_{j=0}^{n-1}(-1)^j ( T_{2n-j} + T_{4n-j} ) - T_{3n}
    ,\;
    \nonumber\\
    & \quad \to\quad \mu_{\rm SCFT}^\ast
    = (\overbrace{-1,\cdots,-1}^{n},\overbrace{\cdots,-2,0,-2,0}^{n},\overbrace{0,\cdots,0}^{n-1},-1,\overbrace{\cdots,-1,1,-1,1}^{n})
    \,.
    \label{CI genericnkminus1 expls}
\end{align}

\medskip
\noindent {\bf Superconformal index.}
The superconformal index of the $T_A[M_3^{(-1)}]$ theory is computed as
\begin{align}
    \CI_{S^2 \times S^1}^{T_A[M_3^{(-1)}]}
    (\q)
    &=
    \sum_{l,k \in \mathbb{Z}^{2n}}
    \prod_{i=1}^{2n}
    \CJ_{\q}(l_i,-k_i+l_{i-1}-2l_i+l_{i+1})
    \CJ_{\q}(l_i,k_i)
    \,.
\end{align}
By shifting $k_i \to k_i - l_i $, then flipping the signs $l_i \to - l_i$, with further using a property $\CJ_{\q}(\mathfrak{m},e) = \CJ_{\q}(-e,-\mathfrak{m})$, we express it as
\begin{align}
    \CI_{S^2 \times S^1}^{T_A[M_3^{(-1)}]} (\q)
    =
    \sum_{l,k \in \mathbb{Z}^{2n}}
    \prod_{i=1}^{2n}
    \CJ_{\q}(k_i + l_{i-1} -l_i + l_{i+1} , l_i )
    \CJ_{\q}( -k_i - l_i , l_i )
    \,,
\end{align}
with $l_0=l_{2n+1} \equiv 0$. This exactly matches with the index of $k=+1$ case
\begin{align}
    \CI_{S^2 \times S^1}^{T_A[M_{3}^{(-1)}]} (\q)
    =
    \CI_{S^2 \times S^1}^{T_A[M_{3}^{(1)}]} (\q)
    \,,
\end{align}
and therefore we claim that the superconformal index trivialises at the TQFT point~\eqref{eq: A2n k=-1 m}.

\medskip
\noindent {\bf Half-index.}
The half-index of $T_A[M_3^{(-1)}]$ is also calculated as
\begin{align}
    \CI_{D^2\times S^1}^{T_A[M_{3}^{(-1)}]} (\q)
    &=
    \sum_{l,k \in \mathbb{Z}_{\geq 0}^{2n}}
    \frac{
    \q^{\frac{1}{2} \big(
    \sum_{i=1}^{2n} 2l_i^2 - \sum_{i=1}^{2n-1} 2l_i l_{i+1}
    + \sum_{i=1}^{2n} k_i^2 + \sum_{i=1}^{2n} l_i k_i
    \big) }
    }
    {
    \prod_{i=1}^{2n}
    (\q)_{l_i} (\q)_{k_i}
    }
    (-\q^{1/2})^{\sum_{i=1}^{2n} k_i }
    \,.
    \label{eq: A2n k=-1 half index}
\end{align}
Meanwhile, the trace of the monodromy operator $M(\q)$ of the $A_{2n}$ Argyres--Douglas theory is expected to compute a vacuum character of the affine $osp(1|2n)_1$ VOA, which arises from the $k=-1$ twisted compactification \cite{Creutzig:2024ljv,Kim:2024dxu,Cecotti:2015lab}:
\begin{align}
    \Tr \big( M(\q) \big)
    &=
    \Tr \Big(
    \prod_{i\in \text{even}} \Psi_{\q}(X_{\g_i})
    \prod_{j\in \text{odd}} \Psi_{\q}(X_{\g_j})
    \prod_{i'\in \text{even}} \Psi_{\q}(X_{-\g_{i'}})
    \prod_{j'\in \text{odd}} \Psi_{\q}(X_{-\g_{j'}})
    \Big)
    \nonumber\\
    &=
    \sum_{l \in \mathbb{Z}_{\geq 0}^{2n}}
    \frac{
    \q^{\sum_{i=1}^{2n} l_i^2 - \sum_{i=1}^{2n-1} l_i l_{i+1} }
    }{ \prod_{i=1}^{2n} (\q)_{l_i}^2 }
    \,,
    \label{eq: A2n k=-1 monodromy}
\end{align}
where $X_\g$'s are the quantum torus algebra variables whose commutation is determined by the Dirac pairing $X_\g \, X_{\g'} = \q^{\langle \g,\g' \rangle}X_{\g'} \, X_\g$, and the function $\Psi_{\q}(X)$ is defined as
\begin{align}
    \Psi_{\q}(X) = \sum_{l \geq 0} \frac{\q^{\frac{l^2}{2}}}{(\q)_l} X^l
    \,.
\end{align}
The trace selects the algebra variable singlets, so that the affine $osp(1|2n)_1$ vacuum character can be written as
\begin{align}
    \tilde\chi_0^{ osp(1|2n)_1 }(\q)
    &=
    \sum_{l,k \in \mathbb{Z}_{\geq 0}^{2n}}
    \frac{
    \q^{\frac{1}{2} \big(
    \sum_{i=1}^{2n} 2l_i^2 - \sum_{i=1}^{2n-1} 2l_i l_{i+1}
    + \sum_{i=1}^{2n} k_i^2 + \sum_{i=1}^{2n} l_i k_i
    \big) }
    (-\q^{1/2})^{\sum_{i=1}^{2n} k_i }
    }
    {
    \prod_{i=1}^{2n}
    (\q)_{l_i} (\q)_{k_i}
    }
    \,,
\end{align}
This perfectly coincides with the half-index $\eqref{eq: A2n k=-1 half index}$, 
\begin{align}
    \CI_{D^2\times S^1}^{T_A[M_3^{(-1)}]} (\q)
    =
    \tilde\chi_0^{ osp(1|2n)_1 }(\q)
    \,.
\end{align}
Thus, we claim that the 3d TQFT $T_A[M_3^{(-1)}]$ can support the affine $ osp(1|2n)_1$ VOA on its holomorphic boundary, as expected~\cite{Cecotti:2015lab,Kim:2024dxu}.

\subsubsection{Generic \texorpdfstring{$k$}{k} }
Finally, let us explicitly construct the DGG theories $T_A[M_3^{(k)}]$ for generic $n$ and $k$ coprime to $2n+3$, for a specific choice of polarisation. Let us first consider the case $k>0$. The internal edges in \eqref{eq: A2n internal edge} are easily generalised as
\begin{align}
    C_i =
    Z_i'
    +
    (1-\d_{i,\k(i)}) Z'_{i + \l(i)}
    +
    Z_{1+[i+n-1]_{4kn}}
    +
    Z_{1+[i-n-1]_{4kn}}
    \;,\;\;\;
    \text{for}
    \;\;
    i=1,\cdots,4kn
    \,,
\end{align}
with $\l(i) = (-1)^{\lceil i/n \rceil}$, $\k(i) = \Big\lceil \frac{i}{n} \Big\rceil \, n 
    - \Big[
    \frac{\l(i)-1}{2}
    \Big]_n$, then the coefficient matrices are
\begin{align}
    g_{ij}^{(0)}
    =
    \d_{j,1+[i+n-1]_{4kn}}
    +
    \d_{j,1+[i-n-1]_{4kn}}
    \;,\;\;
    g_{ij}^{(1)} 
    = 
    \d_{j,i} + (1-\d_{i,\k(i)}) \d_{j,i+\l(i)}
    \;,\;\;
    g_{ij}^{(2)} 
    =
    0\,.\nn
\end{align}
We find, for a polarisation choice $\s = (\overbrace{1,\cdots,1}^{2n},\overbrace{0,\cdots,0}^{2n(2k-1)})$, that the NZ $B$ matrix satisfies $|\text{det}(B)|=1$ so that the CS level matrix $K$ of $T[ M_{3}^{(k>0)} ]$ is given by
\begin{align}
    K_{(k-1)\a+i , (k-1)\b+j} &= 
    \d_{\a,1}\d_{\b,1} \mathfrak{A}_{ij}
    +
    (\d_{\a,\b} - \d_{\a,1} \d_{\b,1}) \mathfrak{B}_{ij}
    +
    \nonumber\\
    &+ \d_{\a+1,\b} \mathfrak{C}_{ij} + \d_{\a,\b+1} \mathfrak{C}_{ji}
    +\d_{\a,k}\d_{\b,1} \mathfrak{D}_{ij} + \d_{\a,1} \d_{\b,k} \mathfrak{D}_{ji}
    \,,
    \label{eq: CS generic positive k}
\end{align}
where the indices run as $\a,\b=1,\cdots,k$ and $i,j = 1,\cdots, 4n$, with four $4n \times 4n$ block matrices
\begin{align}
    &\mathfrak{A} =
    \left(
    \begin{array}{c|c|c|c}
        {\bf 0} & \Xi_- & {\bf 1} & {\bf 0} \\
        \hline
        \Xi_+ & {\bf 0} & {\bf 0} & {\bf 0} \\
        \hline
        {\bf 1} & {\bf 0} & {\bf 1} & \Xi_+^{-1} \\
        \hline
        {\bf 0} & {\bf 0} & \Xi_-^{-1} & {\bf 1}
    \end{array}
    \right)
    \;\;,\;\;\;\quad
    \mathfrak{B} =
    \left(
    \begin{array}{c|c|c|c}
        {\bf 1} & \Xi_+^{-1} & {\bf 0} & {\bf 0} \\
        \hline
        \Xi_-^{-1} & {\bf 1} & \Xi_-^{-1} & {\bf 0} \\
        \hline
        {\bf 0} & \Xi_+^{-1} & {\bf 1} & \Xi_+^{-1} \\
        \hline
        {\bf 0} & {\bf 0} & \Xi_-^{-1} & {\bf 1}
    \end{array}
    \right)
    ,\;
    \nonumber\\
    &\mathfrak{C} =
    \left(
    \begin{array}{c|c|c|c}
        {\bf 0} & {\bf 0} & {\bf 0} & {\bf 0} \\
        \hline
        {\bf 0} & {\bf 0} & {\bf 0} & {\bf 0} \\
        \hline
        {\bf 0} & {\bf 0} & {\bf 0} & {\bf 0} \\
        \hline
        \Xi_-^{-1} & {\bf 0} & {\bf 0} & {\bf 0} \\
    \end{array}
    \right)
    \;\;,\;\;\;\;\qquad\quad
    \mathfrak{D} =
    \left(
    \begin{array}{c|c|c|c}
        {\bf 0} & {\bf 0} & {\bf 0} & {\bf 0} \\
        \hline
        {\bf 0} & {\bf 0} & {\bf 0} & {\bf 1} \\
        \hline
        {\bf 0} & {\bf 0} & {\bf 0} & -\Xi_+^{-1} \\
        \hline
        {\bf 0} & {\bf 0} & {\bf 0} & {\bf 0} \\
    \end{array}
    \right)\,.
\end{align}
The $4nk$ internal edges are all easy type corresponding to the superpotential terms \eqref{eq: half BPS op} which brings us to the TQFT point $T_A[M_3^{(k>0)}]$:
\begin{align}
    (\m_{\rm TQFT}^\ast)_j =
    \Bigg\{
    \begin{array}{cc}
        -2 & 1\leq j \leq 2n \\
         -1 & \;\;\;\; 2n < j \leq 3n , \text{or}\;\; 4kn-n < j \leq 4kn \\
         (-1)^{j + \lceil j/n \rceil} & \text{otherwise}
    \end{array}
    \,.
    \label{eq: generic nk m}
\end{align}
Note that the CS level matrix \eqref{eq: CS generic positive k} and the mixing parameter \eqref{eq: generic nk m} recover the previous results in \eqref{eq: A2n K} and \eqref{eq: A2n m} for the $k=1$ case. 

\medskip
\noindent
{\bf SCFT points.} The chord distance $s_I$ of the $I$-th internal edges $C_I$ is given by
\begin{align}
    s_I =
    \left\{
    \begin{array}{cc}
        2[I] &\text{if }\, 1\leq [I] \leq \lceil \frac{n}{2} \rceil\,, \\
        2(n-[I])+3 &\text{if }\, \lceil \frac{n}{2} \rceil < [I] \leq n\,, \\
        2[I]-2n+1 &\text{if }\, n < [I] \leq 2n-\lceil \frac{n}{2} \rceil\,, \\
        4n+2 -2[I]\quad &\text{if }\, 2n-\lceil \frac{n}{2} \rceil < [I] \leq 2n\,.
    \end{array}
    \right.
\end{align}
where we defined $[I]:=\big((I-1)\mod 2n\big) + 1$. For instance, we have:
\begin{align}
    & (n,k)=(2,2) 
    \;\;\to\;\;
    (s_1,\cdots,s_{16}) = (2, 3, 3, 2, 2, 3, 3, 2, 2, 3, 3, 2, 2, 3, 3, 2)
    \,,
    \nonumber\\
    & (n,k)=(5,1) 
    \;\;\to\;\;
    (s_1,\cdots,s_{20})
    =
    (2, 4, 6, 5, 3, 3, 5, 6, 4, 2, 2, 4, 6, 5, 3, 3, 5, 6, 4, 2)
    \,.
\end{align}
For $k\neq \pm(n+1) \mod N$, we expect the rank-0 SCFT $T[M_3^{(k>0)}]$ upon drilling an internal edge $C_I$ of $s_I = \pm 2k \mod N$, which frees up the symmetry $U(1)_{\CA^{(I)}}$ generated by
\begin{align}
    \CA^{(I)} = \sum_{j=1}^{4nk} (\m_{\CA^{(I)}})_j T_j
    \;,\qquad \;\;\;
    (\m_{\CA^{(I)}})_j = (B^{-1})_{jI}
    \,,
    \label{eq: free up sym}
\end{align}
that determines the fixed point as $\m_{\rm SCFT}^\ast = \m_{\rm TQFT}^\ast + \m_{\CA^{(I)}}$ from the TQFT point \eqref{eq: generic nk m}. Here, the inverse of $B$ can be written in terms of $k\times k$ blocks
\begin{align}
    B^{-1} =
    \left(
    \begin{array}{c|c|c|c}
        \mathfrak{F} & & & \\
        \hline
         & \mathfrak{G} & & \\
         \hline
         & & \ddots & \\
         \hline
         & & & \mathfrak{G}
    \end{array}
    \right)
    +
    \left(
    \begin{array}{c|c|c|c}
        {\bf 0} & & & \\
        \hline
        \vdots & & & \\
        \hline
        {\bf 0} & & & \\
         \hline
        \mathfrak{F}' & {\bf 0} & \cdots & {\bf 0}
    \end{array}
    \right)
\end{align}
where each block is a $4n\times 4n$ matrix among the below three with $\D_\pm := (\Xi_\pm)^{-1}$
\begin{align}
    \mathfrak{F} =
    \left(
    \begin{array}{c|c|c|c}
        {\bf 0} & {\bf 1} & {\bf 0} & {\bf 0} \\
        \hline
        {\bf 1} & {\bf 0} & {\bf 0} & {\bf 0} \\
        \hline
        \D_+ & {\bf 0} & -\D_+ & {\bf 0} \\
        \hline
        {\bf 0} & {\bf 0} & {\bf 0} & -\D_-
    \end{array}
    \right)
    \;,
    \mathfrak{F}' =
    \left(
    \begin{array}{c|c|c|c}
        {\bf 0} & {\bf 0} & {\bf 0} & {\bf 0} \\
        \hline
        {\bf 0} & {\bf 0} & {\bf 0} & {\bf 0} \\
        \hline
        {\bf 0} & {\bf 0} & {\bf 0} & {\bf 0} \\
        \hline
        {\bf 0} & \D_- & {\bf 0} & {\bf 0}
    \end{array}
    \right)
    \;,
    \mathfrak{G} =
    \left(
    \begin{array}{c|c|c|c}
        -\D_+ & {\bf 0} & {\bf 0} & {\bf 0} \\
        \hline
        {\bf 0} & -\D_- & {\bf 0} & {\bf 0} \\
        \hline
        {\bf 0} & {\bf 0} & -\D_+ & {\bf 0} \\
        \hline
        {\bf 0} & {\bf 0} & {\bf 0} & -\D_-
    \end{array}
    \right)
    \,.
\end{align}
Likewise, for a negative $\Wk$-th twist, say $\Wk = - k$ for some positive $k$, the triangulation of $M_3^{(\Wk)}$ can be established by the orientation reversal $Z_i' \leftrightarrow Z_i''$ of the tetrahedra in the triangulation of $M_3^{(k)}$. For the polarisation $\s = (\overbrace{1,\cdots,1}^{2n},\overbrace{2,\cdots,2}^{2n(2k-1)})$ we similarly find the CS level matrix $\widetilde{K}$
\begin{align}
    \widetilde{K}_{(k-1)\a+i,(k-1)\b+j} &= 
    \d_{\a,1}\d_{\b,1} \widetilde{\mathfrak{A}}_{ij}
    +
    (\d_{\a,\b}-\d_{\a,1}\d_{\b,1}) \widetilde{\mathfrak{B}}_{ij}
    \nonumber\\
    &+
    \d_{\a,1}\Theta(\b-2) \widetilde{\mathfrak{C}}_{ij}
    +
    \d_{\b,1}\Theta(\a-2) \widetilde{\mathfrak{C}}_{ji}
    \nonumber\\
    &+
    \Theta(\b-\a-1)\Theta(\a-2) \widetilde{\mathfrak{D}}_{ij}
    +
    \Theta(\a-\b-1)\Theta(\b-2) \widetilde{\mathfrak{D}}_{ji}
    \,,
    \label{eq: negative K for generic}
\end{align}
with the Heaviside step function $\Theta(x)$, and four $4n\times 4n$ block matrices
\begin{align}
    &\widetilde{\mathfrak{A}} =
    \left(
    \begin{array}{c|c|c|c}
        {\bf 1} & {\bf 0} & {\bf 1} & {\bf 0} \\
        \hline
        {\bf 0} & {\bf 1} & {\bf 0} & {\bf 1} \\
        \hline
        {\bf 1} & {\bf 0} & 2\cdot{\bf 1} & -\Xi_- \\
        \hline
        {\bf 0} & {\bf 1} & -\Xi_+ & 2\cdot {\bf 1}
    \end{array}
    \right)
    \;\;,\;\;\;\qquad
    \widetilde{\mathfrak{B}} =
    \left(
    \begin{array}{c|c|c|c}
        2\cdot {\bf 1} & -\Xi_- & -{\bf 1} & \Xi_- \\
        \hline
        -\Xi_+ & 2\cdot {\bf 1} & {\bf 0} & -{\bf 1} \\
        \hline
        -{\bf 1} & {\bf 0} & 2\cdot {\bf 1} & -\Xi_- \\
        \hline
        \Xi_+ & -{\bf 1} & -\Xi_+ & 2\cdot {\bf 1}
    \end{array}
    \right)
    ,\;
    \nonumber\\
    &\widetilde{\mathfrak{C}} =
    \left(
    \begin{array}{c|c|c|c}
        -{\bf 1} & {\bf 0} & {\bf 1} & {\bf 0} \\
        \hline
        {\bf 0} & -{\bf 1} & {\bf 0} & {\bf 1} \\
        \hline
        -{\bf 1} & \Xi_- & {\bf 1} & -\Xi_- \\
        \hline
        {\bf 0} & -{\bf 1} & {\bf 0} & {\bf 1} \\
    \end{array}
    \right)
    \;\;,\;\;\;\qquad
    \widetilde{\mathfrak{D}} =
    \left(
    \begin{array}{c|c|c|c}
        {\bf 1} & -\Xi_- & -{\bf 1} & \Xi_- \\
        \hline
        {\bf 0} & {\bf 1} & {\bf 0} & -{\bf 1} \\
        \hline
        -{\bf 1} & \Xi_- & {\bf 1} & -\Xi_- \\
        \hline
        {\bf 0} & -{\bf 1} & {\bf 0} & {\bf 1} \\
    \end{array}
    \right)\,,
\end{align}
and $4nk$ superpotential terms following \eqref{eq: half BPS op} that fix the TQFT point as
\begin{align}
    (\m_{\rm TQFT}^\ast)_j = - \Theta(2n-j)\,,
    \;\;\;
    \text{for}\; j=1,\cdots, 4kn
    \,.
\end{align}
Similarly, drilling $C_I$ of $s_I = \pm 2k \mod N$ gives rise to a symmetry as in \eqref{eq: free up sym} but the inverse of $B$ now reads
\begin{align}
    B^{-1} =
    \left(
    \begin{array}{c|cccc}
        \tilde{\mathfrak{F}} & \mathfrak{W} &  & \cdots & \mathfrak{W} \\
        \hline
        \mathfrak{K} & \tilde{\mathfrak{G}} & \tilde{\mathfrak{W}} & \cdots & \tilde{\mathfrak{W}} \\
        & \tilde{\mathfrak{K}} & \tilde{\mathfrak{G}} & \ddots & \vdots \\
        \vdots & \vdots & \ddots & \tilde{\mathfrak{G}} & \tilde{\mathfrak{W}} \\
        \mathfrak{K} & \tilde{\mathfrak{K}} & \cdots & \tilde{\mathfrak{K}} & \tilde{\mathfrak{G}}
    \end{array}
    \right)
    \,,
\end{align}
with the below six $4n\times 4n$ matrices for the blocks
\begin{align}
    &\tilde{\mathfrak{F}}
    =
    \left(
    \begin{array}{c|c|c|c}
        -\D_+ & {\bf 0} & \D_+ & {\bf 0} \\
        \hline
        {\bf 0} & -\D_- & {\bf 0} & \D_- \\
        \hline
        -\D_+ & {\bf 0} & \D_+ & -{\bf 1} \\
        \hline
        {\bf 0} & -\D_- & -{\bf 1} & \D_-
    \end{array}
    \right)
    \,,\quad
    \tilde{\mathfrak{G}}
    =
    \left(
    \begin{array}{c|c|c|c}
        \D_+ & -{\bf 1} & -\D_+ & {\bf 1} \\
        \hline
        -{\bf 1} & \D_- & {\bf 0} & -\D_- \\
        \hline
        -\D_+ & {\bf 0} & \D_+ & -{\bf 1} \\
        \hline
        {\bf 1} & -\D_- & -{\bf 1} & \D_-
    \end{array}
    \right)
    \,,
    \nonumber\\
    &\mathfrak{W}
    =
    \left(
    \begin{array}{c|c|c|c}
        -\D_+ & {\bf 0} & \D_+ & {\bf 0} \\
        \hline
        {\bf 0} & -\D_- & {\bf 0} & \D_- \\
        \hline
        -\D_+ & {\bf 1} & \D_+ & -{\bf 1} \\
        \hline
        {\bf 0} & -\D_- & {\bf 0} & \D_-
    \end{array}
    \right)
    \,,\quad
    \tilde{\mathfrak{W}}
    =
    \left(
    \begin{array}{c|c|c|c}
        \D_+ & -{\bf 1} & -\D_+ & {\bf 1} \\
        \hline
        {\bf 0} & \D_- & {\bf 0} & -\D_- \\
        \hline
        -\D_+ & {\bf 1} & \D_+ & -{\bf 1} \\
        \hline
        {\bf 0} & -\D_- & {\bf 0} & \D_-
    \end{array}
    \right)
    \,,
    \nonumber\\
    &\mathfrak{K}
    =
    \left(
    \begin{array}{c|c|c|c}
        \D_+ & {\bf 0} & -\D_+ & {\bf 0} \\
        \hline
        {\bf 0} & \D_- & {\bf 1} & -\D_- \\
        \hline
        -\D_+ & {\bf 0} & \D_+ & -{\bf 0} \\
        \hline
        {\bf 0} & -\D_- & -{\bf 1} & \D_-
    \end{array}
    \right)
    \,,\quad
    \tilde{\mathfrak{K}}
    =
    \left(
    \begin{array}{c|c|c|c}
        \D_+ & {\bf 0} & -\D_+ & {\bf 0} \\
        \hline
        -{\bf 1} & \D_- & {\bf 1} & -\D_- \\
        \hline
        -\D_+ & {\bf 0} & \D_+ & {\bf 0} \\
        \hline
        {\bf 1} & -\D_- & -{\bf 1} & \D_-
    \end{array}
    \right) 
    \,.
\end{align}
These completely determine the rank-0 SCFT points and the corresponding axial symmetries. For generic values $n$ and $k$, the computation of supersymmetric partition functions with the 3d $A$-model methods becomes computationally prohibitive, but we have nevertheless carried out limited checks of the existence of these rank-0 SCFTs (such as {\it e.g.}~the computation of the SCI at low order for $nk\leq 4$). In any case, these 3d $\CN=4$ SCFTs are predicted from the geometry and the 3d/3d correspondence. It would be very exciting to explore them further with stronger computational methods. 
 
\section*{Acknowledgements}
 We thank  Arash Ardehali, Chris Beem, Dongmin Gang, Heeyeon Kim, Minsung Kim, Pietro Longhi, Rajath Radhakrishnan, 
 and Jaewon Song 
for useful discussions and correspondence. CC also acknowledges very useful discussions with Claude (Anthropic) Opus and Fable models. 
The work of CC is supported in part by a Universty Research Fellowship of the Royal Society.
The work of AK is supported by the School of Mathematics at the University of Birmingham.
The research of SK is supported by a KIAS Individual Grant PG09102 at Korea Institute for Advanced Study. SK also acknowledges the hospitality of the School of Mathematics at the University of Birmingham during his stay. 

\appendix

\section{Gaiotto curves and BPS quivers}\label{app:Gaiotto curve and quiver}

In this appendix, we review aspects of the class-$\CS$ construction which are most relevant for our discussion. In particular, we explain how to read BPS quivers for the 4d SQFT $T[\CC]$  from the curve $\CC$. This was originally discussed in~\cite{Gaiotto:2009hg} and interpreted from a geometric engineering perspective in~\cite{Alim:2011ae}. Here we consider a general Gaiotto curve $\CC$ for the type-$A_1$ 6d SCFT, while in the main text we focus on the case of the genus-0 curve with one irregular puncture. 

\medskip
\noindent
{\bf Gaiotto curve.} 
The Gaiotto curve (or UV curve) $\CC$ is a Riemann curve with punctures~\cite{Gaiotto:2009we}. It comes equipped with a holomorphic quadratic differential $\phi(z)= \phi_2(z) dz^2$, where $z$ is a local coordinate on $\CC$, with a prescribed asymptotic behavior near each puncture $p_i \in \CC$,
\begin{align}
    \phi(z)
    \sim
    \frac{dz^2}{z^{n_i + 2}}~,
\end{align}
determined by the integer $n_i \geq 0$; here the puncture $p_i$ is located at $z=0$. The 4d $\CN=2$ SQFT $T[\CC]$ is obtained as the partial compactification of two M5-branes on $\CC$, and $\phi(z)$ parameterises local deformations of this configuration, which correspond to extended Coulomb-branch parameters. The 4d low-energy effective field theory is encoded by the Seiberg--Witten (SW) curve $\Sigma$ and the SW differential $\l$~\cite{Seiberg:1994rs, Seiberg:1994aj}, which are obtained as a double cover of the Gaiotto curve. Locally, we have~\cite{Gaiotto:2009we}:
\begin{align}
    \Sigma = 
    \big\{ (z,y) \, | \, y^2 = \phi_2(z)\, \big\}~,\qquad
    \quad
    \l = \sqrt{\phi}~.
\end{align}
In the following, we will take $\CC$ of genus zero with (at least) one puncture at $z=\infty$.

\medskip
\noindent
{\bf The $A_l$ Argyres--Douglas theory.} This AD $\CN=2$ SCFT, often denoted by $(A_1, A_l)$, is realised by a genus-zero surface $\CC$ with a single puncture at $z=\infty$ with $\phi(z)= z^{l+1} dz^2$. The SW differential has $U(1)_r$ charge $r[\l]=1$, hence we have:
\be\label{r of z in A1Al}
r[z]= {2\ov l+3}~.
\ee
A generic deformation of $\phi$ reads:
\be
\phi= \left(z^{l+1}+ \varepsilon_{l-1} z^{l-1}+\cdots + \varepsilon_1 z+\varepsilon_0 \right)dz^2
\ee
and corresponds to turning on Coulomb-branch VEVs, flavour masses and/or relevant deformations of the SCFT, with conformal dimensions $r=\Delta$ given by:
\be\label{CB spectrum A1Al}
r[\varepsilon_s]={2\ov l+3}(l+1-s)~.
\ee
For $l=2 n+1$, the spectrum includes a mass parameter $\varepsilon_{n}$ of dimension $r[\varepsilon_n]=1$, which corresponds to a rank-one flavour symmetry and to the presence of a Higgs branch. For $l=2n$,  the case of interest in this paper, the theory has no flavour symmetry and no Higgs branch. In either case, $l=2n+1$ or $l=2n$, the SCFT has rank $n$, with the Coulomb branch spectrum given by~\eqref{CB spectrum A1Al} for $s=0, \cdots, n-1$. Correspondingly, the SW curve $\Sigma$ has genus $n$.

\medskip
\noindent
{\bf  Ideal triangulation of the Gaiotto curve.} 
At a generic point on the Coulomb branch of the 4d $\CN=2$ theory $T[\CC]$, one finds a spectrum of massive BPS states, which correspond to 1-cycles $\g\in  H_1(\Sigma)$ in the SW curve $\S$ calibrated by $\l$. Recall that the first homology of $\Sigma$ together with its antisymmetric intersection pairing corresponds to the charge lattice of the SQFT together with the Dirac pairing,%
\footnote{To include flavour symmetries, one considers $\Sigma$ as an affine curve with poles for $\l$~\protect\cite{Seiberg:1994aj}.} 
hence we denote the electro-magnetic charge of a BPS state by $\g$ as well. 
The projection of the calibrated one-cycle $\g$ onto $\CC$ satisfies the flow equation:
\begin{align}
    \sqrt{\phi} |_\g
    =
    e^{i\theta} dt~,
    \label{eq: flow eq}
\end{align}
where $t\in \mathbb{R}$ parametrises $\g$, and $\theta$ is a constant phase aligned with the central charge $Z_\g$ of the would-be BPS particle as $\theta = \text{arg}(Z_\g)$. The sign ambiguity in the square-root appearing in~\eqref{eq: flow eq} corresponds to choosing either the BPS or the anti-BPS particle with charge $\gamma$ or $-\gamma$, respectively. 

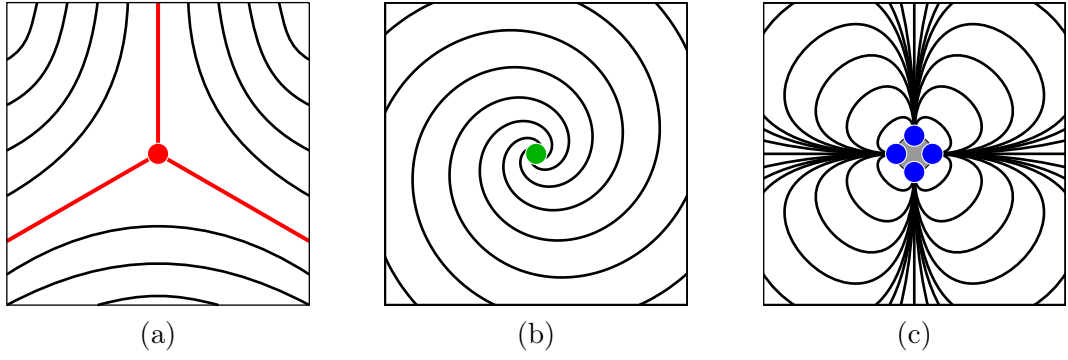
\begin{figure}[tbp]
\centering
\begin{tikzpicture}[scale=2, rotate=0]
\usetikzlibrary{calc}

    \coordinate (A0) at (0,0);
    \coordinate (A1) at (0,1);
    \coordinate (A2) at (-1,{-tan(30)});
    \coordinate (A3) at (1,{-tan(30)});

    \coordinate (P1) at (0.2,1);
    \coordinate (P2) at (0.4,1);
    \coordinate (P3) at (0.6,1);
    \coordinate (P4) at (0.8,1);
    \coordinate (P1') at (1,{-tan(30)+0.3});
    \coordinate (P2') at (1,{-tan(30)+0.6});
    \coordinate (P3') at (1,{-tan(30)+0.9});
    \coordinate (P4') at (1,{-tan(30)+1.2});

    \coordinate (Q1) at (-0.2,1);
    \coordinate (Q2) at (-0.4,1);
    \coordinate (Q3) at (-0.6,1);
    \coordinate (Q4) at (-0.8,1);
    \coordinate (Q1') at (-1,{-tan(30)+0.3});
    \coordinate (Q2') at (-1,{-tan(30)+0.6});
    \coordinate (Q3') at (-1,{-tan(30)+0.9});
    \coordinate (Q4') at (-1,{-tan(30)+1.2});

    \coordinate (R1) at (-1,{-tan(30)-0.22});
    \coordinate (R2) at (-1,-1);
    \coordinate (R3) at (-0.4,-1);
    \coordinate (R1') at (1,{-tan(30)-0.22});
    \coordinate (R2') at (1,-1);
    \coordinate (R3') at (0.4,-1);

    \draw[red, line width=1.5pt] (A0) -- (A1);
    \draw[red, line width=1.5pt] (A0) -- (A2);
    \draw[red, line width=1.5pt] (A0) -- (A3);
    \filldraw[red, draw=white] (A0) circle (0.07);

    \draw[line width=1pt] (P1) to[out=-88, in=150] (P1');
    \draw[line width=1pt] (P2) to[out=-83, in=153] (P2');
    \draw[line width=1pt] (P3) to[out=-80, in=153] (P3');
    \draw[line width=1pt] (P4) to[out=-77, in=150] (P4');

    \draw[line width=1pt] (Q1) to[out=180+88, in=180-150] (Q1');
    \draw[line width=1pt] (Q2) to[out=180+83, in=180-153] (Q2');
    \draw[line width=1pt] (Q3) to[out=180+80, in=180-153] (Q3');
    \draw[line width=1pt] (Q4) to[out=180+77, in=180-150] (Q4');

    \draw[line width=1pt] (R1) to[out=33, in=147] (R1');
    \draw[line width=1pt] (R2) to[out=28, in=152] (R2');
    \draw[line width=1pt] (R3) to[out=15, in=165] (R3');

    \draw[line width=0.5 pt] (1,1) -- (-1,1) -- (-1,-1) -- (1,-1) -- (1,1);
    
    \begin{scope}[shift={(2.5,0)}]
    \clip (-1,-1) rectangle (1,1);
    \draw[black, line width=1pt, domain=0:20, samples=200, smooth] 
    plot ({0.009*(\x*\x)*cos(\x r)}, {0.009*(\x*\x)*sin(\x r)});
    \draw[black, line width=1pt, domain=0:20, samples=200, smooth] 
    plot ({0.009*(\x*\x)*cos(\x r+90)}, {0.009*(\x*\x)*sin(\x r+90)});
    \draw[black, line width=1pt, domain=0:20, samples=200, smooth] 
    plot ({0.009*(\x*\x)*cos(\x r+180)}, {0.009*(\x*\x)*sin(\x r+180)});
    \draw[black, line width=1pt, domain=0:20, samples=200, smooth] 
    plot ({0.009*(\x*\x)*cos(\x r+270)}, {0.009*(\x*\x)*sin(\x r+270)});
    \filldraw[green!70!black, draw=white] (0,0) circle (0.07);
    \draw[line width=1 pt] (1,1) -- (-1,1) -- (-1,-1) -- (1,-1) -- (1,1);
    \end{scope}

    \begin{scope}[shift={(5,0)}]
        \clip (-1,-1) rectangle (1,1);

    \coordinate (I0) at (0,0);
    \coordinate (I1) at (0,0.12);
    \coordinate (I2) at (-0.12,0);
    \coordinate (I3) at (0,-0.12);
    \coordinate (I4) at (0.12,0);

    \draw[line width=1 pt] (1,1) -- (-1,1) -- (-1,-1) -- (1,-1) -- (1,1);
    
    \draw[line width=1pt] (I1) to[out=90, in=0,looseness=60] (I4);
    \draw[line width=1pt] (I1) to[out=90, in=0,looseness=41] (I4);
    \draw[line width=1pt] (I1) to[out=90, in=0,looseness=34] (I4);
    \draw[line width=1pt] (I1) to[out=90, in=0,looseness=26] (I4);
    \draw[line width=1pt] (I1) to[out=90, in=0,looseness=19] (I4);
    \draw[line width=1pt] (I1) to[out=90, in=0,looseness=12] (I4);
    \draw[line width=1pt] (I1) to[out=90, in=0,looseness=5] (I4);
    \draw[line width=1pt] (I1) to (0,1.5);

    \draw[line width=1pt] (I4) to[out=0, in=-90,looseness=60] (I3);
    \draw[line width=1pt] (I4) to[out=0, in=-90,looseness=41] (I3);
    \draw[line width=1pt] (I4) to[out=0, in=-90,looseness=34] (I3);
    \draw[line width=1pt] (I4) to[out=0, in=-90,looseness=26] (I3);
    \draw[line width=1pt] (I4) to[out=0, in=-90,looseness=19] (I3);
    \draw[line width=1pt] (I4) to[out=0, in=-90,looseness=12] (I3);
    \draw[line width=1pt] (I4) to[out=0, in=-90,looseness=5] (I3);
    \draw[line width=1pt] (I4) to (1.5,0);

    \draw[line width=1pt] (I3) to[out=-90, in=180,looseness=60] (I2);
    \draw[line width=1pt] (I3) to[out=-90, in=180,looseness=41] (I2);
    \draw[line width=1pt] (I3) to[out=-90, in=180,looseness=34] (I2);
    \draw[line width=1pt] (I3) to[out=-90, in=180,looseness=26] (I2);
    \draw[line width=1pt] (I3) to[out=-90, in=180,looseness=19] (I2);
    \draw[line width=1pt] (I3) to[out=-90, in=180,looseness=12] (I2);
    \draw[line width=1pt] (I3) to[out=-90, in=180,looseness=5] (I2);
    \draw[line width=1pt] (I3) to (0,-1.5);

    \draw[line width=1pt] (I2) to[out=180, in=90,looseness=60] (I1);
    \draw[line width=1pt] (I2) to[out=180, in=90,looseness=41] (I1);
    \draw[line width=1pt] (I2) to[out=180, in=90,looseness=34] (I1);
    \draw[line width=1pt] (I2) to[out=180, in=90,looseness=26] (I1);
    \draw[line width=1pt] (I2) to[out=180, in=90,looseness=19] (I1);
    \draw[line width=1pt] (I2) to[out=180, in=90,looseness=12] (I1);
    \draw[line width=1pt] (I2) to[out=180, in=90,looseness=5] (I1);
    \draw[line width=1pt] (I2) to (-1.5,0);

    \filldraw[gray!80!white, draw=black] (I0) circle (0.12);
    \filldraw[blue, draw=white] (I1) circle (0.07);
    \filldraw[blue, draw=white] (I2) circle (0.07);
    \filldraw[blue, draw=white] (I3) circle (0.07);
    \filldraw[blue, draw=white] (I4) circle (0.07);
    \end{scope}

    \node at (0,-1.2) {(a)};
    \node at (2.5,-1.2) {(b)};
    \node at (5,-1.2) {(c)};
\end{tikzpicture}
\vspace{-10pt}
\caption{\label{fig: localflow} Local behaviors of flow diagram nearby (a): a zero of $\phi$; (b): a regular puncture;  (c): an irregular puncture with $n_i=4$. In (a), the red dot denotes a zero with three red trajectories emanating from it. In (b), a regular puncture is denoted by a green dot where the logarithmic spiral flow lines terminate. In (c), a shaded infinitesimal disc has been cut out, resulting in an ideal circle boundary carrying $n_i$ marked points colored in blue. 
}
\end{figure}

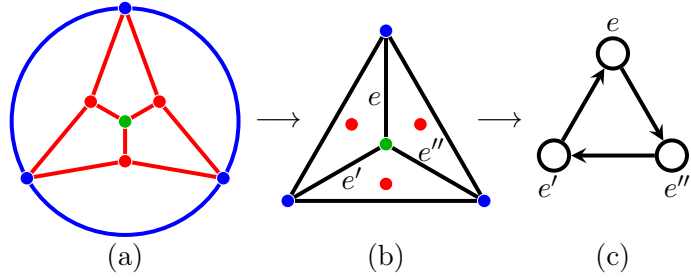
\begin{figure}[tbp]
\centering
\begin{tikzpicture}[scale=1.5, rotate=0,>=stealth]
\usetikzlibrary{calc}

    \coordinate (R0) at (0,0);
    \coordinate (Q1) at ({0.35*cos(150)},{0.35*sin(150)});
    \coordinate (Q2) at ({0.35*cos(30)},{0.35*sin(30)});
    \coordinate (Q3) at ({0.35*cos(-90)},{0.35*sin(-90)});
    \coordinate (P1) at (0,1);
    \coordinate (P2) at ({cos(210)},{sin(210)});
    \coordinate (P3) at ({cos(-30)},{sin(-30)});

    \draw[blue!100!black, line width=1.5pt] (0,0) circle (1);
    \draw[red, line width=1.5pt] (P1) -- (Q1);
    \draw[red, line width=1.5pt] (P1) -- (Q2);
    \draw[red, line width=1.5pt] (P2) -- (Q1);
    \draw[red, line width=1.5pt] (P2) -- (Q3);
    \draw[red, line width=1.5pt] (P3) -- (Q2);
    \draw[red, line width=1.5pt] (P3) -- (Q3);
    \draw[red, line width=1.5pt] (R0) -- (Q1);
    \draw[red, line width=1.5pt] (R0) -- (Q2);
    \draw[red, line width=1.5pt] (R0) -- (Q3);

    \filldraw[fill=green!70!black, draw=white] (0,0) circle (0.06);
    \filldraw[red, draw=white] (Q1) circle (0.06);
    \filldraw[red, draw=white] (Q2) circle (0.06);
    \filldraw[red, draw=white] (Q3) circle (0.06);
    \filldraw[blue, draw=white] (P1) circle (0.06);
    \filldraw[blue, draw=white] (P2) circle (0.06);
    \filldraw[blue, draw=white] (P3) circle (0.06);

    \coordinate (R0') at (0+2.3,0-0.2);
    \coordinate (Q1') at ({0.35*cos(150)+2.3},{0.35*sin(150)-0.2});
    \coordinate (Q2') at ({0.35*cos(30)+2.3},{0.35*sin(30)-0.2});
    \coordinate (Q3') at ({0.35*cos(-90)+2.3},{0.35*sin(-90)-0.2});
    \coordinate (P1') at (0+2.3,1-0.2);
    \coordinate (P2') at ({cos(210)+2.3},{sin(210)-0.2});
    \coordinate (P3') at ({cos(-30)+2.3},{sin(-30)-0.2});

    \draw[black, line width=1.5pt] (R0') -- (P1');
    \draw[black, line width=1.5pt] (R0') -- (P2');
    \draw[black, line width=1.5pt] (R0') -- (P3');
    \draw[black, line width=1.5pt] (P1') -- (P2');
    \draw[black, line width=1.5pt] (P2') -- (P3');
    \draw[black, line width=1.5pt] (P3') -- (P1'); 

    \filldraw[fill=green!70!black, draw=white] (R0') circle (0.06);
    \filldraw[red, draw=white] (Q1') circle (0.06);
    \filldraw[red, draw=white] (Q2') circle (0.06);
    \filldraw[red, draw=white] (Q3') circle (0.06);
    \filldraw[blue, draw=white] (P1') circle (0.06);
    \filldraw[blue, draw=white] (P2') circle (0.06);
    \filldraw[blue, draw=white] (P3') circle (0.06);

    \node at (0+2.2,0.4-0.2) {$e$};
    \node at (-0.3+2.3,-0.25-0.25) {$e'$};
    \node at (0.4+2.3,-0.1-0.15) {$e''$};

    \coordinate (B1) at ({0.6*cos(90)+4.3},{0.6*sin(90)});
    \coordinate (B2) at ({0.6*cos(210)+4.3},{0.6*sin(210)});
    \coordinate (B3) at ({0.6*cos(-30)+4.3},{0.6*sin(-30)});

    \node[circle, draw, minimum size=4mm, line width=1.5pt] (g1) at (B1) {};
    \node[circle, draw, minimum size=4mm, line width=1.5pt] (g2) at (B2) {};
    \node[circle, draw, minimum size=4mm, line width=1.5pt] (g3) at (B3) {};
    
    \draw[->, line width=1.5pt] (g1) -- (g3);
    \draw[->, line width=1.5pt] (g3) -- (g2);
    \draw[->, line width=1.5pt] (g2) -- (g1);

    \node at ({0.6*cos(90)+4.3},{0.6*sin(90)+0.25}) {$e$};
    \node at ({0.6*cos(210)-0.05+4.3},{0.6*sin(210)-0.27}) {$e'$};
    \node at ({0.6*cos(-30)+0.05+4.3},{0.6*sin(-30)-0.27}) {$e''$};
    
    \node at (1.35,0) {$\longrightarrow$};
    \node at (3.3,0) {$\longrightarrow$};
    \node at (0,-1.2) {(a)};
    \node at (2.3,-1.2) {(b)};
    \node at (4.3,-1.2) {(c)}; 
\end{tikzpicture}
\vspace{-10pt}
\caption{\label{fig: flowquiver} An example of flow diagram (a), corresponding ideal triangulation (b), and the associated quiver (c). Each red dot corresponds to zero of $\phi$, while green and blue dot correspond to a marked point from regular and irregular punctures, respectively. An ideal circle boundary from the irregular puncture is colored in blue in (a). In (b), each edge is a representative of each one-parameter family of flows, i.e., each divided chamber in (a). Also, each triangle in (b) contains exactly one zero. By labeling three internal edges in (b) as $(e,e',e'')$, a quiver can be obtained as shown in (c). }
\end{figure}

Each zero of $\phi(z)$ sits at the intersection of three solutions to~\eqref{eq: flow eq} which divide the local neighbourhood into three regions, with families of flow lines as depicted in figure~\ref{fig: localflow}(a).
Meanwhile, the behaviour near a puncture at $p_i$ depends on the non-negative integer $n_i$. A {\it regular} puncture has $n_i=0$, in which case the local behavior give rise to a logarithmic spiral trajectory as shown in~figure \ref{fig: localflow}(b). At an {\it irregular} puncture with $n_i > 0$, the local behavior displays Stokes phenomena as pictured in figure~\ref{fig: localflow}(c). One typically cuts out an infinitesimal disc around the irregular puncture, resulting in an ideal circle boundary with $n_i$ {\it marked points} at which the $n_i$ Stokes rays end. 
By putting together these local behaviors for a generic BPS angle $\theta$, the global structure of the flow diagram on $\CC$ consists of two types of trajectories:
\begin{itemize}
    \item[1.] A {\it separating trajectory} connects a zero of $\phi$ to a regular puncture or to one of the $n_i$ marked points associated to an irregular puncture.
    
    \item[2.] A {\it generic trajectory} connects two endpoints from either regular punctures or marked points of irregular punctures. These trajectories always form one-parameter families. 

\end{itemize}
An example of this behaviour is shown in figure~\ref{fig: flowquiver}(a). 
Surprisingly, this global flow diagram naturally provides an ideal triangulation of the Gaiotto curve, which is constructed by drawing a single representative trajectory for each one-parameter family of generic trajectories, as shown in figure~\ref{fig: flowquiver}(b). Each triangle always encloses a single zero of $\phi$.

\medskip
\noindent
{\bf BPS quiver from the triangulation.} 
One can associate a {\it quiver} to any given ideal triangulation of a Riemann surface~\cite{Fomin:2007rcq}. The procedure can be summarised as follows:
\begin{enumerate}
    \item Assign a node to each internal edge $E$ in the triangulation that does not lie on the boundary.

    \item For each pair of internal edges $(e,e')$, find all the triangles that have both $e$ and $e'$ as their edges.

    \item For each such triangle, draw an arrow from the quiver node associated with $e$ to the node associated with $e'$ if $e$ directly precedes  $e'$ when going around the triangle in the counter-clockwise direction. Otherwise, draw an arrow in the opposite direction.
    
\end{enumerate}
See~figure \ref{fig: flowquiver}c for an example. Remarkably, the extracted quiver coincides with the {\it BPS quiver} of the class-$\CS$ theory $T[\CC]$. We refer to~\cite{Alim:2011ae} for an explanation of this fact from the perspective of geometric engineering in type IIB string theory on a Calabi-Yau threefold.

Importantly, as we continuously vary the BPS angle $\theta$ in~\eqref{eq: flow eq}, the flow diagram changes accordingly. In particular, as $\theta$ passes a critical value $\theta = \theta_c$ at which a BPS state exists --- that is such that $\theta_c= {\rm arg}(Z_{\gamma})$ for some BPS (anti-)particle of charge $\gamma$ --- the flow diagram discontinuously jumps, and so does the corresponding triangulation. This jump is realised as a {\it flip} of an internal edge of the triangulation, as displayed in figure~\ref{fig: flip} in the main text. In the BPS quiver, this flip corresponds to a {\it mutation} at the corresponding node. This is key to constructing the three-manifold corresponding to a twisted compactification of the 4d $\CN=2$ SCFT, as we explain in the main text.

\medskip
\noindent {\bf Mutations.} Recall that a mutation of the BPS quiver at the $i$-th node flips the direction of the arrows connected to that node.%
\footnote{It also introduces the `meson' arrows and a superpotential term. In this paper, we only consider very simple BPS quivers without superpotential.} 
The mutation also acts on the charges $\gamma_j$. There are two types of mutations to consider, which are inverse of each other. The {\it right mutation} $\mu^+_i$ of a BPS quiver at the $i$-th node with assigned charge $\g_i$ transforms the charge basis $\{\g_j\}$ into new basis $\{\tilde{\g}_j\}$ according to:
\be
    \tilde{\g}_j
    =
    \mu_i^+ (\g_j)
    =
    \Big\{
    \begin{array}{cc}
            (-1)^{\delta_{ij}} \, \g_j &\;\; \text{if} \;\; \langle \g_j,\g_i \rangle \leq 0~,\\
        \g_j + \langle \g_j,\g_i \rangle\, \g_i & \;\;\text{if}\;\; \langle \g_j,\g_i \rangle > 0~,
    \end{array}
    \label{eq: mutation}
\ee
where the Dirac pairing $\langle \g_j,\g_i \rangle$ gives us the net number of quiver arrows from $i$ to $j$ (in particular, $\langle \g_j,\g_i \rangle =-n<0$ means that there are $n$ arrows from $j$ to $i$)~\cite{Alim:2011ae}. Similarly, a {\it left mutation} $\mu^-_i$ at the $i$-th node acts on the charges as
\be
    \tilde{\g}_j
    =
    \mu_i^- (\g_j)
    =
    \Big\{
    \begin{array}{cc}
            (-1)^{\delta_{ij}} \, \g_j &\;\; \text{if} \;\; \langle \g_j,\g_i \rangle \geq 0~,\\
        \g_j - \langle \g_j,\g_i \rangle\, \g_i & \;\;\text{if}\;\; \langle \g_j,\g_i \rangle < 0~.
    \end{array}
    \label{eq: mutation bis}
\ee
Note that the right mutation shifts the charges $\gamma_j$ of the nodes $j$ connected to $i$ by outgoing arrows $i\rightarrow j$, while the left mutation shifts the charges of the nodes $j$ connected to $i$ by incoming arrows $j\rightarrow i$. The stability parameters $\zeta_i$ (that is, the Fayet--Iliopoulos parameters of the 1d $\CN=4$ supersymmetric quantum mechanics) determine the mutation; here, $\zeta_i>0$ or $\zeta_i<0$ implies that we should perform a right mutation or a left mutation, respectively. When applying the mutation method as sketched in figure~\ref{fig: cone mutation} in the main text, a clockwise rotation corresponds to a right mutation. 


\section{Galois orbits of the \texorpdfstring{$S$}{S}-matrix of \texorpdfstring{$M(2,2n+3)$}{M(2,2n+3)}}\label{app:galois}

In this appendix, we further explore the Galois orbits of modular tensor categories (MTC) related to the Virasoro minimal model $M(2,2n+3)$; we refer to~{\it e.g.}~\cite{Coste:1993af, DiFrancesco:1997nk, bakalov2001lectures} for general background on this subject. 

\subsection{Galois conjugates from characteristic polynomial}
By definition, Galois conjugate MTCs share the same fusion algebra $\CA$ of topological lines $\mathscr{L}_\alpha$ indexed by $\alpha=0, \cdots, n$,
\be
\mathscr{L}_\alpha \mathscr{L}_\beta = \sum_{\gamma} {\CN_{\a\b}}^\gamma \mathscr{L}_\gamma~,
\ee
which is also the algebra of chiral primaries of the boundary VOA. The famous Verlinde formula
\be\label{verlinde}
{\CN_{\a\b}}^\gamma = \sum_{\delta} {S_{\alpha \delta} S_{\beta\delta} S^{-1}_{\gamma\delta}\ov S_{0\gamma}}
\ee
is the statement that the modular $S$ matrix diagonalises the fusion rules. Let $(\CN_\alpha)_{\beta\gamma}\equiv {\CN_{\a\b}}^\gamma$ denote the fusion matrix of $\SL_\alpha$ with all other lines $\SL_\beta$. We then have $\CN_\alpha= S\Lambda_\alpha S^{-1}$ with the diagonal matrix
\be\label{eigenvalues lambdada}
\Lambda_\alpha =\diag(\lambda_\alpha^{(0)}, \cdots, \lambda_\alpha^{(n)})~, \qquad \qquad \lambda_\alpha^{(\delta)}= {S_{\alpha\delta}\ov S_{0\delta}}
\ee
for any MTC with $n+1$ lines, and where we denote by $\SL_0=1$ the trivial line. It follows from~\eqref{verlinde} that, for any fixed $\delta$, the map
\be\label{CA ring hom}
\CA \longrightarrow \R\; :\; \SL_\alpha \mapsto \lambda_\alpha^{(\delta)}  
\ee
is a ring homomorphism. In particular, the quantity
\be
d_\alpha \equiv \lambda_{\alpha}^{(0)}= {S_{\alpha 0}\ov S_{00}}
\ee
is called the quantum dimension of $\SL_\alpha$.%
\footnote{It is a positive real number for a unitary MTC, while it can be a negative in the more general case (though one always imposes that $d_\alpha \neq 0$).}

The ring homomorphism property~\eqref{CA ring hom} imposes strong constraints on the quantum dimensions. For the minimal model $M(2,2n+3)$, the fusion of the first non-trivial line $\SL_1$ with the other lines takes the form $\SL_1\SL_\alpha = \SL_{\alpha-1}+\SL_{\alpha+1}$ for $\alpha\leq n-1$, together with $\SL_1\SL_n = \SL_{n-1}+\SL_{n}$. This gives us recursion relations on the quantum dimensions:
\be
d_1 d_\alpha = \begin{cases}
    d_{\alpha-1}+ d_{\alpha+1}\qquad & \text{if }\, \alpha \leq n-1~,\\
    d_{n-1}+ d_n& \text{if }\, \alpha=n~.
\end{cases}
\ee
Thus we have $d_0=1$ and all other $d_\alpha$ for $\alpha>1$ written in terms of $d_1$:
\be
  d_2= d_1^2-1~, \quad d_3= d_1^3-2 d_1~, \quad \text{etc.} 
\ee
This recursion relation is isomorphic to the one defining the Chebyshev polynomials of the second kind, $U_\alpha(x)$ with $2x=d_1$, therefore we find:
\be\label{dalpha formula}
d_{\alpha}= U_\alpha\big({d_1\ov 2}\big)~, \qquad \alpha=0~, \cdots~, n~.
\ee
The consistency of the truncation at level $\alpha=n$ simultaneously determines $d_1$ to be a root of a degree-$(n+1)$ polynomial:
\be\label{P for d1}
P(\lambda) = U_{n+1}\big({\lambda\ov 2}\big)-U_{n}\big({\lambda \ov 2}\big)=0~, \qquad d_1= \lambda~.
\ee
Equivalently, this polynomial is the characteristic polynomial of the fusion matrix $\CN_1$. 
Using the properties of the Chebyshev polynomials, one finds the explicit solutions:
\be\label{dalphak sol}
d_\alpha(k)= (-1)^\a \,  {\sin\Big({2\pi k (\alpha+1)\ov 2n+3} \Big)\Big/\sin\Big({2\pi k \ov 2n+3} \Big)}
\ee
which are indexed by some integer $k$. For $k=1$, this reproduce the quantum dimensions of the minimal model $M(2,2n+3)$, which corresponds to choosing $d_1$ to be the smallest (negative) root of~\eqref{P for d1}. The other choices of $k$ correspond to choosing  another root $d_1(k)$ of~\eqref{P for d1} as the quantum dimension of the first non-trivial object, the other $d_\alpha(k)$ being then determined, at fixed $k$, by~\eqref{dalpha formula}.

The formula~\eqref{dalphak sol} agrees with the modular $S$ matrix given in~\eqref{ST VOAk}. Similarly, the other eigenvalues in~\eqref{eigenvalues lambdada} are determined as
\be
\lambda^{(\delta)}_{\alpha}= U_\alpha\big({\lambda^{(\delta)}_1\ov 2}\big)~, \qquad \alpha=0~, \cdots~, n~,
\ee
with $\lambda^{(\delta)}_1$ being the $n+1$ roots of~\eqref{P for d1}. Once we choose $d_1= \lambda_1^{(0)}$, the ordering of the other roots is fixed by the requirement that $S$ be a symmetric matrix. This fully determines $S$ up to an overall prefactor. 
The allowed values of $k$ are such that $\gcd(k,N)=1$ with $N\equiv 2n+3$. These values (mod $N$) index elements of the Galois group
\be
{\rm Gal}(\Q(\zeta_{N})/\Q) \cong \Z_N^\times~,
\ee
where $\zeta_N\equiv e^{2\pi i \ov N}$ is a primitive $N$-th root of unity, and with the Galois action corresponding to $\zeta_N \rightarrow \zeta_N^k$. The complex conjugation $k\rightarrow -k$ corresponds to complex-conjugate MTCs. We directly see from~\eqref{dalphak sol} that choosing $k$ not coprime to $N$ corresponds to a root of~\eqref{P for d1} such that some of the $d_\alpha$'s vanish, which is not allowed in a MTC.

\medskip\noindent
{\bf Example $n=1$.} For $M(2,5)$, $N=5$ is a prime hence we have 4 distinct MTCs in the Galois orbit, or two up to complex conjugation. Indeed, the fusion algebra is the Fibonacci fusion
\be
\SL_1 \SL_1=\SL_0 + \SL_1~, 
\ee
and the polynomial~\eqref{P for d1} is $P(d_1)= d_1^2-d_1-1$, which has  two roots:
\be
d_1(k) = \begin{cases}
    {1-\sqrt{5}\ov 2}\qquad & \text{for $k=\pm 1$ mod $5$,}\\
    {1+\sqrt{5}\ov 2}\qquad & \text{for $k=\pm 2$ mod $5$,}\\
\end{cases} 
\ee
corresponding to the Yang-Lee (YL) and to the Fibonacci (Fib), respectively. More precisely, we have
\be
M(2,5)^{(k)} \cong \begin{cases}
 \text{YL}\quad &\text{for $k=1$ mod $5$,}\\
 \overline{\text{Fib}}\quad &\text{for $k=2$ mod $5$,}\\
 \text{Fib}\quad &\text{for $k=3$ mod $5$,}\\
 \overline{\text{YL}}\quad &\text{for $k=4$ mod $5$,}\\
\end{cases}
\ee
up to tensoring with an invertible (rank-1) MTC. 

\medskip\noindent
{\bf Example $n=2$.} For $N=7$ with Galois group $\Z_7^\times$, we have 3 pairs of complex-conjugate Galois conjugate MTCs with three lines and with the same fusion as $M(2,7)$. Indeed, we now have
\be
P(d_1)= d_1^3-d_1^2-2 d_1+1~,
\ee
which has three distinct roots, corresponding to the three possibilities. We then have $d_2= d_1^2-1$.

\medskip\noindent
{\bf Example $n=3$.}  $N=9$ is not prime and the Galois group $\Z_9^\times$ corresponds to $k=1,2,4,5,7,8$, giving us three Galois conjugates up to complex conjugation. We have
\be
P(d_1)= (d_1-1) \left(d_1^3-3 d_1-1\right)~, \qquad d_2= d_1^2-1~, \qquad d_3= d_1^3-2d_1~,
\ee
where the root $d_1=1$ is excluded since it would imply $d_2=0$. One can check that the solutions $d_1(k=1,2,4)$ correspond to the other three roots.

\subsection{Galois permutation matrix and cyclotomic fields}\label{app:cyclotimic}

From the explicit expression~\eqref{ST VOAk} for $S$ at fixed $k$, namely
\be\label{S ab appendix}
S_{\a\b}^{(k)} =  \genfrac{(}{)}{}{}{k}{2n+3}  {2 (-1)^{n k+ \a+\b}\ov \sqrt{2n+3}} \sin\left({2\pi k (\a+1)(\b+1) \ov 2n+3}\right)~,
\ee
we can use the orthogonality properties of the sine function to derive the Galois permutation matrix appearing in~\eqref{Gk appearing}. In~\eqref{S ab appendix} and in what follows, $\genfrac{(}{)}{}{}{k}{N}\in \{\pm 1\}$ denotes the Jacobi symbol. One finds:
\be\label{Gk explicit}
(G_{(k)})_{\alpha \beta} 
=  \genfrac{(}{)}{}{}{k}{2n+3} (-1)^{n(k-1)+\alpha+\beta} \varepsilon_k(\alpha) \, \delta_{\sigma_k(\alpha), \beta}~,
\ee
where the permutation $\sigma_k$ and additional signs $\varepsilon_k(\alpha)$ are defined as follows. Let us define
\be
k[\alpha] \equiv k(\alpha+1) \; \; \text{mod } 2n+3~.
\ee
Then, we have
\be
\sigma_k(\alpha) = \begin{cases}
    k[\alpha]-1 \quad &\text{if }\, 1\leq k[\alpha] \leq n+1~,\\
    2n+2-k[\alpha] \quad &\text{if }\, n+2\leq k[\alpha] \leq 2n+2~,\\
\end{cases}
\ee
and the signs
\be
\varepsilon_k(\alpha) = \begin{cases}
    1 \quad &\text{if }\, 1\leq k[\alpha] \leq n+1~,\\
    -1 \quad &\text{if }\, n+2\leq k[\alpha] \leq 2n+2~.\\
\end{cases}
\ee
Finally, let us recall that the Galois action on the $S$ matrix is best explained by realising that~\eqref{S ab appendix} is valued in the cyclotomic field $\Q(\zeta_{4N})$ with $N=2n+3$~\cite{Coste:1993af}. Indeed, denoting by $\zeta_M= e^{2\pi i /M}$ the $M$-th root of unity, we have:
\be
S_{\a\b}^{(1)} = -i \sqrt{N} {(-1)^{n+ \a+\b}\ov N} \left(\zeta_N^{(\a+1)(\b+1)}-\zeta_N^{-(\a+1)(\b+1)}\right)
\ee
for the $S$-matrix of the minimal model $M(2,N)$, which almost looks cyclotomic except for the prefactor $i\sqrt{N}$. Now, it is a fun number-theoretic fact that $i\sqrt{N} \in \Q(\zeta_{4N})$ for $N=2n+3$, and furthermore $i\sqrt{N}\in \Q(\zeta_N)$ if $n$ is even. Let us consider the prime decomposition of $N$ as
\be
N= \prod_p p^{a_p} = \tilde{N}^2 N_{\rm sf}~, \qquad \tilde{N}= \prod_p p^{b_p}~, \quad N_{\rm sf}= \prod_p p^{\delta_p}~, 
\ee
where $N_{\rm sf}$ is the square-free part of $N$, with $a_p= 2b_p+ \delta_p \in \Z_{\geq 0}$ and $\delta_p\in \{0,1\}$ for every prime number $p$. The key identity involves the Jacobi symbols:
\be
\sum_{j=1}^{N_{\rm sf}-1} \genfrac{(}{)}{}{}{j}{N_{\rm sf}} \zeta_{N_{\rm sf}}^j= \kappa_n \sqrt{N_{\rm sf}}~, \qquad \kappa_n = \begin{cases}
    1 \quad &\text{if $n$ is odd,}\\
    i \quad &\text{if $n$ is even.}
\end{cases}
\ee
Hence $i\sqrt{N} \in \Q(\zeta_{N_{\rm sf}}) \subseteq  \Q(\zeta_{N})$ if $n$ is even. We can then consider the Galois action
\be\label{gk action}
g_k \; :\; \zeta_N \mapsto g_k(\zeta_N)= \zeta_N^k~,
\ee
for any $k$ coprime to $N$, which gives us:
\be
g_k(i \sqrt{N})=\tilde{N} g_k(i \sqrt{N_{\rm sf}})=
\genfrac{(}{)}{}{}{k}{N_{\rm sf}} i \sqrt{N} =\genfrac{(}{)}{}{}{k}{N} i \sqrt{N}~,
\ee
which explains the Jacobi symbol appearing in~\eqref{S ab appendix}. We then find $S^{(1)}\in \Q(\zeta_N)$ and 
\be\label{gk on S}
g_k(S^{(1)})= S^{(k)}~.
\ee
For $n$ odd, we have to deal with the additional factor of $i=e^{2\pi i\ov 4}\in \Q(\zeta_4)$, which means that the full $S$ matrix is valued in the composite cyclotomic field $\Q(\zeta_{4N})= \Q(\zeta_N, \zeta_4)$. The Galois action remains~\eqref{gk action} together with
\be\label{gk on i}
g_k(i) \rightarrow (-1)^{k-1} i~,
\ee
which is a choice of lift of the $g_k$ action to $\Q(\zeta_{4N})$. Given this choice, the action~\eqref{gk on S}  precisely reproduces the formula~\eqref{S ab appendix}. The ambiguity in choosing~\eqref{gk on i} was not fixed by considering $S$ by itself, but by imposing the full modular group relations after having fixed the $T$ matrix.

\bibliographystyle{JHEP}
\bibliography{ref}
\end{document}